\documentclass[journal,comsoc]{IEEEtran}
\usepackage[T1]{fontenc}

\usepackage[dvips]{graphicx}
\usepackage{array}    
\usepackage{amssymb,amsthm, mathtools}%
\usepackage{amsmath}

\usepackage{float}
\usepackage{subfig}
\usepackage[english]{babel}
\usepackage{import}
\usepackage[table, dvipsnames]{xcolor}
\usepackage{xcolor}
\usepackage{tikz}
\usepackage{pgfplots}
\usetikzlibrary{shapes.multipart}

\usetikzlibrary{patterns,calc,fit,arrows,chains,positioning}

\newdimen{\Offset}

\pgfplotsset{
	width=0.62\textwidth,
	height=0.45\textwidth,
}

\usepackage{cite}

\newtheoremstyle{mytheoremstyle} 
{\topsep}                    
{0pt}                    
{\itshape}                   
{}                           
{\bfseries}                   
{.}                          
{.5em}                       
{}  

\theoremstyle{mytheoremstyle}

\newtheorem{theorem}{Theorem}
\newtheorem{lemma}{Lemma}

\newtheorem{remark}{Remark}

\usepackage{algorithm}
\usepackage{algpseudocode}

\algsetblockdefx{beginIni}{endIni}{1}{}
{\textbf{Initialize:}}%
{\textbf{Initialize}}

\algsetblockdefx{beginChEst}{endChEst}{2}{}
[2]{#1 \textbf{channel estimation} #2}
[2]{#1 \textbf{channel estimation} #2}

\algsetblockdefx{beginChEstData}{endChEstData}{2}{}
[2]{#1 \textbf{channel estimation} #2}
[2]{#1 \textbf{channel estimation} #2}

\algsetblockdefx{beginLinComb}{endLinComb}{2}{}
{\textbf{Linear combining}}
{\textbf{Linear combining}}

\algsetblockdefx{beginDataDec}{endDataDec}{3}{}
[1]{\textbf{Data decoding} #1}
{\textbf{Data decoding}}

\newcommand{\url}[1]{#1}

\usepackage{picture}
\graphicspath{{./Figures/}}

\def\Bw{B_\textsc{w}}

\def\var{\bb{V}\rm{ar}}

\def\pulrp{\rho^{\textsc{rp}}} 
\def\pulsp{\rho^{\textsc{sp}}}

\def\qulrp{q^{\textsc{rp}}} 
\def\qulsp{q^{\textsc{sp}}}

\def\pTul{\varrho}

\def\Tp{\tau_p}
\def\Tc{\tau_c}
\def\Tul{\tau_c}
\def\Tuld{\tau_d}

\newcommand{\bb}[1]{\mathbb{#1}}
\renewcommand{\bf}[1]{\mathbf{#1}}
\renewcommand{\rm}[1]{\mathrm{#1}}
\renewcommand{\cal}[1]{\mathcal{#1}}
\newcommand{\bfs}[1]{\boldsymbol{#1}}
\renewcommand{\sf}[1]{\mathsf{#1}}

\newcommand{\iter}[2]{{\textsc{#1}\scriptscriptstyle{(#2)}}}

\newcolumntype{L}[1]{>{\raggedright\let\newline\\\arraybackslash\hspace{0pt}}m{#1}}
\newcolumntype{C}[1]{>{\centering\let\newline\\\arraybackslash\hspace{0pt}}m{#1}}
\newcolumntype{R}[1]{>{\raggedleft\let\newline\\\arraybackslash\hspace{0pt}}m{#1}}

\pgfplotsset{xindexrange/.style 2 args={
		x filter/.code={
			\ifnum\coordindex<#1\fi
			\ifnum\coordindex>#2\fi
		}
	}}

\definecolor{mittelblau}{RGB}{0, 126, 198}
\definecolor{hellblau}{RGB}{100, 149, 237}
\definecolor{violettblau}{cmyk}{0.9, 0.6, 0, 0}
\definecolor{rot}{RGB}{238, 28 35}
\definecolor{apfelgruen}{RGB}{140, 198, 62}
\definecolor{gelb}{RGB}{1, 221, 0}
\definecolor{orange}{RGB}{244, 111, 33}
\definecolor{pink}{RGB}{237, 0, 140}
\definecolor{lila}{RGB}{128, 10, 145}
\definecolor{hellgrau}{RGB}{224, 224, 224}
\definecolor{mittelgrau}{RGB}{128, 128, 128}
\definecolor{dunkelgrau}{RGB}{80,80,80}
\definecolor{anthrazit}{RGB}{19, 31, 31}

\begin{document}

\title{Massive-MIMO Iterative Channel Estimation and Decoding (MICED) in the Uplink}

\author{Daniel~Verenzuela,~\IEEEmembership{Student~Member,~IEEE,}
		Emil~Bj\"ornson,~\IEEEmembership{Senior~Member,~IEEE,}
		Xiaojie~Wang,~\IEEEmembership{Student~Member,~IEEE,}
        Maximilian~Arnold,~\IEEEmembership{Student~Member,~IEEE,}
        and~Stephan~ten~Brink,~\IEEEmembership{Senior~Member,~IEEE}
\thanks{D. Verenzuela and E. Bj\"{o}rnson are with the Department
	of Electrical Engineering (ISY), Link\"{o}ping University, Link\"{o}ping, SE-58183 Sweden (e-mail: daniel.verenzuela@liu.se; emil.bjornson@liu.se).}
\thanks{X. Wang, M. Arnold and S. ten Brink are with the Institute of Telecommunications, University of Stuttgart, Stuttgart 70659, Germany (e-mail: xiaojie.wang@inue.uni-stuttgart.de;  maximilian.arnold@inue.uni-stuttgart.de;  tenbrink@inue.uni-stuttgart.de).}
\thanks{This paper has received funding from ELLIIT and the Swedish Foundation for Strategic Research (SSF).}  
\thanks{The simulations were performed on resources provided by Linköping University (LiU) at National Supercomputer Centre (NSC). We thank Peter Kjellström, Mats Kronberg, Peter Münger, and Kent Engström at the NSC for their assistance with technical support.
	}}

\maketitle

\begin{abstract}
	
Massive MIMO uses a large number of antennas to increase the spectral efficiency (SE) through spatial multiplexing of users, which requires accurate channel state information. It is often assumed that regular pilots (RP), where a fraction of the time-frequency resources is reserved for pilots, suffices to provide high SE. However, the SE is limited by the pilot overhead and pilot contamination. An alternative is superimposed pilots (SP) where all resources are used for pilots and data. This removes the pilot overhead and reduces pilot contamination by using longer pilots. However, SP suffers from data interference that reduces the SE gains. This paper proposes the Massive-MIMO Iterative Channel Estimation and Decoding (MICED) algorithm where partially decoded data is used as side-information to improve the channel estimation and increase SE. We show that users with precise data estimates can help users with poor data estimates to decode. Numerical results with QPSK modulation and LDPC codes show that the MICED algorithm increases the SE and reduces the block-error-rate with RP and SP compared to conventional methods. The MICED algorithm with SP delivers the highest SE and it is especially effective in scenarios with short coherence blocks like high mobility or high frequencies.

\end{abstract}

\IEEEpeerreviewmaketitle

\section{Introduction}
\label{sec:introduction}

Next generation wireless networks need to accommodate a large amount of data and number of devices while fulfilling a variety of requirements like high data rates and low energy consumption. Massive MIMO is a multiuser multiple-input multiple-output (MIMO) technology able to serve several user equipments (UEs) on the same time-frequency resources by means of spatial multiplexing in which the base station (BS) utilizes a large number of antennas. This technology has received large attention from both academia and industry for its ability to greatly increase the spectral efficiency (SE) compared to current cellular networks \cite{Hien_bounds,16_Marzetta_MAMIMO_book,17_Bjornson_MAMIMO_book}.
	
To enable the spatial multiplexing of UEs in Massive MIMO, the signals from all BS antennas are processed coherently for which accurate channel state information (CSI) is required. A standard approach for channel estimation, called regular pilots (RP), is to reserve some time-frequency resources for UEs to send known orthogonal signals in the uplink (UL), called pilots. This method can provide channel estimates of sufficient quality at the expense of having a pilot overhead, due to the fact that not all time-frequency resources can be used for data transmission. Typically, the SE has the form $\rm{SE} = \rm{prelog} \times \log_2(1 + \rm{SINR})$ where $\rm{SINR}$ stands for the signal-to-interference plus noise ratio, and $\rm{prelog}$ refers to the fraction of time-frequency resources used for data transmission. Having a pilot overhead means that $\rm{prelog} < 1$, and as more pilots are used the SE decreases linearly with the $\rm{prelog}$. On the other hand, using more pilot symbols results in better channel estimates and thereby higher $\rm{SINR}$. Thus, there is a non-trivial trade-off between channel estimation quality and pilot overhead to obtain the highest SE \cite{Emil_pilot_SE}. 

In a multicell Massive MIMO system, many UEs are expected to be active which means that there are not enough time-frequency resources to assign orthogonal pilot signals to each UE. Thus, some pilots would need to  be reused causing interference in the channel estimation process which, in turn, results in lower coherent MIMO signal gain and the presence of coherent interference that reduces the SE. This phenomenon is known as pilot contamination \cite{16_Marzetta_MAMIMO_book,17_Bjornson_MAMIMO_book}. 

The problem of pilot contamination has been well studied in the Massive MIMO literature resulting in many methods for its mitigation. For instance, (semi) blind channel estimation methods making use of angle domain representation and amplitude interference rejection have been proposed in \cite{Mueller2014b,Yin2016a,Julia_pilot_decont, Hien_EVD_pilot}. Other approaches in \cite{MAMI_low_nr_ant, coord_ch_est,18_Bjornson_unlimited_cap,17_Bjornson_MAMIMO_book}, exploit the structure of the spatial correlation matrices to mitigate pilot contamination. In particular, \cite{18_Bjornson_unlimited_cap} has shown that the capacity of multicell Massive MIMO systems grows without bound with the number of BS antennas if the correlation matrices are known and advanced processing is used. A simpler method to reduce the pilot contamination is to increase the pilot overhead to afford longer pilots that are reused more sparsely in the spatial domain by introducing a pilot reuse factor \cite{T_marzetta_total_EE,T_marzetta_non_asym, Emil_pilot_SE, Emil_pilot_cluster}. 

In the aforementioned methods, the transmission of pilots and data is done on disjoint time-frequency resources. An alternative approach is to send a superposition of pilots and data to support longer pilot sequences and eliminate the pilot overhead. This is called superimposed pilot (SP) transmission \cite{Hoeher99_ch_est_SP, SP_stat_fading_MIMO_2017}. In Massive MIMO, the SP method has been proposed to mitigate the pilot contamination effect by allowing the use of longer pilots that can support sparser pilot reuse \cite{SIP_part1_KU_SA, VT_SP_approx_2016}. However, when pilot and data symbols are superimposed, there is interference from data symbols in the channel estimation process. This interference reduces the coherent gain and creates coherent interference that limits the SE gains of SP over RP \cite{Verenzuela2018a}. 

In summary, RP channel estimation provides sufficiently accurate CSI to obtain high SINR, which translates into high SE in Massive MIMO. However, this comes at the expense of having a pilot overhead, which in turn, limits the maximum achievable SE through the $\rm{prelog}$ factor. On the other hand, SP channel estimation gives comparable SE to that of RP and it is instead limited by the data interference. Thus, a potential way to further improve the SE in Massive MIMO is to use SP with data-aided channel estimation to reduce the data interference.

\subsection{Contributions}
\label{subsec:contributions}

This paper evaluates the potential improvements of data-aided channel estimation in the UL of multicell Massive MIMO systems with RP and SP methods while considering the effect of channel coding and spatially correlated fading among BS antennas.

The idea of using partially decoded data to improve channel estimation with SP has been proposed a couple of decades ago for single antenna systems \cite{Hoeher99_ch_est_SP}. Extensions to single-user point-to-point MIMO systems \cite{Zhu2003a,Khalighi2008a} show an improvement in terms of bit-error-rate (BER) with SP, which translates into higher SE compared to RP since SP removes the pilot overhead. In the case of Massive multiuser MIMO, \cite{SIP_part1_KU_SA,VT_SP_approx_2016} depict that iterative data-aided channel estimation with SP has the potential to increase the SE compared to RP systems. However, this was only shown for uncoded data estimates where the effect of channel coding was not considered. Recently, \cite{Ma2014a} used partially decoded data to improve channel estimation in the UL of a multicell single-input multiple-output (SIMO) system (only one UE served per cell) with RP, while considering i.i.d. Rayleigh fading and maximum ratio (MR) combining (also known as maximal ratio combining \cite{D_Tse_wireless_book}). The results show that the pilot contamination effect can be reduced by means of iterative data-aided channel estimation which in turn reduces the BER. However, in \cite{Ma2014a} the main analysis considers a single UE per cell, disregarding the effect of inter-user interference.\footnote{Here, the term inter-user refers to UEs that are served by the same BS via spatial multiplexing.} An extension to the Massive multiuser MIMO case was done in \cite{Ma2017a_MultiUE} showing that SP outperforms RP in scenarios with high mobility and high number of spatially multiplexed UEs. However, in \cite{Ma2017a_MultiUE} although the paper considers a multi-cell setup, the authors approximate the intercell interference as i.i.d. Gaussian noise, which removes all the structure that intercell interference has and effectively reduces the model to a single-cell setup. Thus, the effect of data-aided channel estimation considering channel coding in a multicell Massive MIMO system remains to be investigated. In addition, theoretical analysis on the impact of spatial correlation, inter-user and intercell interference in the aforementioned system is missing in the literature.

In this article, \textit{data-aided channel estimation} refers to the use of partially decoded bits as side information to increase the channel estimation quality. That is, initial channel estimates from pilots are used to obtain soft data estimates which, in turn, are utilized to revise the channel estimates. This process is done in an iterative form using previous soft data estimates to update the channel estimates. The revised data-aided channel estimates are then used to decode the data symbols to reduce errors.

The Massive-MIMO Iterative Channel Estimation and Decoding (MICED) algorithm is proposed to harvest the benefits of data-aided channel estimation in multicell Massive MIMO systems. To obtain insights into the benefits of the MICED algorithm, closed-form expressions for the error correlation matrices of data-aided channel estimates are computed assuming Gaussian data symbols. These expressions are analyzed to indicate how the mean squared error (MSE) of data-aided channel estimates behaves in terms of the data estimation quality and number of time-frequency resources. Note that in contrast to \cite{Mueller2014b,Yin2016a,Julia_pilot_decont, Hien_EVD_pilot} the MICED algorithm does not rely on asymptotic results, angle domain representations, or separability of power levels between UEs, which are conditions that might not be satisfied in practice. For example, having similar received power levels between UEs in the UL is often desired to mitigate the near-far effect of pathloss and to have signals with a low dynamic range which is important for the use of low-resolution analog-to-digital converters in the BSs \cite{S_Jacobsson_Durisi_ADCs_UL}.

To evaluate the SE, MR and single-cell MMSE (S-MMSE) combining are assumed to assess the differences between maximizing coherent combination and suppressing inter-user interference through linear signal processing. The S-MMSE method is based on \cite[Ch.~4]{17_Bjornson_MAMIMO_book} while considering no CSI exchange among BSs.

The MICED algorithm is also implemented with finite-alphabet modulated symbols indicating how the redundancy of channel coding can be used to obtain estimates of complex modulated data symbols. Finally, numerical analysis with quadrature phase shift keying (QPSK) modulation and low-density parity check (LDPC) codes is used to evaluate the performance of the MICED algorithm compared to conventional pilot-based channel estimation in terms of the block-error-rate (BLER) and the achievable SE.  The results show that the MICED algorithm increases the SE and reduces the BLER compared to pilot-based channel estimation with both RP and SP. The highest SE is found when implementing the MICED algorithm with SP since there is no pilot overhead and the data interference is mitigated by the data-aided channel estimation process. The use of SP with the MICED algorithm increases the SE in scenarios with high mobility and high carrier frequencies. In addition, it allows for aggressive spatial multiplexing that can enable other services such as machine type communications.

\subsection*{Notation}
Bold lower and upper case letters denote column vectors and matrices respectively. The trace, matrix inversion, transpose, conjugate, and conjugate transpose operations are denoted as $\rm{tr}(\cdot)$, $(\cdot)^{-1}$, $(\cdot)^T$, $(\cdot)^*$, $(\cdot)^H$ respectively. The sets of natural, real, and complex numbers are denoted as $\bb{N}$, $\bb{R}$, and $\bb{C}$ respectively. The element on the $i^{th}$ row and $j^{th}$ column of a matrix $\bf{X}$ is denoted as $[\bf{X}]_{ij}$, the $j^{th}$ column of $\bf{X}$ is denoted as $[\bf{X}]_{j}$. The matrix composed of the first $n$ columns of $\bf{X}$ is denoted as $[\bf{X}]_{1:n}$. The $j^{th}$ element of vector $\bf{x}$ is denoted as $[\bf{x}]_j$. The identity matrix of size $N$ is denoted as $\bf{I}_N$. For $\bf{A}$, $\bf{B}\;\in \bb{C}^{N\times N}$, the ordering notations $\bf{A}\succeq\bf{B}$ and $\bf{A}\succ\bf{B}$ indicate that $\bf{A}- \bf{B}$ is a positive semi-definite and definite matrix, respectively.

\section{System Model}
\label{sec:system_model}

Consider the UL of a multicell Massive MIMO system where each base station (BS) has $M$ antennas and serves $K$ single-antenna user equipments (UEs) via spatial multiplexing. The BS serving the UEs in cell $l\in \Phi$ is denoted as $\rm{BS}_l$ where $\Phi$ is a set containing all the cell indices. The UE $k$ in cell $l$ is denoted as $\rm{UE}_{lk}$. A standard block fading channel model is assumed where the channel is considered static over a time period of $T_c$ [s] and frequency-flat within a bandwidth of $B_c$ [Hz] \cite{16_Marzetta_MAMIMO_book}. The total system bandwidth is $\Bw$ and is equally divided between all coherence blocks, such that $\Bw/B_c$ is an integer.\footnote{This can be accomplished, for example, by utilizing orthogonal frequency division multiplexing (OFDM) modulation \cite{16_Marzetta_MAMIMO_book}.} The time-frequency block in which the channel is considered time-invariant and frequency-flat is called a coherence block and it is comprised of $\Tc = T_c B_c$ complex samples. The block fading model assumes that the channel is constant within a coherence block and changes independently from one coherence block to another to account for the effect of frequency selectivity and time variations \cite{16_Marzetta_MAMIMO_book}.   

 The communication channel is modeled as a random variable that has an independent realization in each coherence block. Let $\bf{h}_{llk}\sim \cal{CN}(\bf{0}, \bf{R}_{llk})$ be the channel between $\rm{BS}_l$ and $\rm{UE}_{lk}$ where $\bf{R}_{llk}$ is the spatial correlation matrix and $\beta_{llk} = \rm{tr}(\bf{R}_{llk})/M$ is the average channel gain. The received signals at $\rm{BS}_l$ in the UL is
\begin{equation}
\bf{Y}_l = \sum_{\ell\in \Phi}\sum_{k'=1}^{K} \underbrace{\bf{h}_{l\ell k'}
	\vphantom{K_{K_K}}
	}_{M\times 1} \underbrace{\bf{x}_{\ell k'}^T
	\vphantom{K_{K_K}}
	}_{1 \times \Tul} + \bf{N}\quad\in \bb{C}^{M\times \Tul} 
\label{eq:rec_sig_UL}
\end{equation}
where $\bf{N} =[ \bf{n}_{1} ,\ldots, \bf{n}_{\Tul}]$ is the thermal noise with i.i.d. columns distributed as $\bf{n}_{j}\sim\cal{CN}(\bf{0},\sigma^2\bf{I}_M)$ with $\sigma^2$ being the average noise energy per symbol. The signal transmitted from $\rm{UE}_{lk}$ is
\begin{align*}
\bf{x}_{lk} = \begin{cases}
\left[\sqrt{\qulrp_{lk}}\bfs{\phi}_{lk}^T\quad  \sqrt{\pulrp_{lk}}\bf{s}_{lk}^T\right]^T & \text{ with RP}\\
\sqrt{\qulsp_{lk}}\bfs{\varphi}_{lk} + \sqrt{\pulsp_{lk}}\bf{s}_{lk} & \text{ with SP}
\end{cases}
\end{align*}
where $\qulrp_{lk}$, $\pulrp_{lk}$, $\qulsp_{lk}$, and $\pulsp_{lk}$ are the pilot and data energy per symbol with RP, and SP respectively.  Note that with SP, the available transmission energy per symbol $\pTul_{lk}$ is divided between pilot and data symbols such that $\qulsp_{lk} +  \pulsp_{lk} = \pTul_{lk}$. On the other hand, with RP, the same energy per symbol is used for pilot and data symbols since they are transmitted disjointly. Thus,
\begin{align*}
&\qulrp_{lk} = \pulrp_{lk} = \pTul_{lk} & \text{with RP,}\\
&\qulsp_{lk}=\Delta_{lk} \pTul_{lk},\; \pulsp_{lk} = (1 - \Delta_{lk})\pTul_{lk} & \text{with SP,}
\end{align*}
where $\Delta_{lk} \in [0,1]$ is the proportion of power used for pilots.
\begin{figure}[!t]	
	\centering
\usetikzlibrary{shapes}

\pgfmathsetmacro{\thsp}{0.58}
\pgfmathsetmacro{\tvsp}{0.44}
\pgfmathsetmacro{\th}{0.58}
\pgfmathsetmacro{\tv}{0.44}

\begin{tikzpicture}[line width=1pt]
%


\begin{scope}

\node[draw, fit={({\th},0) ({2.5*\th},{\tv})},fill=white,opacity=1,text opacity=1, inner sep=0pt,style={pattern=grid,pattern color=black!20!white}, label=center:{$\Tuld$}] (1RP00) {};
\node[draw, fit={({\th},{\tv}) ({2.5*\th},{2*\tv})},fill=white,opacity=1,text opacity=1, inner sep=0pt,style={pattern=grid,pattern color=black!20!white}, label=center:{$\Tuld$}] (1RP01) {};
\node[draw, fit={({3.5*\th},{\tv}) ({5*\th},{2*\tv})},fill=white,opacity=1,text opacity=1, inner sep=0pt,style={pattern=grid,pattern color=black!20!white}, label=center:{$\Tuld$}] (1RP11) {};

\node[preaction={fill, red!60!white}, draw, fit={(0,0) ({\th},{\tv})},inner sep=0pt, label=center:{$\Tp$}, style={pattern=grid,pattern color=black!20!white}] (1RP00p) {};
\node[preaction={fill, red!60!white},draw, fit={(0,{\tv}) ({\th},{2*\tv})},inner sep=0pt, label=center:{$\Tp$},style={fill=red,pattern=grid,pattern color=black!20!white}] (1RP01p) {};
\node[preaction={fill, red!60!white},draw, fit={({2.5*\th},{\tv}) ({3.5*\th},{2*\tv})},inner sep=0pt, label=center:{$\Tp$},style={fill=red,pattern=grid,pattern color=black!20!white}] (1RP11p) {};

\node[preaction={fill, gray,opacity=0},draw, fit={({0*\th},{-1*\tv}) ({7*\th},{2*\tv})},inner sep=0pt,label={[shift={({-3.2*\th},{1*\tv})}]south east:{  codeword 1}} ] (RPCW1) {};

\node[draw, fit={({8*\th},0) ({9.5*\th},{\tv})},fill=white,opacity=1,text opacity=1, inner sep=0pt,style={pattern=grid,pattern color=black!20!white}, label=center:{$\Tuld$}] (2RP00) {};
\node[draw, fit={({8*\th},{\tv}) ({9.5*\th},{2*\tv})},fill=white,opacity=1,text opacity=1, inner sep=0pt,style={pattern=grid,pattern color=black!20!white}, label=center:{$\Tuld$}] (2RP01) {};
\node[draw, fit={({10.5*\th},{\tv}) ({12*\th},{2*\tv})},fill=white,opacity=1,text opacity=1, inner sep=0pt,style={pattern=grid,pattern color=black!20!white}, label=center:{$\Tuld$}] (2RP11) {};

\node[preaction={fill, red!60!white}, draw, fit={({7*\th},0) ({8*\th},{\tv})},inner sep=0pt, label=center:{$\Tp$}, style={pattern=grid,pattern color=black!20!white}] (2RP00p) {};
\node[preaction={fill, red!60!white},draw, fit={({7*\th},{\tv}) ({8*\th},{2*\tv})},inner sep=0pt, label=center:{$\Tp$},style={fill=red,pattern=grid,pattern color=black!20!white}] (2RP01p) {};
\node[preaction={fill, red!60!white},draw, fit={({9.5*\th},{\tv}) ({10.5*\th},{2*\tv})},inner sep=0pt, label=center:{$\Tp$},style={fill=red,pattern=grid,pattern color=black!20!white}] (2RP11p) {};

\node[preaction={fill, gray,opacity=0},draw, fit={({7*\th},{-1*\tv}) ({14*\th},{2*\tv})},inner sep=0pt,label={[shift={({-3.2*\th},{1*\tv})}]south east:{  codeword 2}} ] (RPCW2) {};

\end{scope}

\node[fit=(1RP00.south) (1RP00.south), draw = {none},label={[xshift=-10pt, yshift=-18pt]$\vdots$}] {};
\node[fit=(1RP11.south) (1RP11.south), draw = {none},label={[xshift=-16pt, yshift=-22pt]$\ddots$}] {};
\node[fit=(1RP11.south) (1RP11.south), draw = {none},label={[xshift=32pt, yshift=-5pt]$\cdots$}] {};

\node[fit=(2RP00.south) (2RP00.south), draw = {none},label={[xshift=-10pt, yshift=-18pt]$\vdots$}] {};
\node[fit=(2RP11.south) (2RP11.south), draw = {none},label={[xshift=-16pt, yshift=-22pt]$\ddots$}] {};
\node[fit=(2RP11.south) (2RP11.south), draw = {none},label={[xshift=32pt, yshift=-5pt]$\cdots$}] {};

\begin{scope}

\node[preaction={fill, blue!30!white}, draw, fit={(0,{-4*\tvsp}) ({2.5*\thsp},{-3*\tvsp})}, yshift=-10pt,fill=white,opacity=1,text opacity=1, inner sep=0pt,style={pattern=grid,pattern color=black!20!white}, label=center:{$\Tul$}] (1SP00) {};
\node[preaction={fill, blue!30!white},draw, fit={(0,{-3*\tvsp}) ({2.5*\thsp},{-2*\tvsp})}, yshift=-10pt,fill=white,opacity=1,text opacity=1, inner sep=0pt,style={pattern=grid,pattern color=black!20!white}, label=center:{$\Tul$}] (1SP01) {};
\node[preaction={fill, blue!30!white},draw, fit={({2.5*\thsp},{-3*\tvsp}) ({5*\thsp},{-2*\tvsp})}, yshift=-10pt,fill=white,opacity=1,text opacity=1, inner sep=0pt,style={pattern=grid,pattern color=black!20!white}, label=center:{$\Tul$}] (1SP11) {};

\node[preaction={fill, gray,opacity=0},draw, fit={({0*\thsp},{-5*\tvsp}) ({6*\thsp},{-2*\tvsp})},inner sep=0pt, yshift=-10pt,label={[shift={({-3.2*\thsp},{1*\tvsp})}]south east:{ codeword 1}} ] (SPCW1) {};

\node[preaction={fill, blue!30!white}, draw, fit={({6*\thsp},{-4*\tvsp}) ({8.5*\thsp},{-3*\tvsp})}, yshift=-10pt,fill=white,opacity=1,text opacity=1, inner sep=0pt,style={pattern=grid,pattern color=black!20!white}, label=center:{$\Tul$}] (2SP00) {};
\node[preaction={fill, blue!30!white},draw, fit={({6*\thsp},{-3*\tvsp}) ({8.5*\thsp},{-2*\tvsp})}, yshift=-10pt,fill=white,opacity=1,text opacity=1, inner sep=0pt,style={pattern=grid,pattern color=black!20!white}, label=center:{$\Tul$}] (2SP01) {};
\node[preaction={fill, blue!30!white},draw, fit={({8.5*\thsp},{-3*\tvsp}) ({11*\thsp},{-2*\tvsp})}, yshift=-10pt,fill=white,opacity=1,text opacity=1, inner sep=0pt,style={pattern=grid,pattern color=black!20!white}, label=center:{$\Tul$}] (2SP11) {};

\node[preaction={fill, gray,opacity=0},draw, fit={({6*\thsp},{-5*\tvsp}) ({12*\thsp},{-2*\tvsp})}, yshift=-10pt,inner sep=0pt,label={[shift={({-3.2*\thsp},{1*\tvsp})}]south east:{ codeword 2}} ] (SPCW2) {};

\end{scope}

\node[fit=(1SP00.south) (1SP00.south), draw = {none},label={[xshift=0pt, yshift=-18pt]$\vdots$}] {};
\node[fit=(1SP11.south) (1SP11.south), draw = {none},label={[xshift=-8pt, yshift=-18pt]$\ddots$}] {};
\node[fit=(1SP11.south) (1SP11.south), draw = {none},label={[xshift=30pt, yshift=-3pt]$\cdots$}] {};

\node[fit=(2SP00.south) (2SP00.south), draw = {none},label={[xshift=0pt, yshift=-18pt]$\vdots$}] {};
\node[fit=(2SP11.south) (2SP11.south), draw = {none},label={[xshift=-8pt, yshift=-18pt]$\ddots$}] {};
\node[fit=(2SP11.south) (2SP11.south), draw = {none},label={[xshift=30pt, yshift=-3pt]$\cdots$}] {};

\node[fit=(RPCW2.east) (RPCW2.east), draw = {none},label={[xshift=10pt, yshift=-10pt]$\cdots$}] {};
\node[fit=(SPCW2.east) (SPCW2.east), draw = {none},label={[xshift=10pt, yshift=-10pt]$\cdots$}] {};

\node[fit=(RPCW1.north) (RPCW1.north), draw = {none},label={[xshift= -15pt, yshift=0pt]{\parbox{3cm}{RP: {\hspace*{8pt} Pilot}  {\hspace*{10pt} Data}  }} }] {};

\node[fit=(RPCW1.north) (RPCW1.north),xshift=-34pt, yshift=10pt, draw,scale=0.4,regular polygon, regular polygon sides=4,fill,red!60!white](){};	
\node[fit=(RPCW1.north) (RPCW1.north),xshift=1pt, yshift=10pt, draw,scale=0.4,regular polygon, regular polygon sides=4,fill=none,black!20!white](){};	

\node[fit=(SPCW1.north) (SPCW1.north), draw = {none},label={[xshift=-9pt, yshift=0pt]{\parbox{2.8cm}{SP: {\hspace*{8pt} Pilot + data}}} }] {};
\node[fit=(SPCW1.north) (SPCW1.north),xshift=-26pt, yshift=10pt, draw,scale=0.4,regular polygon, regular polygon sides=4,fill,blue!30!white](){};

\end{tikzpicture}
	\caption{Time-frequency resource allocation with RP and SP over codewords of equal length.}
	\label{fig:CohblockRPSP}
\end{figure} 
 The pilot symbols are given by $\bfs{\phi}_{lk}\in \bb{C}^{\Tp}$ and $\bfs{\varphi}_{lk}\in \bb{C}^{\Tul}$ with RP and SP respectively.  The UL data symbols are denoted as $\bf{s}_{lk}\in \bb{C}^{\Tuld}$ with both RP and SP. In the case of RP, the pilot and data symbols are sent disjointly, thus $\Tp \leq \Tul$ UL samples of the coherent block are used for pilots and $\Tuld =\Tul - \Tp$ for UL data symbols. With SP, all UL symbols are used for data and pilots, thus $\Tuld =\Tul$. Figure~\ref{fig:CohblockRPSP} illustrates the allocation of samples in the coherence block corresponding to the complex symbols transmitted in codewords of equal size. Note that since RP has a pilot overhead, it needs more coherence blocks to transmit a full codeword compared to SP. Section~\ref{sec:finiteAlphabet} explains in more detail how the information bits are mapped into the complex symbols transmitted over the channel within coherence blocks. 
 
 In this article, all deterministic quantities (e.g., transmission powers, spatial correlation matrices, etc.) are considered known. Since they are deterministic, the signaling overhead for estimating them is negligible. In practice, they can be estimated by aggregating observations from several coherence blocks\cite{Bjorson2019_MaMIMO_2}.

\subsection*{Pilot-based Channel Estimation}
\label{sec:pilotChEst}

In this section, standard pilot-based channel estimation with RP and SP is described. Consider the channel realizations to be estimated based on the UL pilot symbols. Let $\cal{U}_{\Tp}$ be a set of $\Tp$ mutually orthogonal pilot sequences with elements having unit modulus such that for ${\bfs{\phi}_a, \bfs{\phi}_b \in \cal{U}_{\Tp}}$, $|[\bfs{\phi}_a]_j| = |[\bfs{\phi}_b]_j| = 1\,\forall j\in \{1,\ldots,\Tp\}$,   $\bfs{\phi}_a^H\bfs{\phi}_b = 0$ if $a\neq b$, and $\bfs{\phi}_a^H\bfs{\phi}_b = \Tp$ if $a=b$. The choice of pilot sequences having elements with unit modulus is done to have equal energy per symbols, more information on how to generate such sequences can be found in \cite[Sec.~3.1.1]{17_Bjornson_MAMIMO_book}. The pilot sequence used by $\rm{UE}_{lk}$ is $\bfs{\phi}_{lk} \in \cal{U}_{\Tp}$ and $\bfs{\varphi}_{lk} \in \cal{U}_{\Tul}$ with RP and SP respectively. In a large multicell network, the total number of UEs is larger than the available pilots, which means that these pilots need to be reused among cells. The sets containing all UEs that share the same pilot as $\rm{UE}_{lk}$ (itself included) are defined as
\begin{align}
\cal{P}_{lk}^\textsc{rp} &= \left\{(\ell,k')\,:\, \bfs{\phi}_{\ell k'}^H\bfs{\phi}_{lk} \neq 0  \right\},\quad \text{with RP},\\
\cal{P}_{lk}^\textsc{sp} &= \left\{(\ell,k')\,:\, \bfs{\varphi}_{\ell k'}^H\bfs{\varphi}_{lk} \neq 0  \right\},\quad \text{with SP}.
\end{align}
Note that since with SP the pilots are much longer than with RP, the number of elements in $\cal{P}_{lk}^\textsc{sp}$ is far less than that of $\cal{P}_{lk}^\textsc{rp}$. For example, if $\Tul = 100$ and $\Tp = 10$ there will be ten times less UEs sharing pilots with SP than with RP.

To estimate the channel from $\rm{UE}_{lk}$ at $\rm{BS}_l$, the received UL pilot signal is multiplied with the pilot sequence of $\rm{UE}_{lk}$, which is equivalent to a de-spreading operation, to obtain the observations
\begin{align*}
\bf{z}_{llk}^\textsc{rp} &= [\bf{Y}_l]_{1:\Tp} \frac{\bfs{\phi}_{lk}^*}{\Tp\sqrt{\qulrp_{lk} }}
\\
 &= \bf{h}_{llk} + \underbrace{\sum_{(\ell,k')\in \cal{P}_{lk}^\textsc{rp} \backslash(l,k)}\bf{h}_{l\ell k'}\sqrt{\frac{\qulrp_{\ell k'}}{\qulrp_{lk}}}
\vphantom{\frac{\bar{\bf{n}}}{\sqrt{\Tp \qulrp_{lk}} }}
}_{\textrm{Pilot contamination}}  + \underbrace{\frac{\bar{\bf{n}}}{\sqrt{\Tp \qulrp_{lk}}  }
\vphantom{\sum_{K_K\backslash}^{K^K}}
}_{\textrm{Noise}}
\stepcounter{equation}\tag{\theequation}\label{eq:ch_est_obs_RP}
\\
\bf{z}_{llk}^\textsc{sp} 
&=\bf{Y}_l \frac{\bfs{\varphi}_{lk}^*}{\Tul\sqrt{\qulsp_{lk}}} =  \bf{h}_{llk} + \underbrace{\sum_{(\ell,k')\in \cal{P}_{lk}^\textsc{sp} \backslash(l,k)}\bf{h}_{l\ell k'}\sqrt{\frac{\qulsp_{\ell k'}}{\qulsp_{lk}}}
\vphantom{\sum_{K_K\backslash}^{K^K}}
}_{\textrm{Pilot contamination}} 
\\
&
 +  \underbrace{\sum_{\ell\in \Phi}\sum_{k'=1}^{K} \bf{h}_{l\ell k'} \sqrt{\frac{\pulsp_{\ell k'}}{\qulsp_{lk}}}\frac{\bf{s}_{\ell k'}^T\bfs{\varphi}_{lk}^*}{\Tul}
\vphantom{\frac{\bf{N}\bfs{\varphi}_{lk}^*}{\Tul\sqrt{\qulsp_{lk}}}}
}_{\textrm{Data interference}}+  \underbrace{\frac{\bf{N}\bfs{\varphi}_{lk}^*}{\Tul\sqrt{\qulsp_{lk}}}
\vphantom{\sum_{K_K\backslash}^{K^K}}
}_{\textrm{Noise}},
\stepcounter{equation}\tag{\theequation}\label{eq:ch_est_obs_SP}
\end{align*}  
where $\bar{\bf{n}} = [\bf{N}]_{1:\Tp}\bfs{\phi}_{lk}^*/\sqrt{\Tp} \sim \cal{CN}(\bf{0},\sigma^2\bf{I}_M)$ is the equivalent noise after the de-spreading operation with RP. The linear minimum mean-squared-error (LMMSE) channel estimate of $\bf{h}_{llk}$ is summarized in the following lemma.
\begin{lemma}
	\label{lem:LMMSE_ch_est}
	Based on the observations $\bf{z}_{llk}^\textsc{rp}$ and $\bf{z}_{llk}^\textsc{sp}$, the LMMSE estimates of $\bfs{h}_{llk}$ are
	\begin{align}
		\label{eq:LMMSE_RP}
		\hat{\bf{h}}_{llk}^\textsc{rp} &=   
		\bf{R}_{llk} {\bfs{\Psi}_{llk}^\textsc{rp}}^{-1} \bf{z}_{llk}^\textsc{rp}\\
		\hat{\bf{h}}_{llk}^\textsc{sp} &=
		\bf{R}_{llk} {\bfs{\Psi}_{llk}^\textsc{sp}}^{-1} \bf{z}_{llk}^\textsc{sp}
		\label{eq:LMMSE_SP}
	\end{align}
	where
	\begin{align}
	\label{eq:cov_chObs_pilotRP}
		\bfs{\Psi}_{llk}^\textsc{rp} & =\!\!\underbrace{\sum_{(\ell,k')\in \cal{P}_{lk}^\textsc{rp}}\!\bf{R}_{l\ell k'}\frac{\qulrp_{\ell k'}}{\qulrp_{lk}}
			\vphantom{\sum_{K_K^K}}
			}_{\textrm{Pilot contamination}}   + \underbrace{\frac{\sigma^2}{\qulrp_{lk} \Tp}\bf{I}_M
						\vphantom{\sum_{K_K^K}}
			}_{\textrm{Noise}},\\
		\bfs{\Psi}_{llk}^\textsc{sp} &=\!\! \underbrace{\sum_{(\ell,k')\in \cal{P}_{lk}^\textsc{sp}}\!\bf{R}_{l\ell k'}\frac{\qulsp_{\ell k'}}{\qulrp_{lk}}
			\vphantom{\sum_{K_K^K}}
			}_{\textrm{Pilot contamination}}  +  \frac{1}{\Tul}\left(
			\vphantom{\sum_{K}^K}	\right.
			\underbrace{\sum_{\ell\in \Phi}\sum_{k'=1}^{K} \bf{R}_{l\ell k'}\frac{ \pulsp_{\ell k'}}{\qulrp_{lk}} 
			\vphantom{\sum_{K_K^K}}
			}_{\textrm{Data interference}} +  \underbrace{\frac{\sigma^2}{\qulrp_{lk}}\bf{I}_M
			\vphantom{\sum_{K_K^K}}
			}_{\textrm{Noise}}
			\left. \vphantom{\sum_{K}^K}\right).
		\label{eq:cov_chObs_pilotSP}
	\end{align}
	The channel estimates are uncorrelated to the channel estimation errors which are defined as
	\begin{align}
	\tilde{\bf{h}}_{llk}^\textsc{rp} &= \bf{h}_{llk} - \hat{\bf{h}}_{llk}^\textsc{rp},\\
	\tilde{\bf{h}}_{llk}^\textsc{sp} &= \bf{h}_{llk} - \hat{\bf{h}}_{llk}^\textsc{sp},
	\end{align}
	with error correlation matrices given by
	\begin{align}
	\label{eq:errCorrMatRP}
	{\bf{C}}_{llk}^\textsc{rp} &= \bf{R}_{llk} -  \bf{R}_{llk} {\bfs{\Psi}_{llk}^\textsc{rp}}^{-1} \bf{R}_{llk},\\
	\label{eq:errCorrMatSP}
	{\bf{C}}_{llk}^\textsc{sp} &= \bf{R}_{llk} - \bf{R}_{llk} {\bfs{\Psi}_{llk}^\textsc{sp}}^{-1} \bf{R}_{llk}.
	\end{align}
\end{lemma} 
\begin{IEEEproof}	
It follows from employing standard LMMSE estimation techniques to the problem at hand where the LMMSE estimate of a vector $\bf{h}$ from an observation $\bf{z}$ is given by $\bb{E}\left\{\bf{z}\bf{h}^H\right\}\left(\bb{E}\left\{\bf{z}\bf{z}^H\right\}\right)^{-1}\bf{z}$ (assuming all random variables have zero mean) \cite[Ch.~3]{17_Bjornson_MAMIMO_book}, \cite[Ch.~15]{steve_M_kay}.
\end{IEEEproof}

Notice that the MSE of the channel estimates, that is $\rm{tr}(\bf{C}_{llk}^\textsc{rp})/M$ and $\rm{tr}(\bf{C}_{llk}^\textsc{sp})/M$ with RP and SP respectively,\footnote{Recall that the MSE of $\hat{\bf{h}}_{llk}$ is defined as $\bb{E}\{ \| \bf{h}_{llk} - \hat{\bf{h}}_{llk} \|^2\}/M$.} can be reduced by increasing the pilot length. Having more pilot symbols would lower the number of shared pilots (i.e., the number of elements in $\cal{P}_{lk}^\textsc{rp}$  and $\cal{P}_{lk}^\textsc{sp}$) and also decrease the effect of noise. However, the pilot length (being $\Tp \leq \Tul$ with RP and $\Tul$ with SP) is ultimately limited by the size of the coherence block $\Tc$ which is set by physical properties of the channels and cannot be made arbitrarily large. Thus, the estimation errors cannot be alleviated and interference management would be a potential way to improve the channel estimation quality.

\section{Uplink Combining and Achievable SE}
\label{sec:UL_detect_Ach_SEs}
This section introduces the process of coherently combining signals in Massive MIMO with RP and SP, as well as, the definition of an achievable SE for performance evaluation. Linear signal processing is assumed where the combining vector for $\rm{UE}_{lk}$ is defined as 
\begin{align}
\bf{v}_{lk} = 
\begin{cases}
\hat{\bf{h}}_{llk} & \text{ MR }\\
\!\left(\sum\limits_{k'=1}^{K} \!\! \pTul_{lk'}\!\left(\hat{\bf{h}}_{ll k'} \hat{\bf{h}}_{ll k'}^H \!+\! \bf{C}_{ll k'}\right) \! +\! \sigma^2 \bf{I}_M  \!\right)^{\!-1}\!\! \hat{\bf{h}}_{llk}\pTul_{lk} & \text{S-MMSE}
\end{cases}
\label{eq:combining_v}
\end{align} 
where the superscripts indicating RP and SP are dropped to show that these combining methods can be applied with RP and SP alike. MR combining aims at maximizing the received power from $\rm{UE}_{lk}$, whereas S-MMSE balances interference suppression and signal amplification while only using CSI available at $\rm{BS}_l$. This means, that S-MMSE only relies on CSI obtained from UL pilots and does not require sharing CSI among BSs.

After combining the received signal from all BS antennas, the following observations of the data symbols within one coherence block are obtained, with RP and SP respectively:
\begin{align*}	\allowdisplaybreaks
\underbrace{\hat{\bf{y}}_{lk}^{\textsc{rp}^T}}_{1\times \Tuld} &= \underbrace{\bf{v}_{lk}^H}_{1\times M}\underbrace{[{\bf{Y}_{l}}]_{\Tp+1:\Tul}}_{M \times \Tuld}
\stepcounter{equation}\tag{\theequation}\label{eq:data_est_RP}
\\
\underbrace{\hat{\bf{y}}_{lk}^{\textsc{sp}^T}}_{1\times \Tul} &=\underbrace{\bf{v}_{lk}^H}_{1\times M}\underbrace{\bf{Y}_{l}}_{M\times \Tul}
.
\stepcounter{equation}\tag{\theequation}\label{eq:data_est_SP}
\end{align*}

To compute an achievable SE, which is a rigorous lower bound on the ergodic capacity, the data symbols are assumed as i.i.d. $\bf{s}_{lk}\sim \cal{CN}(\bf{0},\bf{I}_{\Tuld})$, recall that $\Tuld =\Tul -\Tp$ with RP and $\Tuld = \Tul$ with SP. Since the data symbols are i.i.d. and the channel is memoryless, it is enough to focus on one arbitrary data symbol, denoted as $s_{lk}$, taken from $\bf{s}_{lk}$. 
The corresponding data observation taken from \eqref{eq:data_est_RP} with RP, or \eqref{eq:data_est_SP} with SP is denoted as $\hat{y}_{lk}$, and can be expressed as
\begin{equation}
	\hat{y}_{lk} =  \underbrace{\bb{E}\left\{\hat{y}_{lk}s_{lk}^*\right\}}_{\textrm{channel gain}}s_{lk} + \underbrace{\hat{y}_{lk} -    \bb{E}\left\{\hat{y}_{lk}s_{lk}^*\right\}s_{lk}}_{\textrm{effective noise}}
	\label{eq:rec_sig_UandForget}
\end{equation} 
where adding and subtracting the first term in \eqref{eq:rec_sig_UandForget} results in an equivalent single-input single-output (SISO) system with deterministic known channel gain and uncorrelated non-Gaussian effective noise. Then, a lower bound on the ergodic capacity is obtained by considering the effective noise to be Gaussian since the Gaussian distribution maximizes the entropy, and therefore, corresponds to the worst-case distribution for the effective noise. This bounding technique is sometimes called the ``use-and-then-forget'' bound and it is commonly used in Massive MIMO literature \cite{16_Marzetta_MAMIMO_book,17_Bjornson_MAMIMO_book,Mollen2016a}. The name ``use-and-then-forget'' follows from using the CSI to construct the combining vector $\bf{v}_{lk}$ but then dismissing it in \eqref{eq:rec_sig_UandForget} to obtain the lower bound on the ergodic capacity. Thus, an achievable SE is given by 
\begin{align}
\rm{SE}_{lk}
&=\frac{\Tuld}{\Tul}\log_2\left(1 + \frac{\left|\bb{E}\left\{\hat{y}_{lk}s_{lk}^*\right\}\right|^2}{\var \left\{\hat{y}_{lk} -    \bb{E}\left\{\hat{y}_{lk}s_{lk}^*\right\}s_{lk} \right\}}\right)
,\label{eq:ach_SE_SP}
\end{align}
the superscripts for RP and SP are removed since the same bound can be applied in both cases. 
\begin{figure*}[t]
\begin{equation*}
\tag{21}\label{eq:coCchObsDataAid}
\bfs{\Psi}_{llk}^\iter{}{i}=  
\bf{R}_{llk} + \underbrace{\sum_{k' = 1}^{K}\bf{R}_{llk'}\bb{E}\left\{\left|\bf{u}_{lk}^{\iter{}{i}^H}\tilde{\bf{x}}_{lk'}^\iter{}{i}  \right|^2 \right\}  \vphantom{\sum_{K_K\backslash}^{K^K}}
}_{\textrm{Intracell interference}}
+ \underbrace{\sum_{\ell\in \Phi\backslash l}\sum_{k'=1}^{K}\bf{R}_{l\ell k'}\bb{E}\left\{\left|\bf{u}_{lk}^{\iter{}{i}^H}\bf{x}_{\ell k'} \right|^2 \right\}
	\vphantom{\sum_{K_K\backslash}^{K^K}}
}_{\textrm{Intercell interference}} + \underbrace{\sigma^2\bb{E}\left\{\left\|\bf{u}_{lk}^{\iter{}{i}}\right\|^2 \right\} \bf{I}_M \vphantom{\sum_{K_K\backslash}^{K^K}}
}_{\textrm{Noise}}
\end{equation*}
\hrulefill
\end{figure*}
\section{MICED - Massive-MIMO Iterative Channel Estimation and Decoding}
\label{sec:dataAidChEst}

In this section, the proposed MICED algorithm is defined, explained and analyzed in terms of channel estimation quality, and SE, while discussing its feasibility in terms of computational complexity. First, the basis of the MICED algorithm is illustrated in Algorithm~\ref{alg:basis_MICED}. Second, a detailed analysis of the data-aided channel estimation process is given. Third, the computational complexity of the MICED algorithm is discussed. Fourth, numerical examples with Gaussian data symbols illustrate the possible gains of the MICED algorithm in terms of MSE of channel estimates and achievable SE.

\begin{algorithm}[t]
	{\small%
	\caption{MICED basic algorithm. The following abbreviations are used: calculate (cal.), and observation (obs.)}	\label{alg:basis_MICED}
	\begin{algorithmic}[1]
		\fontsize{9}{12}\selectfont
		\beginIni
			\State set $i = 0$ and define $i_{\max} \in \bb{N}$
		\beginChEst{\textbf{Pilot-based}}{\text{for $\rm{UE}_{lk}\,\forall k \in \{1,\ldots,K\}$}} \label{alglin:pilotChEst}
			\State cal. channel obs. $\bf{z}_{llk}^\iter{}{0}$ \Comment{use \eqref{eq:ch_est_obs_RP} with RP and \eqref{eq:ch_est_obs_SP} with SP}
			\State cal. channel estimates $\hat{\bf{h}}_{llk}^\iter{}{0}$ \Comment{use Lemma~\ref{lem:LMMSE_ch_est}} \label{alglin:pilotChEstend}
		\beginLinComb   \label{alglin:LinComb}
			\State cal. combining vector $\bf{v}_{lk}^\iter{}{0}$ \Comment{use \eqref{eq:combining_v}}
			\State cal. data obs. $\hat{\bf{y}}_{lk}^\iter{}{0}$ \Comment{use \eqref{eq:data_est_RP} with RP and \eqref{eq:data_est_SP} with SP}
		\beginDataDec{\text{for $\rm{UE}_{lk}\,\forall k \in \{1,\ldots,K\}$}}  \Comment{See Section~\ref{sec:finiteAlphabet} for details}
			\State decode data and extract data estimates $\hat{\bf{s}}_{lk}^\iter{}{0}$ 
			\State set $i = 1$ \label{alglin:Decend}
		\While{$i \leq i_{\max}$ \textbf{or} at least one UE has decoding errors}
			\beginChEstData{\textbf{Data-based}}{\text{for $\rm{UE}_{lk}\,\forall\! k\! \in\! \{1,\ldots,K\}\!\!\!\!\!\!$}}	\label{alglin:dataChEst}	
			\State cal. channel obs. $\bf{z}_{llk}^\iter{}{i}$ \Comment{use \eqref{eq:ChObsIt} with $\hat{\bf{s}}_{lk}^\iter{}{i-1}$}
			\State cal. channel estimates $\hat{\bf{h}}_{llk}^\iter{}{i}$ \Comment{use Lemma~\ref{lem:LMMSE_ch_est} with $\bf{z}_{llk}^\iter{}{i}$}	\label{alglin:dataChEstend}	
			\beginLinComb \label{alglin:dataLinComb}
			\State cal. combining vector $\bf{v}_{lk}^\iter{}{i}$ \Comment{use \eqref{eq:combining_v} with $\hat{\bf{h}}_{llk}^\iter{}{i}$}
			\State cal. data obs. $\hat{\bf{y}}_{lk}^\iter{}{i}$ \Comment{use \eqref{eq:data_obs_i_RP} with RP, \eqref{eq:data_obs_i_SP} with SP} \label{alglin:dataLinCombend}
			\beginDataDec{\text{for $\rm{UE}_{lk}\,\forall k \in \{1,\ldots,K\}$}}  \Comment{See Section~\ref{sec:finiteAlphabet} for details} 	\label{alglin:dataDec}
			\State decode data and extract data estimates $\hat{\bf{s}}_{lk}^\iter{}{i}$ 
			\State set $i = i + 1$			\label{alglin:dataDecend}
		\EndWhile
	
	\end{algorithmic}
}	
\end{algorithm}

\subsection{Basis of the MICED algorithm}
\label{subsec:Basis_MICED}

The main principle of using data estimates to perform channel estimation is to spread the interference effect between UEs. That is, to trade the main interfering sources in the channel estimation for reduced interference coming from all cells. The main source of interference in the channel estimation with RP is pilot contamination, and the use of data estimates reduces this effect. On the other hand, with SP the pilot contamination is reduced by having longer pilots and the main interference in the channel estimation is due to data symbols, thus the aim of using data estimates in this case is to reduce the data intracell (same cell as $\rm{UE}_{lk}$) interference.

At the start of the MICED algorithm, pilot-based channel estimation is performed once (at iteration $i = 0$) as shown in lines~\ref{alglin:pilotChEst}-\ref{alglin:pilotChEstend} of Algorithm~\ref{alg:basis_MICED}. Afterwards, linear combining is performed followed by data decoding (see lines~\ref{alglin:LinComb}-\ref{alglin:Decend} of Algorithm~\ref{alg:basis_MICED}). This initial decoding procedure provides soft estimates of the data symbols which are then used to improve the quality of channel estimates with the aim of achieving lower data decoding errors in the next iterations.

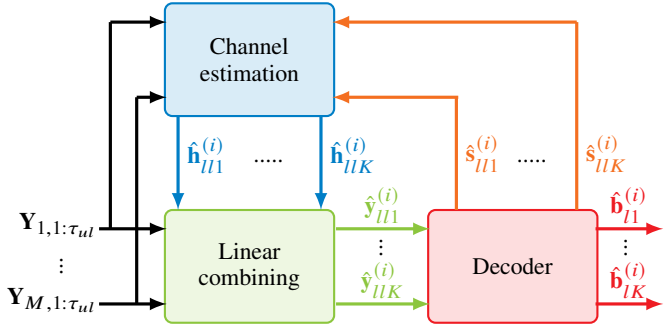
\begin{figure}[!t]	
	\centering

\begin{tikzpicture}
\small
\tikzset{>=latex}
\tikzstyle{box} = [draw,rounded corners=.1cm,minimum height=3em, minimum width=2em, align=center, thick] 
\tikzstyle{vertical_box} = [draw,rounded corners=.1cm,minimum height=2em, minimum width=3em, align=center, thick,rotate=90]

\def\originx{0}
\def\originy{0}

\node[box,minimum height=1.5cm,minimum width=2.25cm,draw=mittelblau,fill=mittelblau!15] (CE) at (\originx,\originy) {Channel \\ estimation};

\node[box,minimum height=1.5cm,minimum width=2.25cm,draw=apfelgruen,fill=apfelgruen!15] (LC) at (\originx,\originy-2.75) {Linear \\ combining};

\node[box,minimum height=1.5cm,minimum width=2.25cm,draw=rot,fill=rot!15] (DE) at (\originx+3.5,\originy-2.75) {Decoder};

\draw[->,black,very thick] (\originx-2,\originy-2.25) node [left,xshift=0.1cm,yshift=+0.05cm] {$\bf{Y}_{1,1:\tau_{ul}}$}  -- (\originx-1.1,\originy-2.25) ;
\draw[-,black,very thick] (\originx-1.85,\originy-2.25)   -- (\originx-1.85,\originy+0.5) ;
\draw[->,black,very thick] (\originx-1.85,\originy+0.5)   -- (\originx-1.1,\originy+0.5) ;
\draw[-,black,very thick] (\originx-1.5,\originy-3.25)   -- (\originx-1.5,\originy-0.5) ;
\draw[->,black,very thick] (\originx-1.5,\originy-0.5)   -- (\originx-1.1,\originy-0.5) ;

\draw[->,black,very thick] (\originx-2,\originy-3.25) node [left,xshift=0.1cm,yshift=+0.05cm] {$\bf{Y}_{M,1:\tau_{ul}}$}  -- (\originx-1.1,\originy-3.25) ;
\node[box,draw=none,rotate=90] () at (\originx-2.5,\originy-2.75) {...};

\draw[->,mittelblau,very thick] (\originx-0.95,\originy-0.75) -- (\originx-0.95,\originy-2) node [midway,  right,yshift=0.1cm] {$\hat{\bf{h}}^{(i)}_{ll1}$};
\draw[->,mittelblau,very thick] (\originx+0.95,\originy-0.75) -- (\originx+0.95,\originy-2) node [midway,  right,yshift=0.1cm] {$\hat{\bf{h}}^{(i)}_{llK}$};
\node[box,draw=none] (channele) at (\originx+0.25,\originy-1.35) {.....};

\draw[->,apfelgruen,very thick] (\originx+1.15,\originy-2.25) -- (\originx+2.4,\originy-2.25) node [midway,  above,yshift=-0.05cm] {$\hat{\bf{y}}^{(i)}_{ll1}$};
\draw[->,apfelgruen,very thick] (\originx+1.15,\originy-3.25) -- (\originx+2.4,\originy-3.25) node [midway,  above,yshift=-0.05cm] {$\hat{\bf{y}}^{(i)}_{llK}$};
\node[box,draw=none,rotate=90] () at (\originx+1.75,\originy-2.5) {...};

\draw[->,rot,very thick] (\originx+4.6,\originy-2.25) -- (\originx+5.5,\originy-2.25) node [midway,  above,yshift=-0.05cm] {$\hat{\bf{b}}^{(i)}_{l1}$};
\draw[->,rot,very thick] (\originx+4.6,\originy-3.25) -- (\originx+5.5,\originy-3.25) node [midway,  above,yshift=-0.05cm] {$\hat{\bf{b}}^{(i)}_{lK}$};
\node[box,draw=none,rotate=90] () at (\originx+5,\originy-2.5) {...};

\draw[-,orange,very thick] (\originx+4.35,\originy-2)   -- (\originx+4.35,\originy+0.5)  node [midway,  right,yshift=-0.5cm] {$\hat{\bf{s}}^{(i)}_{llK}$} ;
\draw[->,orange,very thick] (\originx+4.35,\originy+0.5)   -- (\originx+1.1,\originy+0.5) ;

\draw[-,orange,very thick] (\originx+2.75,\originy-2)   -- (\originx+2.75,\originy-0.5) node [midway,  right,yshift=0cm] {$\hat{\bf{s}}^{(i)}_{ll1}$} ;
 \draw[->,orange,very thick] (\originx+2.775,\originy-0.5)   -- (\originx+1.1,\originy-0.5) ;
\node[box,draw=none] (channele) at (\originx+3.75,\originy-1.35) {.....};
\end{tikzpicture}
	\caption{Block diagram of iterative receiver. The notation $\cdot^\iter{}{i}$ is used to represent the values of variables at the $i^{th}$ iteration of the receiver algorithm.}
	\label{fig:it_dec_diag}
\end{figure}

Once the received signals have been linearly combined, the data observations in \eqref{eq:data_est_RP} and \eqref{eq:data_est_SP} are used to detect which symbols were sent. Due to the effects of interference and noise, some symbols may be detected erroneously leading to a failure in retrieving the desired information. In practice, the redundant information in the channel code is used to detect when the decoding procedure fails, for instance cyclic redundancy check codes are often used for this purpose. In such cases, the data observations in \eqref{eq:data_est_RP} and \eqref{eq:data_est_SP} can also be used to obtain estimates of the data symbols and, in turn, use those to improve the channel estimates as shown in lines~\ref{alglin:dataChEst}-\ref{alglin:dataChEstend} of Algorithm~\ref{alg:basis_MICED}. These improved channel estimates can then be used to perform linear combining again (see lines~\ref{alglin:dataLinComb}-\ref{alglin:dataLinCombend} of Algorithm~\ref{alg:basis_MICED}), and obtain updated data observations as follows:
\begin{align*}	\allowdisplaybreaks
\hat{\bf{y}}_{lk}^{\iter{rp}{i}^T} &= \bf{v}_{lk}^{\iter{rp}{i}^H}\left(\!\vphantom{\sum_{K}^{K}}\right.
[{\bf{Y}_{l}}]_{\Tp+1:\Tul} - \sum_{\substack{k'=1 \\ k'\neq k}}^{K} \hat{\bf{h}}_{llk'}^{\iter{rp}{i}} \sqrt{\pulrp_{lk'}} \hat{\bf{s}}_{lk'}^{\iter{rp}{i-1}^T}
\left.\vphantom{\sum_{K}^{K}}\!\right)
\stepcounter{equation}\tag{\theequation}\label{eq:data_obs_i_RP}
\\
\hat{\bf{y}}_{lk}^{\iter{sp}{i}^T} &=\bf{v}_{lk}^{\iter{sp}{i}^H}\left(\!\vphantom{\sum_{K}^{K}}\right.
\bf{Y}_{l}  - \sum_{\substack{k'=1 \\ k'\neq k}}^{K} \hat{\bf{h}}_{llk'}^{\iter{sp}{i}}
\left(\sqrt{\qulsp_{lk'}}\bfs{\varphi}_{lk'}^T + \sqrt{\pulsp_{lk'}}\hat{\bf{s}}_{lk'}^{\iter{sp}{i-1}^T} \right) 
\left.\vphantom{\sum_{K}^{K}}\!\right)
\stepcounter{equation}\tag{\theequation}\label{eq:data_obs_i_SP}
\end{align*}%
with RP and SP respectively. Note that updated channel estimates and previous data estimates are also used to subtract the intracell interference. Thus, the updated data observations may contain less interference and, in turn, lead to fewer decoding errors. This procedure is done iteratively to improve data and channel estimates in each iteration as shown in Figure.~\ref{fig:it_dec_diag}. Each iteration starts with the channel estimation, followed by linear combining, and finishing with the data decoding. The MICED algorithm ends when the maximum number of iterations is reached or the data from all UEs is successfully decoded.

Note that the MICED algorithm is analyzed using an arbitrary set of coherence blocks since the channel realizations are considered independent across blocks. In real propagation channels, the coherence blocks that are close to each other (either in time or frequency) exhibit some degree of correlation which can also be exploited to improve channel the estimation \cite{Morelli2001a}. However, this analysis  falls outside the scope of this article and it is therefore left for future work.

\begin{figure*}[t]
\allowdisplaybreaks
\begin{align*}
\tag{25}\label{eq:CorMatInt_dataRP}
&\bar{\bfs{\Psi}}_{llk}^\iter{rp}{i} =
\frac{ \sum\limits_{\substack{k' = 1 \\ k'\neq k}}^{K}\bf{R}_{llk'}\pulrp_{lk'} \left(1 - \sigma_{lk'}^{\iter{}{i}^2}\right)+ 
	\sum\limits_{\ell\in \Phi\backslash l}\sum\limits_{k'=1}^{K}\bf{R}_{l\ell k'}\pulrp_{\ell k'}
}{\displaystyle \frac{(\qulrp_{lk})^2 \Tp^2}{\pulrp_{lk}\sigma_{lk}^{\iter{}{i}^2}(\Tuld - K)} +  2\qulrp_{lk}\Tp  + \pulrp_{lk}\sigma_{lk}^{\iter{}{i}^2}\Tuld }
+ 
\frac{  \sum\limits_{  (\ell,k')\in \cal{P}_{lk}^\textsc{rp}\backslash (l,k) } \bf{R}_{l\ell k'} \qulrp_{\ell k'}
}{\displaystyle  \qulrp_{lk}  + \frac{2 \Tuld\pulrp_{lk} \sigma_{lk}^{\iter{}{i}^2}}{\Tp}   + \frac{(\pulrp_{lk}\sigma_{lk}^{\iter{}{i}^2})^2\Tuld(\Tuld + 1)}{\qulrp_{lk} \Tp^2} }
+ \frac{\sigma^2\bf{I}_M}{\qulrp_{lk}\Tp + \pulrp_{lk}\sigma_{lk}^{\iter{}{i}^2}\Tuld } 
\\
&
\bar{\bfs{\Psi}}_{llk}^\iter{sp}{i}= 
\frac{ \sum\limits_{\substack{k' = 1 \\ k'\neq k}}^{K}\!\bf{R}_{llk'}\pulsp_{lk'} \left(1 - \sigma_{lk'}^{\iter{}{i}^2}\right)+\!\! 
	\sum\limits_{\ell\in \Phi\backslash l}\sum\limits_{k'=1}^{K}\!\bf{R}_{l\ell k'}\pulsp_{\ell k'}
}{\Tul \left(\qulsp_{lk} + \pulsp_{lk} \sigma_{lk}^{\iter{}{i}^2}\right) }
+
\frac{ \sum\limits_{ (\ell,k')\in \cal{P}_{lk}^\textsc{sp}\backslash (l,k) } \bf{R}_{l\ell k'} \qulsp_{\ell k'}}{\!\displaystyle \frac{(\qulsp_{lk} + \pulsp_{lk}\sigma_{lk}^{\iter{}{i}^2} )^2}{\qulsp_{lk}} +\frac{\pulsp_{lk}\sigma_{lk}^{\iter{}{i}^2} ( 2\qulsp_{lk} + \pulsp_{lk}\sigma_{lk}^{\iter{}{i}^2} )}{\qulsp_{lk}\Tul} \!}
+ \frac{\sigma^2  \bf{I}_M}{\!\Tul \left(\qulsp_{lk} + \pulsp_{lk} \sigma_{lk}^{\iter{}{i}^2}\right)\!}
\tag{26}\label{eq:CorMatInt_dataSP}
\end{align*}
\hrulefill 
\end{figure*}


\subsection{Analysis of data-aided channel estimation with Gaussian symbols}
\label{subsec:Gauss_analysis}
In practical implementations, the transmitted information is encoded into bits that are then modulated into a finite alphabet of complex symbols. Therefore, the detection is done based on bits rather than complex symbols, which is enclosed within the decoder stage shown in Figure~\ref{fig:it_dec_diag} where $\hat{\bf{b}}_{llk}$ represents the hard bit estimates from $\rm{UE}_{lk}$ $\forall k \in \{1,\ldots,K\}$ at the output of the decoder. In this section, the data symbols are considered as i.i.d. $\bf{s}_{lk}\sim \cal{CN}(\bf{0},\bf{I}_{\Tuld})$ for analytical tractability. This assumption yields theoretical results that give insights into the gains in channel estimation quality that the MICED algorithm can offer. In later sections, the analysis will be extended towards finite-alphabet symbols and the Gaussian assumption will be dropped. 

At the $i^{th}$ iteration of the receiving algorithm, assume that the MMSE data estimate\footnote{These data estimates are obtained from the decoding procedure. Section~\ref{sec:finiteAlphabet} explains in detail how to perform this in practical implementations with finite-alphabet symbols.} of $s_{lk}$ is $\hat{s}_{lk}^\iter{}{i}\sim\cal{CN}(0, \sigma_{lk}^{\iter{}{i}^2})$ such that $s_{lk} =  \hat{s}_{lk}^\iter{}{i} + \tilde{s}_{lk}^\iter{}{i}$,
where $\tilde{s}_{lk}^\iter{}{i}$ is the estimation error that is uncorrelated (i.e., $\bb{E}\{\hat{s}_{lk}^\iter{}{i}\tilde{s}_{lk}^{\iter{}{i}^*} \} = 0$) and independent of the data estimates.

Based on the estimates of the data from UEs within the same cell, the BS can obtain an estimate of the transmitted signal from UEs in cell $l$ as
\begin{align*}
\hat{\bf{x}}_{lk}^\iter{}{i} = \begin{cases}
\left[\sqrt{\qulrp_{lk}}\bfs{\phi}_{lk}^T\quad  \sqrt{\pulrp_{lk}}(\hat{\bf{s}}_{lk}^\iter{}{i})^T\right]^T & \text{ with RP,}\\
\sqrt{\qulsp_{lk}}\bfs{\varphi}_{lk} + \sqrt{\pulsp_{lk}}\hat{\bf{s}}_{lk}^\iter{}{i} & \text{ with SP,}
\end{cases}
\end{align*}
where $\tilde{\bf{x}}_{lk}^\iter{}{i} = \bf{x}_{lk} - \hat{\bf{x}}_{lk}^\iter{}{i}$ is the signal estimation error at the $i^{th}$ iteration.

 By collecting these signal estimates from the $K$ UEs served by $\rm{BS}_l$ within an arbitrary coherence block, and stacking them into the matrix $\hat{\bf{X}}_l^\iter{}{i} = [\hat{\bf{x}}_{l1}^\iter{}{i},\ldots,\hat{\bf{x}}_{lK}^\iter{}{i}]\in \bb{C}^{\Tuld\times K} $, a new observation of the channel $\bf{h}_{llk}$ can be obtained by projecting the received signal in \eqref{eq:rec_sig_UL} with $\bf{u}_{lk}^\iter{}{i} = \left[\hat{\bf{X}}_l^\iter{}{i} \left(\hat{\bf{X}}_l^{\iter{}{i}^H}  \hat{\bf{X}}_l^\iter{}{i} \right)^{-1}\right]_k$ (note that $\Tuld > K$ is assumed), which yields
\stepcounter{equation}
\begin{equation}
\bf{z}_{llk}^\iter{}{i} = \bf{Y}_l\bf{u}_{lk}^{\iter{}{i}^*}
\label{eq:ChObsIt}    
\end{equation}
with correlation matrix (i.e., $\bfs{\Psi}_{llk}^\iter{}{i} = \bb{E}\{\bf{z}_{llk}^\iter{}{i}\bf{z}_{llk}^{\iter{}{i}^H}\}$) given in \eqref{eq:coCchObsDataAid} at the top of the page. 
The superscripts denoting RP and SP are removed to indicate that the correlation matrix in both cases has the same formulation, thus the difference lies in what goes into the expectations.

The use of $\bf{u}_{lk}^{\iter{}{i}}$ to obtain the channel observation $\bf{z}_{llk}^\iter{}{i}$ is aimed at reducing the intracell interference in the channel estimation process. Thus, the more accurate the data estimates are, the less intracell interference will be present in the data-aided channel estimates.
\begin{remark}
	\label{rem:diffDataq}
	The quality of data estimates differs among UEs due to the large-scale fading, transmission power, and interference conditions. Thus, a particular case of interest for the MICED algorithm is when some UEs have high data estimation quality and others not. In this case, the high quality data estimates can be used to improved the data decoding of the UEs with low data estimation quality.  
\end{remark}

 Notice that the expectations in \eqref{eq:coCchObsDataAid} are non-trivial to compute since they involve inverse moments of non-central complex Wishart matrices. To obtain insights into the performance and behavior of data-aided channel estimation, the following theorem introduces closed-form expressions that bound the correlation matrices in \eqref{eq:coCchObsDataAid} in the positive semi-definite sense. 
\begin{theorem}
Given $\bfs{\Psi}_{llk}^\iter{}{i}$ in \eqref{eq:coCchObsDataAid}, it holds that
\begin{align*}
&\bfs{\Psi}_{llk}^\iter{rp}{i}\succeq  \bf{R}_{llk} 
\\
&+ \bf{R}_{llk} \frac{ \pulrp_{lk} \left(1 - \sigma_{lk}^{\iter{}{i}^2}\right)}{\displaystyle \frac{(\qulrp_{lk})^2 \Tp^2}{\pulrp_{lk}\sigma_{lk}^{\iter{}{i}^2}(\Tuld - K)} +  2\qulrp_{lk}\Tp  + \pulrp_{lk}\sigma_{lk}^{\iter{}{i}^2}\Tuld } + \bar{\bfs{\Psi}}_{llk}^\iter{rp}{i}
\stepcounter{equation}\tag{\theequation}\label{eq:cov_chObs_dataRP}
\\
&\bfs{\Psi}_{llk}^\iter{sp}{i}\succeq \bf{R}_{llk} + \bf{R}_{llk}\frac{ \pulsp_{lk} \left(1 - \sigma_{lk}^{\iter{}{i}^2}\right)}{\Tul \left(\qulsp_{lk} + \pulsp_{lk} \sigma_{lk}^{\iter{}{i}^2}\right) }
+ \bar{\bfs{\Psi}}_{llk}^\iter{sp}{i}
\stepcounter{equation}\tag{\theequation}\label{eq:cov_chObs_dataSP}
\end{align*}
where $\bar{\bfs{\Psi}}_{llk}^\iter{rp}{i}$ and $\bar{\bfs{\Psi}}_{llk}^\iter{sp}{i}$ are given in \eqref{eq:CorMatInt_dataRP} and \eqref{eq:CorMatInt_dataSP} respectively, at the top of next page.
\label{th:LB_covMatDataAid}
\end{theorem}
\begin{IEEEproof}
 The proof is shown in Appendix~\ref{app:proofCovMat}.
\end{IEEEproof}
Recall that the MSE of the channel estimates is given by $\rm{tr}(\bf{C}_{llk})/M$ where $\bf{C}_{llk}$ is the correlation matrix of the channel estimation error that follows the same formulation as in \eqref{eq:errCorrMatRP} with RP and \eqref{eq:errCorrMatSP} with SP. Hence, a lower bound on the MSE of data-aided channel estimates can be obtained by replacing the correlation matrix of channel observations in \eqref{eq:cov_chObs_pilotRP} and \eqref{eq:cov_chObs_pilotSP} with those in the right-hand-side of \eqref{eq:cov_chObs_dataRP} and \eqref{eq:cov_chObs_dataSP} respectively. This lower bound is then characterized by the behavior of $\bar{\bfs{\Psi}}_{llk}^\iter{}{i}$, defined in \eqref{eq:CorMatInt_dataRP} and \eqref{eq:CorMatInt_dataSP} with RP and SP respectively, in the subspace spanned by $\bf{R}_{llk}$ (see \eqref{eq:errCorrMatRP} and \eqref{eq:errCorrMatSP}). In addition, notice that since linear combining is considered for data detection, the terms in \eqref{eq:CorMatInt_dataRP} and \eqref{eq:CorMatInt_dataSP} will combine coherently. Thus, reducing $\rm{tr}( \bar{\bfs{\Psi}}_{llk}^\iter{}{i})$ leads to lower MSE of the data-aided channel estimates and coherent interference. 

To obtain better insights into the benefits that the MICED algorithm may bring, the influence of the data estimation quality and number data symbols for each term in \eqref{eq:CorMatInt_dataRP} and \eqref{eq:CorMatInt_dataSP} is analyzed in detail. First, notice that $\rm{tr}( \bar{\bfs{\Psi}}_{llk}^\iter{}{i})$ is a decreasing function of $\sigma_{lk'}^\iter{}{i}$ for $k'\neq k$ (see the first term in \eqref{eq:CorMatInt_dataRP} and  \eqref{eq:CorMatInt_dataSP}). This means that the influence of the intracell interference decreases with the quality of the data estimates. Second, by inspecting the derivative with respect to $\sigma_{lk}^{\iter{}{i}^2}$ of the denominator in the first term of \eqref{eq:CorMatInt_dataRP} (note that this term is a scalar) it can be shown that for 
\addtocounter{equation}{2}
\begin{equation}
 \sigma_{lk}^{\iter{}{i}^2} \geq  \frac{\qulrp_{lk} \Tp}{\pulrp_{lk} \sqrt{\Tuld(\Tuld - K)}}
 \end{equation} 
 the term $\rm{tr}( \bar{\bfs{\Psi}}_{llk}^\iter{rp}{i})$ is a decreasing function of $\sigma_{lk}^{\iter{}{i}^2}$. Moreover, $\rm{tr}( \bar{\bfs{\Psi}}_{llk}^\iter{sp}{i})$ is also a decreasing function of $\sigma_{lk}^{\iter{}{i}^2}$. Thus, the higher quality the data estimates have, the lower influence the interference has on the data-aided channel estimates. This confirms the intuition provided in Remark~\ref{rem:diffDataq} that accurate data estimates of UEs within a given cell can be useful to improve the data decoding of other UEs in the same cell. Third, consider the influence of $\Tuld$ with RP, by inspecting the denominator of the first term  in \eqref{eq:CorMatInt_dataRP} it follows that for 
 \begin{equation}
 \Tuld \geq  \frac{\qulrp_{lk} \Tp}{\pulrp_{lk}  \sigma_{lk}^{\iter{}{i}^2}} + K
 \label{eq:tau_d_condition_RP}
 \end{equation}
 the term $\rm{tr}( \bar{\bfs{\Psi}}_{llk}^\iter{rp}{i})$ is a decreasing function of $\Tuld$ since the second term in \eqref{eq:CorMatInt_dataRP} also decreases with $\Tuld$. The condition \eqref{eq:tau_d_condition_RP} can be interpreted as the minimum value of $\Tuld$ from which it is feasible to implement the MICED algorithm with RP. Moreover, when the number of data symbols increases beyond this condition, the interference effect is reduced. On the other hand, with SP, the trace of the first and last terms in \eqref{eq:CorMatInt_dataSP} is a decreasing function of $\Tul$, whereas, the trace of the second term in \eqref{eq:CorMatInt_dataSP} has a more involved dependency on $\Tul$ since the number of elements in $\cal{P}_{lk}^\textsc{sp}$ decreases with $\Tul$. This means that the data interference and noise are reduced with higher $\Tul$ while the effect of pilot contamination, in turn, is reduced by having more sparse pilot reuse factors as $\Tul$ increases. 
 
In the case of RP, comparing \eqref{eq:cov_chObs_pilotRP} with \eqref{eq:cov_chObs_dataRP} and \eqref{eq:CorMatInt_dataRP} shows that by using the data estimates the pilot contamination is traded for interference from all UEs that decreases with the quality of data estimates and number of data symbols, which in turn might be substantially smaller. Whereas with SP,  comparing \eqref{eq:cov_chObs_pilotSP} with \eqref{eq:cov_chObs_dataSP} and \eqref{eq:CorMatInt_dataSP} shows that the intracell interference from data symbols decreases with the quality of data estimates and the size of the coherence block.

In summary, the purpose of utilizing data estimates to revise the channel estimation is to trade a few terms that cause high interference with many terms that cause low interference.

	\subsection{Computational complexity}
	\label{subsec:complexity}
	In the past few years, several real-time testbeds for Massive MIMO have been built to evaluate its performance in real propagation scenarios \cite{VanderPerre2018_DSPMaMIMO}. In particular, the Lund University Massive MIMO testbed (LuMaMi) \cite{Malkowsky2017_LuMaMIMO} runs a real-time Massive MIMO system with RP, $M=100$, and $K=12$, using a 20 MHz bandwidth with 1200 subcarriers and an OFDM symbol length of 71.4 $\mu$s. In the LuMaMi testbed, the main contributor to the usage of the processing resources is the QR-decomposition which is used to invert the Gramian matrix (e.g., $(\hat{\bf{H}}^H\hat{\bf{H}})^{-1}$ where $\hat{\bf{H}}$ is a $M\times K$ channel estimates matrix). This matrix invertion is employed for interference suppression techniques in the spatial domain like zero-forcing (ZF) or regularized ZF (RZF). However, the latency evaluation in \cite{Malkowsky2016_Lat_LuMaMi} shows that the overall time for performing UL channel estimation and transmitting precoded signals in the downlink\footnote{Note that this time includes the computation of matrix inversions to perform ZF or RZF in the downlink.} (called precoding turnaround time) is 132~$\mu$s. Furthermore, the largest contributor to the latency is OFDM modulation/demodulation whereas the impact of channel estimation and precoding is negligible in comparison.

	To implement the MICED algorithm, the additional computational complexity comes from re-estimating the channel, performing the linear combining, and decoding the data in each iteration. Moreover, to obtain the data-aided channel estimates another $K\times K$ matrix inversion needs to be made corresponding to the Gramian of signal estimates (i.e., $(\hat{\bf{X}}_l^{\iter{}{i}^H}  \hat{\bf{X}}_l^\iter{}{i} )^{-1}$ see Section~\ref{subsec:Gauss_analysis}). This would for sure add an important burden to the signal processing. However, this can be addressed through parallel computing techniques similar to the ones used in the LuMaMi testbed \cite{Malkowsky2016_Lat_LuMaMi,Malkowsky2017_LuMaMIMO} where even if the extra processing duplicates the delay, it would still be less than 285~$\mu$s which is their constraint for the precoding turnaround time. Thus, based on the existing developments in digital signal processing applied to Massive MIMO systems \cite{VanderPerre2018_DSPMaMIMO,Malkowsky2017_LuMaMIMO} it is indeed possible to implement the MICED algorithm in practice for at least a few tens of iterations.

 \begin{table}[!t]
 	\centering { \small
 		\renewcommand{\arraystretch}{1.1}
 		\caption{Simulation parameters.}
 		\label{tab:sim_param}
 		\centering
 		\begin{tabular}{C{0.2\textwidth}|c}
 			\bfseries Parameter  & \bfseries Value\\
 			\hline
 			System bandwidth & $\Bw = 20$ [MHz] \\ 
 			Maximum transmission power per UE & $10\log_{10}(\varrho_\textsc{max}\Bw) = 20$ [dBm]\\ 	 		
	 		Proportion of pilot power with SP & $\Delta = 0.3$\\ 	
 			Noise power & $10 \log_{10}\left(\sigma^2\Bw\right) = -94$ [dBm] \\ 		
 			Inter-BS distance & $0.15$ [km] \\	
 			Pathloss exponent & $\alpha = 3.76$  \\
 			Pathloss at $1$ km & $\omega = 148.1$ [dB]  \\ 			
 			Shadow fading std. deviation& $\sigma_{\rm{sf}}=10$ [dB]\\ 					
 			Angular std. deviation & $\sigma_\rm{ang} = 10^\circ$\\
 			\hline 			
 		\end{tabular}
 	}
 \end{table}

\subsection{Numerical example with Gaussian symbols}
\label{sec:num_ex}
To illustrate the possible gains of the MICED algorithm, numerical results considering Gaussian data symbols are presented in Figures~\ref{fig:SE_MSE_Gauss}~and~\ref{fig:Gauss_SE}. The simulation setup is based on a hexagonal cell grid with $K$ UEs uniformly distributed in each cell, and large-scale fading modeled as $\beta_{llk}=\omega^{-1} d_{llk}^{-\alpha }F_{llk}$. The term $\omega$ is the fixed pathloss at a reference distance of 1 km to account for propagation effects independent of the distance, for example, antenna gains, and wall penetration losses. The distance between $\rm{UE}_{lk}$ and $\rm{BS}_l$ is denoted by $d_{llk}$, and the shadow fading is defined by $10 \log_{10}(F_{llk})\sim \cal{N}(0,\sigma_{\rm{sf}}^2)$. \footnote{This stands in contrast to the simulation setup in \cite{Ma2017a_MultiUE} where the large-scale fading and intercell interference are fixed.} The spatial correlation matrices are computed based on the Gaussian local scattering model with angular standard deviation $\sigma_\rm{ang}$ defined in \cite[Ch.~2]{17_Bjornson_MAMIMO_book}. Statistical channel inversion power control is considered such that $\varrho_{lk} = \min \{\varrho / \beta_{llk}, \varrho_\textsc{max}\}$ where $\varrho$ is a design parameter to set the average transmission energy per symbol and $\varrho_\textsc{max}$ is the maximum transmission energy per symbol for each UE. In the case of SP, the proportion between pilot and data power is fixed as $\Delta_{lk} = \Delta$.\footnote{ Note that $\Delta$ has been selected to maximize the SE with pilot only channel estimation based on numerical results that are omitted in this paper for brevity. See \cite{SPPwctrl_Verenzuela2018a} for more details on power control optimization with SP.} A summary of the main simulation parameters are given in Table~\ref{tab:sim_param}. To calculate the SE per UE in Figure~\ref{fig:Gauss_SE}, the achievable SE in Section~\ref{sec:UL_detect_Ach_SEs} is used.
\begin{figure}[!t]		
	\centering
	{\captionsetup{width=0.45\textwidth}

\begin{tikzpicture}
	\begin{axis}[
	width=0.8\textwidth,
	height=50pt,
	axis line style={draw=none},
	tick style={draw=none},
	xticklabels=\empty,
	yticklabels=\empty,
	xmin = 0,
	xmax = 500,
	ymin = 0,
	ymax = 1,
	line width = 1pt,
	legend columns = 2,
	legend cell align={left},
	legend style={at={(0.36,0)},anchor=north,font=\small,line width=1pt,draw=black,mark size=0.1pt},
	]	

	\addlegendentry{\textsc{rp} pilot $\Tp = K$}	
	\addlegendimage{rot,densely dotted,mark=square*, mark options={solid, line width = 0.7pt,fill=white},  mark size=2.2pt}
	\addlegendentry{\textsc{rp} \textsc{miced} $\Tp = K$}	
	\addlegendimage{apfelgruen,dashdotted,mark=square*, mark options={solid, line width = 0.7pt},  mark size=2.2pt}
	
	\addlegendentry{\textsc{rp} pilot $\Tp = 3K$}	
	\addlegendimage{lila,densely dashdotdotted,mark=triangle*, mark options={solid, line width = 0.7pt,fill=white},  mark size=2.2pt}	
	\addlegendentry{\textsc{rp} \textsc{miced} $\Tp = 3K$}	
	\addlegendimage{pink,loosely dashdotdotted,mark=triangle*, mark options={solid, line width = 0.7pt}        ,   mark size=2.2pt}

	\addlegendentry{\textsc{sp} pilot}
	\addlegendimage{anthrazit,solid,mark=*, mark options={solid, line width = 0.7pt,fill=white},  mark size=2.2pt}
	
	\addlegendentry{\textsc{sp} \textsc{miced}}
	\addlegendimage{mittelblau,dashed,mark=*, mark options={solid, line width = 0.7pt},  mark size=2.2pt}

	\end{axis}
	\end{tikzpicture}%
		\\
			\subfloat[MSE of channel estimates vs $\sigma_{\textsc{est}}^2$ such that $\sigma_{lk}^{\iter{}{i}^2} = \sigma_{\textsc{est}}^2$ ${\forall k\in \{1,\ldots,K\}}$, and $\Tul = 190$.]{

\begin{tikzpicture}
	\begin{semilogyaxis}[
	width=0.47\textwidth,
	height=0.32\textwidth,
	y label style = {at={(-0.03,0.5)},anchor=north},
	xlabel={Data estimation quality ($\sigma_{\textsc{est}}^2$)},
	ylabel={MSE of channel est.},
	xmin = 0,
	xmax = 1,
	ymin = 0.009,
	ymax = 0.3,
	line width = 1pt,
	grid=both,
	legend columns = 3,
	legend style={at={(axis cs: 1.02,0.5)},font=\footnotesize,line width=1pt,draw=black,mark size=0.1pt},
	]	

%
%

\draw (axis cs:0.7,0.22) node[draw=none,fill =white,align = center] {\footnotesize{Pilot-only}};
\draw[-,color = gray] (axis cs:0.82,0.22) -- (axis cs:0.9,0.06);	
\draw[-,color = gray] plot [smooth , tension=3 ] coordinates{ (axis cs: 0.95,0.03) (axis cs: 0.9,0.065) (axis cs: 0.95,0.15)   };


\draw (axis cs:0.4,0.015) node[draw=none,fill =white,align = center] {\footnotesize{\textsc{miced}}};
\draw[-,color = gray] (axis cs:0.55,0.015) -- (axis cs:0.9,0.0125);	
\draw[-,color = gray] plot [smooth , tension=4 ] coordinates{ (axis cs: 0.93,0.0125) (axis cs: 0.9,0.0145) (axis cs: 0.93,0.017)   };

\addplot [rot,only marks,mark=square*, mark options={solid, line width = 0.7pt,fill=white},  mark size=2.2pt]
table[x index={0}, y index={1}] {Figures/MSE_ch_vs_dataEst.txt};									

\addplot [rot,densely dotted,mark=none]
table[x index={0}, y index={2}] {Figures/MSE_ch_vs_dataEst.txt};				


\addplot [apfelgruen,only marks,mark=square*, mark options={solid, line width = 0.7pt},  mark size=2.2pt]
table[x index={0}, y index={3}] {Figures/MSE_ch_vs_dataEst.txt};									

\addplot [apfelgruen,dashdotted,mark=none]
table[x index={0}, y index={4}] {Figures/MSE_ch_vs_dataEst.txt};			


\addplot [lila,only marks,mark=triangle*, mark options={solid, line width = 0.7pt,fill=white},  mark size=2.5pt] 					   
table[x index={0}, y index={5}] {Figures/MSE_ch_vs_dataEst.txt};

\addplot [lila,densely dashdotdotted,mark=none] 					   
table[x index={0}, y index={6}] {Figures/MSE_ch_vs_dataEst.txt};	
	


\addplot [pink,loosely dashdotdotted,mark=none] 					   
table[x index={0}, y index={8}] {Figures/MSE_ch_vs_dataEst.txt};

	
\addplot [anthrazit,only marks,mark=*, mark options={solid, line width = 0.7pt,fill=white},  mark size=2.2pt] 					   
table[x index={0}, y index={9}] {Figures/MSE_ch_vs_dataEst.txt};

\addplot [anthrazit,solid,mark=none, mark options={solid, line width = 0.7pt,fill=white},  mark size=2.2pt] 					   
table[x index={0}, y index={10}] {Figures/MSE_ch_vs_dataEst.txt};


	\addplot [mittelblau,only marks,mark=*, mark options={solid, line width = 0.7pt},  mark size=2.2pt] 					   
	table[x index={0}, y index={11}] {Figures/MSE_ch_vs_dataEst.txt};
	
\addplot [mittelblau,dashed,mark=none, mark options={solid, line width = 0.7pt},  mark size=2.2pt] 					   
table[x index={0}, y index={12}] {Figures/MSE_ch_vs_dataEst.txt};


		\addplot [pink,only marks,mark=triangle*, mark options={solid, line width = 0.7pt}        ,   mark size=2.2pt] 	  			  	   
		table[x index={0}, y index={7}] {Figures/MSE_ch_vs_dataEst.txt};

	\end{semilogyaxis}
	\end{tikzpicture}%
			\label{fig:MSE_vs_Estq}}
\\
		\subfloat[MSE of channel estimates vs $\Tul$ with ${\sigma_{lk}^{\iter{}{i}^2} = 0.6\;\forall k\in \{1,\ldots,K\}.}$]{

\begin{tikzpicture}
	\begin{semilogyaxis}[
	width=0.47\textwidth,
	height=0.32\textwidth,
	y label style = {at={(-0.03,0.5)},anchor=north},
	xlabel={Size of coherence block ($\Tul\vphantom{\sigma_{\textsc{est}}^2}$)},
	ylabel={MSE of channel est.},
	xmin = 50,
	xmax = 500,
	ymin = 0.009,
	ymax = 0.3,
	xtick distance = 100,
	line width = 1pt,
	grid=both,
	legend columns = 3,
	legend style={at={(axis cs: 510,0.5)},font=\footnotesize,line width=1pt,draw=black,mark size=0.1pt},
	]	
%
%

\draw (axis cs:320,0.09) node[draw=none,fill =white,align = center] {\footnotesize{Pilot-only}};
\draw[-,color = gray] plot [smooth , tension=3 ] coordinates{ (axis cs: 490,0.03) (axis cs: 460,0.07) (axis cs: 490,0.16)   };
\draw[-,color = gray] (axis cs:380,0.09) -- (axis cs:460,0.07);	

\draw (axis cs:130,0.015) node[draw=none,fill =white,align = center] {\footnotesize{\textsc{miced}}};
\draw[-,color = gray] (axis cs:190,0.015) -- (axis cs:328,0.013);	
\draw[-,color = gray] plot [smooth , tension=4 ] coordinates{ (axis cs: 340,0.014) (axis cs: 320,0.017) (axis cs: 340,0.021)   };

\addplot [rot,only marks,mark=square*, mark options={solid, line width = 0.7pt,fill=white},  mark size=2.2pt]
table[x index={0}, y index={1}] {Figures/MSE_ch_vs_tau_c.txt};									

\addplot [rot,densely dotted,mark=none]
table[x index={0}, y index={2}] {Figures/MSE_ch_vs_tau_c.txt};

\addplot [apfelgruen,only marks,mark=square*, mark options={solid, line width = 0.7pt},  mark size=2.2pt]
table[x index={0}, y index={3}] {Figures/MSE_ch_vs_tau_c.txt};									

\addplot [apfelgruen,dashdotted,mark=none]
table[x index={0}, y index={4}] {Figures/MSE_ch_vs_tau_c.txt};			

\addplot [lila,only marks,mark=triangle*, mark options={solid, line width = 0.7pt,fill=white},  mark size=2.5pt] 					   
table[x index={0}, y index={5}] {Figures/MSE_ch_vs_tau_c.txt};

\addplot [lila,densely dashdotdotted,mark=none] 					   
table[x index={0}, y index={6}] {Figures/MSE_ch_vs_tau_c.txt};

\addplot [pink,only marks,mark=triangle*, mark options={solid, line width = 0.7pt}        ,   mark size=2.2pt] 	  			  	   
table[x index={0}, y index={7}] {Figures/MSE_ch_vs_tau_c.txt};

\addplot [pink,loosely dashdotdotted,mark=none] 					   
table[x index={0}, y index={8}] {Figures/MSE_ch_vs_tau_c.txt};	

\addplot [anthrazit,only marks,mark=*, mark options={solid, line width = 0.7pt,fill=white},  mark size=2.2pt] 					   
table[x index={0}, y index={9}] {Figures/MSE_ch_vs_tau_c.txt};

\addplot [anthrazit,solid,mark=none, mark options={solid, line width = 0.7pt,fill=white},  mark size=2.2pt] 					   
table[x index={0}, y index={10}] {Figures/MSE_ch_vs_tau_c.txt};

\addplot [mittelblau,only marks,mark=*, mark options={solid, line width = 0.7pt},  mark size=2.2pt] 					   
table[x index={0}, y index={11}] {Figures/MSE_ch_vs_tau_c.txt};

\addplot [mittelblau,dashed,mark=none, mark options={solid, line width = 0.7pt},  mark size=2.2pt] 					   
table[x index={0}, y index={12}] {Figures/MSE_ch_vs_tau_c.txt};

	\end{semilogyaxis}
	\end{tikzpicture}%
			\label{fig:MSE_vs_tau_c}}	
	}\caption{MSE of channel estimates per UE versus $\sigma_{\textsc{est}}^2$ and $\Tul$ for $\varrho = \sigma^2$ (SNR~${=0}$ dB), $M = 100$, and $K =10$. The markers correspond to Monte-Carlo simulations and the lines to the closed-form expressions in Lemma~\ref{lem:LMMSE_ch_est} and Theorem~\ref{th:LB_covMatDataAid} (recall that the latter are lower bounds). }
	\label{fig:SE_MSE_Gauss}
\end{figure}
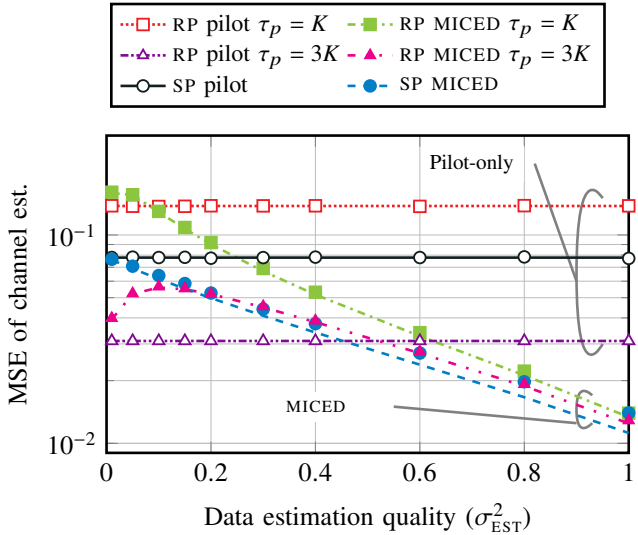
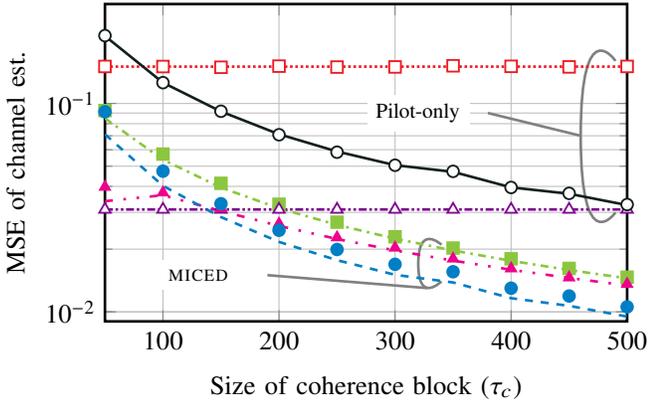
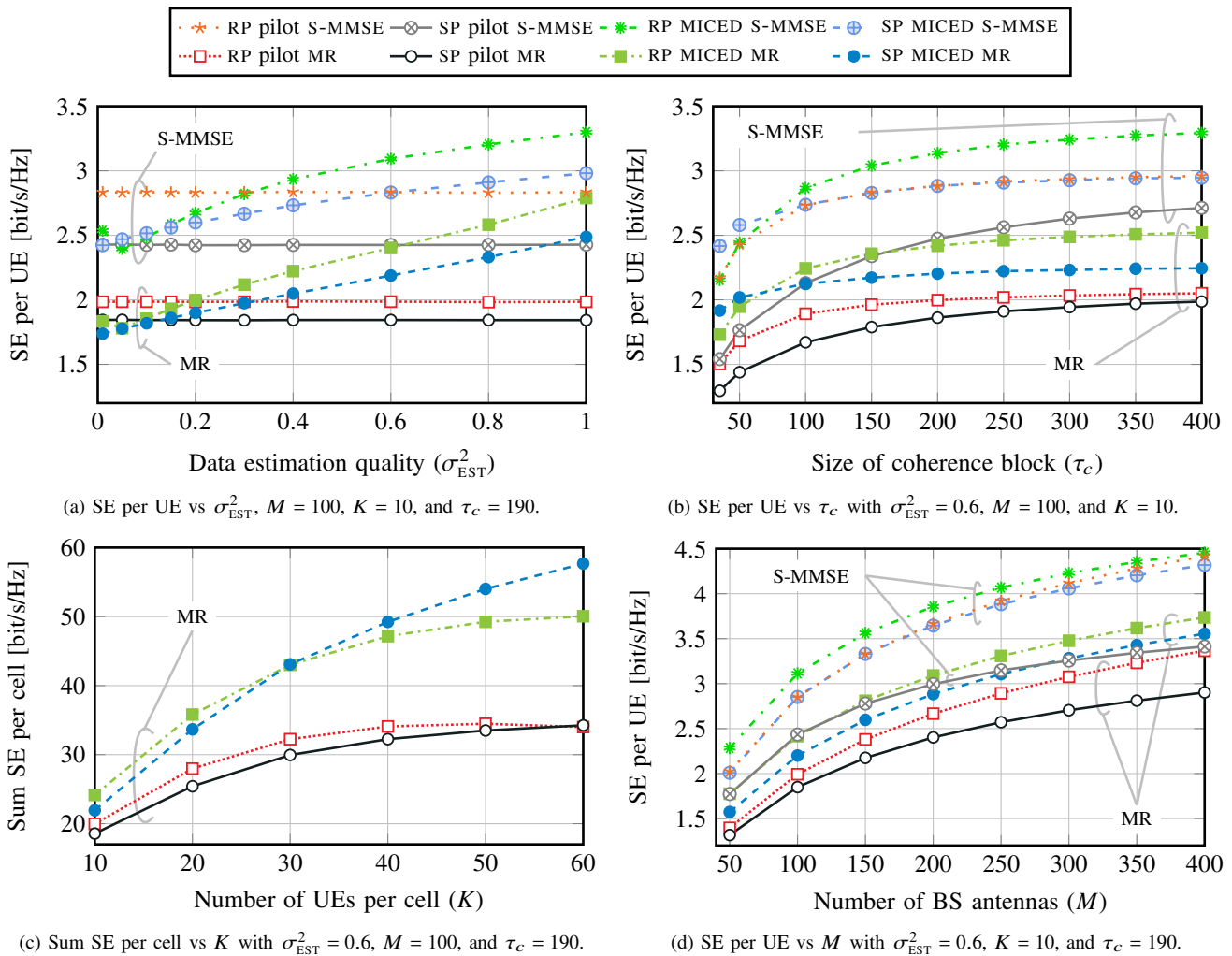
\begin{figure*}[!t]	
	\centering
	{\captionsetup{width=0.45\textwidth}	

\begin{tikzpicture}
	\begin{axis}[
	width=0.8\textwidth,
	height=50pt,
	axis line style={draw=none},
	tick style={draw=none},
	xticklabels=\empty,
	yticklabels=\empty,
	xmin = 0,
	xmax = 500,
	ymin = 0,
	ymax = 1,
	line width = 1pt,
	legend columns = 2,
	transpose legend,
	legend cell align={left},
	legend style={at={(0.5,0)},anchor=north,font=\small,line width=1pt,draw=black,mark size=0.1pt},
	]	

	\addlegendentry{\textsc{rp} pilot \textsc{s-mmse}}	
	\addlegendimage{orange, loosely dotted, mark=star, mark options={solid, line width = 0.7pt},  mark size=2.5pt}
	\addlegendentry{\textsc{rp} pilot \textsc{mr}}	
	\addlegendimage{rot,densely dotted,mark=square*, mark options={solid, line width = 0.7pt,fill=white},  mark size=2.2pt}	
	\addlegendentry{\textsc{sp} pilot \textsc{s-mmse}}	
	\addlegendimage{mittelgrau,solid,mark=otimes*, mark options={solid, line width = 0.7pt,fill=white},  mark size=2.5pt}	
	\addlegendentry{\textsc{sp} pilot \textsc{mr}}	
	\addlegendimage{anthrazit,solid,mark=*, mark options={solid, line width = 0.7pt,fill=white},  mark size=2.2pt}	
	\addlegendentry{\textsc{rp} \textsc{miced} \textsc{s-mmse}}	
	\addlegendimage{gelb,loosely dashdotted,mark=10-pointed star, mark options={solid, line width = 0.7pt},  mark size=2.5pt}
	\addlegendentry{\textsc{rp} \textsc{miced} \textsc{mr}}	
	\addlegendimage{apfelgruen,dashdotted,mark=square*, mark options={solid, line width = 0.7pt},  mark size=2.2pt}	
	\addlegendentry{\textsc{sp} \textsc{miced} \textsc{s-mmse}}	
	\addlegendimage{hellblau,loosely dashed,mark=oplus*, mark options={solid, line width = 0.7pt,fill=hellgrau},  mark size=2.5pt}
	\addlegendentry{\textsc{sp} \textsc{miced} \textsc{mr}}	
	\addlegendimage{mittelblau,dashed,mark=*, mark options={solid, line width = 0.7pt},  mark size=2.2pt}	
	\end{axis}
	\end{tikzpicture}%
		\\[-5pt]
		\subfloat[SE per UE vs $\sigma_{\textsc{est}}^2$, $M =100$, $K = 10$, and ${\Tul = 190}$.]{ 

\begin{tikzpicture}
	\begin{axis}[
	width=0.47\textwidth,
	height=0.32\textwidth,
	y label style = {at={(-0.02,0.5)},anchor=north},
	xlabel={Data estimation quality ($\sigma_{\textsc{est}}^2$)}, 
	ylabel={SE per UE [bit/s/Hz]},
	xmin = 0,
	xmax = 1,
	ymin = 1.2,
	ymax = 3.5,
	ytick distance = 0.5,
	line width = 1pt,
	grid=both,
	legend columns = 2,
	legend style={at={(axis cs: 0.98,1.8)},font=\footnotesize,line width=1pt,draw=black,mark size=0.1pt},
	]

\draw (axis cs:0.2,3.25) node[draw=none,fill =white,align = center] {\footnotesize{S-MMSE}};
\draw[-,color = lightgray] (axis cs:0.12,3.15) -- (axis cs:0.08,2.9);	
\draw[-,color = lightgray] plot [smooth , tension=4 ] coordinates{ (axis cs: 0.1,2.85) (axis cs: 0.07,2.6) (axis cs: 0.1,2.37)   };

\draw (axis cs:0.2,1.5) node[draw=none,fill =white,align = center] {\footnotesize{MR}};
\draw[-,color = lightgray] (axis cs:0.15,1.5) -- (axis cs:0.09,1.65);	
\draw[-,color = lightgray] plot [smooth , tension=4 ] coordinates{ (axis cs: 0.1,2.05) (axis cs: 0.08,1.85) (axis cs: 0.1,1.7)   };

\addplot [rot,densely dotted,mark=square*, mark options={solid, line width = 0.7pt,fill=white},  mark size=2.2pt]
table[x index={0}, y index={3}] {Figures/GaussSE_vs_dataEst.txt};		

\addplot [mittelgrau,solid,mark=otimes*, mark options={solid, line width = 0.7pt,fill=white},  mark size=2.5pt]
table[x index={0}, y index={14}] {Figures/GaussSE_vs_dataEst.txt};	

\addplot [anthrazit,solid,mark=*, mark options={solid, line width = 0.7pt,fill=white},  mark size=2.2pt]
table[x index={0}, y index={15}] {Figures/GaussSE_vs_dataEst.txt};				

\addplot [gelb,loosely dashdotted,mark=10-pointed star, mark options={solid, line width = 0.7pt},  mark size=2.5pt]
table[x index={0}, y index={5}] {Figures/GaussSE_vs_dataEst.txt};									

\addplot [apfelgruen,dashdotted,mark=square*, mark options={solid, line width = 0.7pt},  mark size=2.2pt]
table[x index={0}, y index={6}] {Figures/GaussSE_vs_dataEst.txt};		

	\addplot [hellblau,loosely dashed,mark=oplus*, mark options={solid, line width = 0.7pt,fill=hellgrau},  mark size=2.5pt]
	table[x index={0}, y index={17}] {Figures/GaussSE_vs_dataEst.txt};									

\addplot [mittelblau,dashed,mark=*, mark options={solid, line width = 0.7pt},  mark size=2.2pt]
table[x index={0}, y index={18}] {Figures/GaussSE_vs_dataEst.txt};			

	\addplot [orange, loosely dotted, mark=star, mark options={solid, line width = 0.7pt},  mark size=2.5pt]
	table[x index={0}, y index={2}] {Figures/GaussSE_vs_dataEst.txt};		

	\end{axis}
	\end{tikzpicture}%
			\label{fig:SE_vs_Estq} }
		\subfloat[SE per UE vs $\Tul$ with ${\sigma_{\textsc{est}}^2 = 0.6}$, ${M =100}$, and $K = 10$.]{

\begin{tikzpicture}
	\begin{axis}[
	width=0.47\textwidth,
	height=0.32\textwidth,
	y label style = {at={(-0.02,0.5)},anchor=north},
	xlabel={Size of coherence block ($\Tul\vphantom{\sigma_{\textsc{est}}^2}$)},
	ylabel={SE per UE [bit/s/Hz]},
	xmin = 30,
	xmax = 400,
	ymin = 1.2,
	ymax = 3.5,
	ytick distance = 0.5,
	line width = 1pt,
	grid=both,
	legend columns = 2,
	legend style={at={(axis cs: 395,1.8)},font=\footnotesize,line width=1pt,draw=black,mark size=0.1pt},
	]	

%
%

%
%
%

\draw (axis cs:85,3.3) node[draw=none,fill =white,align = center] {\footnotesize{S-MMSE}};
\draw[-,color = lightgray] (axis cs:140,3.3) -- (axis cs:374,3.4);	
\draw[-,color = lightgray] plot [smooth , tension=3.5 ] coordinates{ (axis cs: 380,3.38) (axis cs: 370,3) (axis cs: 380,2.7)   };

\draw (axis cs:300,1.5) node[draw=none,fill =white,align = center] {\footnotesize{MR}};
\draw[-,color = lightgray] (axis cs:320,1.5) -- (axis cs:385,1.92);	
\draw[-,color = lightgray] plot [smooth , tension=3 ] coordinates{ (axis cs: 390,2.55) (axis cs: 380,2.25) (axis cs: 390,1.95)   };
\addplot [rot,densely dotted,mark=square*, mark options={solid, line width = 0.7pt,fill=white},  mark size=2.2pt]
table[x index={0}, y index={3}] {Figures/GaussSE_vs_tau_c.txt};		

\addplot [mittelgrau,solid,mark=otimes*, mark options={solid, line width = 0.7pt,fill=white},  mark size=2.5pt]
table[x index={0}, y index={14}] {Figures/GaussSE_vs_tau_c.txt};	

\addplot [anthrazit,solid,mark=*, mark options={solid, line width = 0.7pt,fill=white},  mark size=2.2pt]
table[x index={0}, y index={15}] {Figures/GaussSE_vs_tau_c.txt};				

\addplot [gelb,loosely dashdotted,mark=10-pointed star, mark options={solid, line width = 0.7pt},  mark size=2.5pt]
table[x index={0}, y index={5}] {Figures/GaussSE_vs_tau_c.txt};									

\addplot [apfelgruen,dashdotted,mark=square*, mark options={solid, line width = 0.7pt},  mark size=2.2pt]
table[x index={0}, y index={6}] {Figures/GaussSE_vs_tau_c.txt};		

\addplot [hellblau,loosely dashed,mark=oplus*, mark options={solid, line width = 0.7pt,fill=hellgrau},  mark size=2.5pt]
table[x index={0}, y index={17}] {Figures/GaussSE_vs_tau_c.txt};									

\addplot [mittelblau,dashed,mark=*, mark options={solid, line width = 0.7pt},  mark size=2.2pt]
table[x index={0}, y index={18}] {Figures/GaussSE_vs_tau_c.txt};			

\addplot [orange, loosely dotted, mark=star, mark options={solid, line width = 0.7pt},  mark size=2.5pt]
table[x index={0}, y index={2}] {Figures/GaussSE_vs_tau_c.txt};		

	\end{axis}
	\end{tikzpicture}%
			\label{fig:SE_vs_tau_c}}			
		\\[-5pt] 
		\subfloat[Sum SE per cell vs $K$ with  ${\sigma_{\textsc{est}}^2 = 0.6}$, $M=100$, and $\Tul = 190$.]{

\begin{tikzpicture}
	\begin{axis}[
	width=0.47\textwidth,
	height=0.32\textwidth,
	y label style = {at={(-0.02,0.5)},anchor=north},
	xlabel={Number of UEs per cell ($K$)},
	ylabel={Sum SE per cell [bit/s/Hz]},
	xmin = 10,
	xmax = 60,
	ymin = 17,
	ymax = 60,
	ytick distance = 10,
	line width = 1pt,
	grid=both,
	legend columns = 2,
	legend style={at={(axis cs: 59,27)},font=\footnotesize,line width=1pt,draw=black,mark size=0.1pt},
	]

\draw (axis cs:20,50) node[draw=none,fill =white,align = center] {\footnotesize{MR}};
\draw[-,color = lightgray] (axis cs:20,48) -- (axis cs:15,33.5);	
\draw[-,color = lightgray] plot [smooth , tension=3 ] coordinates{ (axis cs: 16,21) (axis cs: 14,27) (axis cs: 16,33)   };


\addplot [rot,densely dotted,mark=square*, mark options={solid, line width = 0.7pt,fill=white},  mark size=2.2pt]
table[x index={0}, y index={3}] {Figures/GaussSE_vs_K.txt};		


\addplot [anthrazit,solid,mark=*, mark options={solid, line width = 0.7pt,fill=white},  mark size=2.2pt]
table[x index={0}, y index={15}] {Figures/GaussSE_vs_K.txt};				


\addplot [apfelgruen,dashdotted,mark=square*, mark options={solid, line width = 0.7pt},  mark size=2.2pt]
table[x index={0}, y index={6}] {Figures/GaussSE_vs_K.txt};		


\addplot [mittelblau,dashed,mark=*, mark options={solid, line width = 0.7pt},  mark size=2.2pt]
table[x index={0}, y index={18}] {Figures/GaussSE_vs_K.txt};			


	\end{axis}
	\end{tikzpicture}%
			\label{fig:SE_vs_K}}
		\subfloat[SE per UE vs $M$ with ${\sigma_{\textsc{est}}^2 = 0.6}$, $K = 10$, and  $\Tul = 190$.]{ 

\begin{tikzpicture}
	\begin{axis}[
	width=0.47\textwidth,
	height=0.32\textwidth,
	y label style = {at={(-0.02,0.5)},anchor=north},
	xlabel={Number of BS antennas ($M$)},
	ylabel={SE per UE [bit/s/Hz]},
	xmin = 40,
	xmax = 400,
	ymin = 1.2,
	ymax = 4.5,
	ytick distance = 0.5,
	line width = 1pt,
	grid=both,
	legend columns = 2,
	legend style={at={(axis cs: 395,2.1)},font=\footnotesize,line width=1pt,draw=black,mark size=0.1pt},
	]	

%
%

%
%
%

\draw (axis cs:110,4.2) node[draw=none,fill =white,align = center] {\footnotesize{S-MMSE}};
\draw[-,color = lightgray] (axis cs:150,4.2) -- (axis cs:230,4.05);	
\draw[-,color = lightgray] plot [smooth , tension=3.5 ] coordinates{ (axis cs: 235,4.08) (axis cs: 230,3.9) (axis cs: 235,3.7)   };
\draw[-,color = lightgray] (axis cs:212,3.1) -- (axis cs:150,4.2);	
\draw[-,color = lightgray] plot [smooth , tension=3.1 ] coordinates{ (axis cs: 215,2.99) (axis cs: 210,3.04) (axis cs: 215,3.09)   };

\draw (axis cs:350,1.5) node[draw=none,fill =white,align = center] {\footnotesize{MR}};
\draw[-,color = lightgray] (axis cs:350,1.65) -- (axis cs:325,2.66);	
\draw[-,color = lightgray] plot [smooth , tension=3.2 ] coordinates{ (axis cs: 330,2.7) (axis cs: 320,2.95) (axis cs: 330,3.2)   };
\draw[-,color = lightgray] (axis cs:350,1.65) -- (axis cs:375,3.43);	
\draw[-,color = lightgray] plot [smooth , tension=3 ] coordinates{ (axis cs: 380,3.45) (axis cs: 370,3.6) (axis cs: 380,3.75)   };

\addplot [rot,densely dotted,mark=square*, mark options={solid, line width = 0.7pt,fill=white},  mark size=2.2pt]
table[x index={0}, y index={3}] {Figures/GaussSE_vs_M.txt};		

\addplot [anthrazit,solid,mark=*, mark options={solid, line width = 0.7pt,fill=white},  mark size=2.2pt]
table[x index={0}, y index={15}] {Figures/GaussSE_vs_M.txt};				

\addplot [gelb,loosely dashdotted,mark=10-pointed star, mark options={solid, line width = 0.7pt},  mark size=2.5pt]
table[x index={0}, y index={5}] {Figures/GaussSE_vs_M.txt};									

\addplot [apfelgruen,dashdotted,mark=square*, mark options={solid, line width = 0.7pt},  mark size=2.2pt]
table[x index={0}, y index={6}] {Figures/GaussSE_vs_M.txt};		

\addplot [hellblau,loosely dashed,mark=oplus*, mark options={solid, line width = 0.7pt,fill=hellgrau},  mark size=2.5pt]
table[x index={0}, y index={17}] {Figures/GaussSE_vs_M.txt};									

\addplot [mittelblau,dashed,mark=*, mark options={solid, line width = 0.7pt},  mark size=2.2pt]
table[x index={0}, y index={18}] {Figures/GaussSE_vs_M.txt};			

\addplot [orange, loosely dotted, mark=star, mark options={solid, line width = 0.7pt},  mark size=2.5pt]
table[x index={0}, y index={2}] {Figures/GaussSE_vs_M.txt};		

\addplot [mittelgrau,solid,mark=otimes*, mark options={solid, line width = 0.7pt,fill=white},  mark size=2.5pt]
table[x index={0}, y index={14}] {Figures/GaussSE_vs_M.txt};	

	\end{axis}
	\end{tikzpicture}%
			\label{fig:SE_vs_M} }	
	}			
	\caption{Sum SE per cell versus $K$, and SE per UE versus $\sigma_{\textsc{est}}^2$, $\Tul$ and $M$ for $\varrho = \sigma^2$ (SNR~${=0}$~dB). The data estimation quality is selected such that ${\sigma_{lk}^{\iter{}{i}^2} = \sigma_{\textsc{est}}^2}$ ${\forall k\in \{1,\ldots,K\}}$.}
	\label{fig:Gauss_SE}
	\hrulefill 
\end{figure*}

Figure~\ref{fig:SE_MSE_Gauss} depicts the MSE of the channel estimates versus the data estimation quality and size of the coherence block. Note that with RP $\Tuld = \Tul -\Tp$ and with SP $\Tuld = \Tul$, thus $\Tul$ is always proportional to the number of transmitted data symbols. Figure~\ref{fig:MSE_vs_Estq} shows that with RP having $Tp = K$, and SP the MSE of the channel estimates is a decreasing function of the data estimation quality. Moreover, with relatively low values of data estimation quality (i.e., $\sigma_\textsc{est}^2>0.2$), the MSE of the channel estimates improves with respect to their initial value with pilot-based channel estimation only. In practice, one might have good data estimation quality for some UEs and utilize that to reduce the MSE for the channels to other UEs. Figure~\ref{fig:MSE_vs_Estq} also shows the MSE of the channel estimates with RP having $\Tp = 3K$, that is a pilot reuse of 3, which is a standard approach to mitigate the effect of pilot contamination. The same channel quality as in that case can be achieved by the MICED algorithm when the data estimation quality is high enough.

The performance of the MICED algorithm depends on the size of the coherence block. In Figure~\ref{fig:MSE_vs_tau_c}, as $\Tul$ increases with RP, the use of data-aided channel estimation continuously decreases the MSE and the improvement over pilot-based channel estimation increases accordingly which is a result from having more observations. On the other hand, with SP, the MSE of the channel estimates with both pilot-based and data-aided methods decreases at the same pace with $\Tul$ since the number of observations is the same in both cases. However, the data-aided channel estimation reduces the MSE compared to pilot-based channel estimation. In addition, when the coherence block is large enough, the data-aided channel estimation quality is higher compared to the standard pilot-based approach with RP and pilot reuse 3.

Figure~\ref{fig:SE_vs_Estq} shows that when the quality of the data estimates is very low, the use of the MICED algorithm provides little or no improvement in terms of SE per UE since the use of corrupted data estimates fails to reduce the intracell interference. However, when the data estimation quality is above a certain value, the SE per UE becomes an increasing function of the data estimation quality and improves with respect to their initial value with pilot-based channel estimation only. Figure~\ref{fig:SE_vs_tau_c} shows that the SE per UE is an increasing function of $\Tul$ for all methods. In the case with RP, the gap between SE with pilot-based and data-aided channel estimation increases with $\Tul$ since the MICED algorithm utilizes more observations for data-aided channel estimation as $\Tul$ increases, whereas, the number of observations with pilot-based channel estimation remain the same. In addition, with RP the benefit of the MICED algorithm is lower when using S-MMSE processing which is a consequence of the additional interference added in the data-aided channel estimates making the interference suppression less accurate. In the case of SP, the gap between SE with pilot-based and data-aided estimation decreases with $\Tul$, which means that the MICED algorithm provides more benefits when the size of the coherence block is short. 

In summary, when the quality of the data estimates is high enough, the MICED algorithm can lower the MSE of the channel estimation, and in turn, increase the SE per UE. By comparing the SE with RP and SP, the former provides higher SE in most cases except when the number of samples in the coherence block is low and data-aided channel estimation is used. Thus, the MICED algorithm with RP is more beneficial in low mobility scenarios or low carrier frequencies with long coherence blocks, while the MICED algorithm with SP performs best in high mobility scenarios or high carrier frequencies where the size of the coherence block is short. In addition, notice that in Figure~\ref{fig:SE_vs_tau_c} there is a cross point between RP and SP using the MICED algorithm, and this point depends on the number of multiplexed UEs $K$. Note that when $K$ increases and $\Tc$ remains fixed, the pilot overhead with RP limits the SE making SP the preferred choice. The aforementioned cross point between RP and SP can also be observed in Figure~\ref{fig:SE_vs_K} where the sum SE per cell is plotted versus $K$ with MR. The same behavior is found with S-MMSE but the plots are omitted for ease of illustration.
\begin{remark}
	\label{rem:aggr_spatial_mtplx}
	The cross point between the SE with RP and SP with respect to $\Tc$ (see in Figure~\ref{fig:SE_vs_tau_c}), and $K$ (see Figure~\ref{fig:SE_vs_K}) indicates that the MICED algorithm with SP also has the possibility to utilize more aggressive spatial multiplexing that not only increases SE but also facilitates the implementation of machine type communication systems where many UEs need to be served.
\end{remark}

Figure~\ref{fig:SE_vs_M} depicts the SE per UE versus the number of BS antennas. In the case with RP, the benefit of the MICED algorithm is higher when using MR processing and it becomes less significant for S-MMSE when the number of BS antennas grows large. This indicates that due to the additional interference in the data-aided channel estimation with RP the interference suppression capabilities of S-MMSE are less effective compared to performing MR combining and then subracting the estimated intracell interference as shown in \eqref{eq:data_obs_i_RP}. On the other hand, with SP, the benefit of using the MICED algorithm compared to pilot-based channel estimations grows with the number of BS antennas and it is higher for S-MMSE processing. Here, the pilot-based channel estimates have interference from all UEs and the MICED algorithm reduces the intracell interference, which in turn, enhances the interference suppression of S-MMSE.
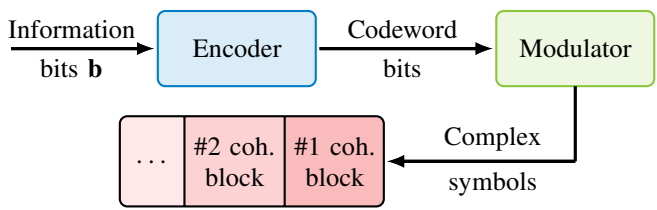
\begin{figure}[!t] 
	\centering 

\begin{tikzpicture}
\tikzset{>=latex}
\tikzstyle{box} = [draw,rounded corners=.1cm,minimum height=3em, minimum width=2em, align=center, thick] 
\tikzstyle{vertical_box} = [draw,rounded corners=.1cm,minimum height=2em, minimum width=3em, align=center, thick,rotate=90] 
\tikzstyle{data}=[rectangle split,rectangle split parts=2,draw,text centered]

\def\originx{0}
\def\originy{0}

\node[box,minimum height=1cm,minimum width=2.1cm,draw=mittelblau,fill=mittelblau!15] (CE) at (\originx,\originy) {Encoder};

\node[box,minimum height=1cm,minimum width=2.1cm,draw=apfelgruen,fill=apfelgruen!15] (LC) at (\originx+4.5,\originy) {Modulator};

\node[rectangle split, rectangle split horizontal, rectangle split parts=3,box,rectangle split part fill={rot!10,rot!20,rot!30},minimum height=1.2cm,minimum width=2.25cm] (DE) at (\originx+0.2,\originy-1.5) { $\; \bf{\dots} \;$  \nodepart{two} \#2 coh.\\  block  \nodepart{three} \#1  coh.\\ block};

\draw[->,black,very thick] (\originx-3,\originy) node [above,xshift=0.8cm] {Information}  -- (\originx-1.1,\originy) ;
\draw[->,black,very thick] (\originx-3,\originy) node [below,xshift=0.8cm] {bits $\bf{b}$}  -- (\originx-1.1,\originy) ;

\draw[->,black,very thick] (\originx+1.1,\originy) node [above,xshift=1.1cm] {Codeword}  -- (\originx+3.4,\originy) ;
\draw[->,black,very thick] (\originx+1.1,\originy) node [below,xshift=1.1cm] {bits }  -- (\originx+3.4,\originy) ;

\draw[-,black,very thick] (\originx+4.5,\originy-0.5)  -- (\originx+4.5,\originy-1.5) ;
\draw[->,black,very thick] (\originx+4.5,\originy-1.5) node [above,xshift=-1.1cm] {Complex} -- (\originx+2,\originy-1.5) ;
\draw[->,black,very thick] (\originx+4.5,\originy-1.5) node [below,xshift=-1.1cm] {symbols} -- (\originx+2,\originy-1.5) ;

\end{tikzpicture}
	\caption{Commun. model over several coherence blocks.} 
	\label{fig:Com_mod} 
\end{figure} 
\begin{figure*}[t]
	\allowdisplaybreaks\begin{align*}	\allowdisplaybreaks
	&(\hat{\bf{y}}_{lk}^\textsc{rp})^T =
\mkern -180mu \underbrace{
		\vphantom{\sum_{\substack{k'=1 \\ \mkern -38 mu (\ell, k' )\neq (l,k) }}^{K}}
		\bf{v}_{lk}^H\hat{\bf{h}}_{llk}\sqrt{\pulrp_{lk}}}_{\parbox{0.5\textwidth}{\centering \footnotesize $\mkern 180mu = g_{lk}^\textsc{rp}$ \textrm{(known equivalent channel)}}}\mkern -180mu\bf{s}_{lk}^T 
	+ \underbrace{\bf{v}_{lk}^H \left(\vphantom{\sum_K^K}\right.
		\tilde{\bf{h}}_{llk}\sqrt{\pulrp_{lk}}\bf{s}_{lk}^T  + \sum_{\ell\in \Phi}\sum_{\substack{k'=1 \\ \mkern -38 mu (\ell, k' )\neq (l,k) }}^{K} \bf{h}_{l\ell k'}\sqrt{\pulrp_{\ell k'}}\bf{s}_{\ell k'}^T  
		- \sum_{k' \neq k} \hat{\bf{h}}_{llk'}\sqrt{\pulrp_{lk'}}\hat{\bf{s}}_{lk'}^T
		+ [\bf{N}]_{\Tp+1:\Tul}
		\left.\vphantom{\sum_K^K}\right) 	 	
	}_{ = \sf{n}_{lk}^\textsc{rp}\, \textrm{(effective noise)}}
	\tag{30}\label{eq:data_est_RPfull}
	\\
	& (\hat{\bf{y}}_{lk}^\textsc{sp})^T  =
	\mkern -180mu\underbrace{\bf{v}_{lk}^H\hat{\bf{h}}_{llk}\sqrt{\pulsp_{lk}}  		
		\vphantom{\sum_{\substack{K \\ K \neq }^K}}
	}_{\parbox{0.5\textwidth}{ \footnotesize $\mkern 250mu= g_{lk}^\textsc{sp}$ \textrm{(known equivalent channel)}}}\mkern -180mu\bf{s}_{lk}^T 
	+  \underbrace{\bf{v}_{lk}^H\left(\vphantom{\sum_K^K}\right.
		\tilde{\bf{h}}_{llk}\bf{x}_{lk}^T + \sum_{\ell\in \Phi}\sum_{\substack{k'=1 \\ \mkern -35 mu (\ell, k' )\neq (l,k) }}^{K}\bf{h}_{l\ell k'} \bf{x}_{\ell k'}^T 
		- \sum_{k' \neq k} \hat{\bf{h}}_{llk'}\hat{\bf{x}}_{lk'}^T
		+  \bf{N}
		\left.\vphantom{\sum_K^K}\right)}_{ = \sf{n}_{lk}^\textsc{sp}\, \textrm{(effective noise)}}
	\tag{31}\label{eq:data_est_SPfull}
	\end{align*}
	\hrulefill
\end{figure*}
\section{Finite alphabet symbols}
\label{sec:finiteAlphabet}
The MICED algorithm was introduced in Section~\ref{sec:dataAidChEst} and evaluated under the assumption of Gaussian data symbols to perform a tractable theoretical analysis. This gave key insights into the cases where the MICED algorithm can provide gains compared to pilot-based channel estimation in terms of channel estimation quality and SE. In this section, the implementation of the MICED algorithm with finite-alphabet modulation is described and evaluated in terms of achievable SE and BLER to further assess its potential benefits in real systems. 

In practical implementations, the information bits sent over a communication system are encoded into finite length codewords by using a predefined channel code. This procedure adds redundant information to combat the errors introduced by the variations of the channel. The bits that make up the codewords are then modulated into complex symbols which in turn are transmitted over the channel. Since typical codewords are made up of long sequences of bits, the resulting number of modulated symbols tends to span several coherence blocks, as illustrated in Figure~\ref{fig:Com_mod}. Moreover, the pilot symbols are inserted into each coherence block along with the modulated data symbols, as shown in Figure~\ref{fig:CohblockRPSP}, resulting in different number of coherence blocks  that contain a full codeword with RP or SP. To successfully decode the received bits and retrieve the information bits at the receiver, the complex symbols containing the bits that form a full codeword need to be received. Based on the observations of the complex symbols obtained at the receiver, log-likelihood ratios (LLR) for each bit in the codeword are computed, and then fed into the decoding algorithm.

Let $N_{mod}$ be the size of the alphabet used by the modulation scheme, and $N_b = \log_2(N_{mod})$ the number of bits per complex modulation symbol. Denote by $\{b_{lk1},\ldots,b_{lkN_b}\}$ a set of bits transmitted by $\rm{UE}_{lk}$, and modulated into an arbitrary complex symbol represented by ${s_{lk} = [\bf{s}_{lk}]_j}$ for any $j\in\{1,\ldots,\Tuld\}$ in a given coherence block. Then, the corresponding observation ${\hat{y}_{lk} = [\hat{\bf{y}}_{lk}]_j}$ obtained from \eqref{eq:data_est_RP} with RP, and \eqref{eq:data_est_SP} with SP, is given by \eqref{eq:data_est_RPfull} and \eqref{eq:data_est_SPfull} respectively, at the top of the next page.
 Notice that to fully utilize the side information of the channel and data estimates, an estimate of the received intracell interference is subtracted, see third term of the effective noise in  \eqref{eq:data_est_RPfull} and \eqref{eq:data_est_SPfull}. The LLR of an arbitrary bit $b_{lkn}$ for $n\in\{1,\ldots,N_b\}$ is given by
\begin{align*}
&\rm{LLR}_{lkn} \!=\! \log\!\left(\!\frac{\rm{Pr}\left( b_{lkn} = 0 | \hat{y}_{lk}\right) }{\rm{Pr}\left( b_{lkn} = 1 | \hat{y}_{lk}\right)}\!\right)
\\
&
			 \!  = \!\log\!\left(\!\frac{ \sum_{i' = 1}^{\frac{N_{mod}}{2}} \rm{Pr}\left(  \hat{y}_{lk} | b_{lkn} \in \cal{H}_{i',b_{lkn} =0} \right) \rm{Pr}\left(b_{lkn} \in \cal{H}_{i',b_{lkn} =0} \right)}{\sum_{i' = 1}^{\frac{N_{mod}}{2}} \rm{Pr}\left(  \hat{y}_{lk} | b_{lkn} \in \cal{H}_{i',b_{lkn} =1} \right) \rm{Pr}\left(b_{lkn} \in \cal{H}_{i',b_{lkn} =1} \right)}\!\right)
\stepcounter{equation}\tag{\theequation}\label{eq:LLR_def}
\end{align*}   
where the set $\cal{H}_{i',b_{lkn} =1} = \{b_{lk1},\ldots,b_{lkn} = 1,\ldots,b_{lkN_b}\}$ is defined as one of the $N_{mod}/2$ possible sets of modulated bits where $b_{lkn} = 1$ and $b_{lkn'} \in \{0,1\}$ for $n'\neq n$. After computing the LLRs for each bit, the decoder utilizes the redundancy in the channel code to correct errors, and obtain the maximum likelihood estimate of the originally transmitted bits. Notice that due to interference and noise, this procedure is not always perfect leading to errors that cannot be corrected by the decoder. For example, the decoder might be able to decode the signal from UEs that are close to the BS (since they would have a high channel gain with respect to the interference and noise), but not for cell-edge UEs that are more susceptible to interference. Similarly, UEs that are subject to strong pilot contamination are more likely to get decoding errors.

The performance of the decoder depends on the effective SNR between the power of the equivalent channel and the effective noise (see \eqref{eq:data_est_RPfull} with RP, and \eqref{eq:data_est_SPfull} with SP), that is denoted as $\rm{SNR}_{lk}^\textsc{eff} = \bb{E}\{|g_{lk}|^2\}/\bb{E}\{|\sf{n}_{lk}|^2\}$. At this stage, the interference mitigation processing has already been done and therefore the effective noise is treated as a noise rather than interference, even though it is made up of interference terms as well as noise. The relation between $\rm{SNR}_{lk}^\textsc{eff}$ and the SNR required to successfully decode the information determines how the decoder performs. Notice, that $\rm{SNR}_{lk}^\textsc{eff}$ depends on the channel estimation accuracy and the linear combining strategy, as well as the interference level and pilot contamination effect. If $\rm{SNR}_{lk}^\textsc{eff}$ is too low, the decoder will fail and there will be some erroneous bits. However, the LLRs at the output of the decoder could still be used, as side information, to estimate the complex modulated data symbols that were sent. More importantly, if the information of some other UEs is decoded successfully, then, perfect knowledge of their complex modulated data symbols will be available.

The data estimates are obtained from LLRs at the output of the decoder which, in turn, require the data observations (see \eqref{eq:data_est_RPfull} and \eqref{eq:data_est_SPfull}) corresponding to all bits in a full codeword. Thus, let $N_\textsc{cw}$ be the number of bits that make a full codeword, and denote by ${\bar{\bf{y}}_{lk} \in \bb{C}^{N_\textsc{cw}/N_b}}$ the observations of the corresponding complex symbols obtained by staking several instances of \eqref{eq:data_est_RPfull} and \eqref{eq:data_est_SPfull}, with RP and SP respectively.\footnote{Recall that one codeword spans several coherence blocks, see Figure~\ref{fig:Com_mod}.} To obtain the estimates of the complex data symbols, the LLRs of each bit (after the decoding procedure) are mapped into complex symbols based on the modulation scheme used. Assume that the complex symbol alphabet is given by the set $\cal{A} = \{s(1),\ldots, s(N_{mod})\}$ where each symbol maps $N_b$ bits such that $\cal{B}_j = \{b_j(1),\ldots,b_j(N_b)\}$ corresponds to the set of bits mapped into a given symbol $s(j)$. At the $i^{th}$ iteration the MMSE estimate of an arbitrary data symbol $s_{lk}$ is
 \addtocounter{equation}{2}\allowdisplaybreaks
\begin{align*}
\hat{s}_{lk}^\iter{}{i} &= \bb{E}\{ s_{lk} | \bar{\bf{y}}_{lk}^\iter{}{i}  \}
=  \sum_{j=1}^{N_{mod}}  s(j)\rm{Pr}(s_{lk} = s(j)| \bar{\bf{y}}_{lk}^\iter{}{i} )
\\
&
=\sum_{j=1}^{N_{mod}}  s(j) \prod_{n = 1}^{N_b} \rm{Pr}(b_{lkn} = b_j(n) | \bar{\bf{y}}_{lk}^\iter{}{i})
\end{align*}
where the conditional probabilities of each bit are given by
\begin{align*}
\rm{Pr}(b_{lkn} = 0 | \bar{\bf{y}}_{lk}^\iter{}{i}) &= \frac{1}{1 + \exp(-\rm{LLR}_{lkn}^\iter{d}{i})},\\
\rm{Pr}(b_{lkn} = 1 | \bar{\bf{y}}_{lk}^\iter{}{i}) &= \frac{1}{1 + \exp(\rm{LLR}_{lkn}^\iter{d}{i})},
\stepcounter{equation} \tag{\theequation} \label{eq:Prob_LLR}
\end{align*}
and $\rm{LLR}_{lkn}^\iter{d}{i}$ represents the LLR after the decoding procedure for the $i^{th}$ iteration. Then, the variance of the data estimates is given by $\sigma_{lk}^{\iter{}{i}^2} = \bb{E}\left\{ |\hat{s}_{lk}^\iter{}{i} |^2\right\}$, which can be estimated by taking a sample mean for all estimated symbols in a codeword.  

It is worth mentioning that the MICED algorithm does not depend on the channel code being used, and thus, it can be implemented with any state of the art decoder. This stands in contrast to \cite{Ma2017a_MultiUE} which focuses on the design of a forward-error-correction channel code.  

\begin{figure*}[!t]		
	\centering
	{
		\captionsetup{width=0.45\textwidth}
		\subfloat[BLER vs $i$ with $M = 100$, and $\varrho = \sigma^2$ (SNR~${=0}$ dB). ]{

\begin{tikzpicture}
	\begin{semilogyaxis}[
	width=0.475\textwidth,
	height=0.35\textwidth,
	y label style = {at={(-0.03,0.5)},anchor=north},
	xlabel={Number of iterations ($i$)},
	ylabel={BLER},
	yticklabels=\empty,
	extra y ticks={0.05 , 0.1, 0.15, 0.2},
	extra y tick labels={0.05, 0.1, 0.15, 0.2},
	xmin = 0,
	xmax = 15,
	ymin = 0.04,
	ymax = 0.2,
	ytick distance = 0.05,
	line width = 1pt,
	grid=both,
	legend columns = 2,
	legend style={at={(axis cs: 15,0.2)},font=\footnotesize,line width=1pt,draw=black,mark size=0.1pt},
	]	
%
%
%
\addlegendentry{\footnotesize \textsc{sp mr}}	
\addlegendimage{mittelblau,dashed,mark=*, mark options={solid, line width = 0.7pt},  mark size=1.5pt}

\addlegendentry{\footnotesize \textsc{sp s-mmse}}	
\addlegendimage{hellblau,loosely dashed,mark=oplus*, mark options={solid, line width = 0.7pt,fill=hellgrau},  mark size=1.8pt}

\addlegendentry{\footnotesize \textsc{rp mr}}	
\addlegendimage{apfelgruen,dashdotted,mark=square*, mark options={solid, line width = 0.7pt},  mark size=1.5pt}

\addlegendentry{\footnotesize \textsc{rp s-mmse}}	
\addlegendimage{gelb,loosely dashdotted,mark=10-pointed star, mark options={solid, line width = 0.7pt},  mark size=1.8pt}

\draw (axis cs:5,0.045) node[draw=none,fill =white,align = center] {\footnotesize{ \textsc{rp s-mmse} pilot reuse-3  }};
\draw[-,color = gray] (axis cs:2.2,0.049) -- (axis cs:1,0.055);	
\draw[-,color = gray] plot [smooth , tension=4 ] coordinates{ (axis cs: 1,0.055) (axis cs: 0.7,0.056) (axis cs: 1,0.059)   };

\addplot [pink,solid,mark=none]
table[x index={0}, y index={8}] {Figures/QPSK_BLER_vs_iter_to_20_R05.txt};

\addplot [apfelgruen,dashdotted,mark=square*, mark options={solid, line width = 0.7pt},  mark size=1.5pt]
table[x index={0}, y index={4}] {Figures/QPSK_BLER_vs_iter_to_20_R05.txt};									

\addplot [gelb,loosely dashdotted,mark=10-pointed star, mark options={solid, line width = 0.7pt},  mark size=1.8pt]
table[x index={0}, y index={5}] {Figures/QPSK_BLER_vs_iter_to_20_R05.txt};


\addplot [mittelblau,dashed,mark=*, mark options={solid, line width = 0.7pt},  mark size=1.5pt]
table[x index={0}, y index={16}] {Figures/QPSK_BLER_vs_iter_to_20_R05.txt};
									
\addplot [hellblau,loosely dashed,mark=oplus*, mark options={solid, line width = 0.7pt,fill=hellgrau},  mark size=1.8pt]
table[x index={0}, y index={17}] {Figures/QPSK_BLER_vs_iter_to_20_R05.txt};

	\end{semilogyaxis}
	\end{tikzpicture}%
			\label{fig:PER_vs_iter}
		}		
		\subfloat[BLER vs SNR with $i = 8$, and $M = 100$.]{

\begin{tikzpicture}
	\begin{semilogyaxis}[
	width=0.475\textwidth,
	height=0.35\textwidth,
	y label style = {at={(-0.03,0.5)},anchor=north},
	xlabel={SNR ($\varrho/\sigma^2$) \text{[dB]}},
	ylabel={BLER},
	xmin = -25,
	xmax = 10,
	ymin = 0.04,
	ymax = 1,
	xtick distance = 5,
	line width = 1pt,
	grid=both,
	legend columns = 1,
	legend style={at={(axis cs: 10,1)},font=\footnotesize,line width=1pt,draw=black,mark size=0.1pt},
	]


	\addlegendentry{\footnotesize \textsc{sp mr} pilot}	
	\addlegendimage{anthrazit,solid,mark=*, mark options={solid, line width = 0.7pt,fill=white},  mark size=1.5pt}

	\addlegendentry{\footnotesize \textsc{rp mr} pilot}	
	\addlegendimage{rot,densely dotted,mark=square*, mark options={solid, line width = 0.7pt,fill=white},  mark size=1.5pt}
	
	

	\addlegendentry{\footnotesize \textsc{sp mr miced}}	
	\addlegendimage{mittelblau,dashed,mark=*, mark options={solid, line width = 0.7pt},  mark size=1.5pt}
	
	\addlegendentry{\footnotesize \textsc{rp mr miced}}	
	\addlegendimage{apfelgruen,dashdotted,mark=square*, mark options={solid, line width = 0.7pt},  mark size=1.5pt}



	\draw (axis cs:-14.8,0.052) node[draw=none,fill =white,align = center] {\footnotesize{ {\hspace*{-8pt} \textsc{rp s-mmse} pilot reuse-3} \hspace*{-8pt} }};%
	\draw[-,color = gray] (axis cs:-15,0.062) -- (axis cs:-10,0.079);	
	\draw[-,color = gray] plot [smooth , tension=4 ] coordinates{ (axis cs: -9,0.081) (axis cs: -10,0.079) (axis cs: -9,0.072)   };

\addplot [pink,  solid, mark=none] 					   
table[x index={0}, y index={8}] {Figures/QPSK_PER_vs_SNR_R05.txt};

\addplot [rot,densely dotted,mark=square*, mark options={solid, line width = 0.7pt,fill=white},  mark size=1.5pt]
table[x index={0}, y index={1}] {Figures/QPSK_PER_vs_SNR_R05.txt};

	
\addplot [apfelgruen,dashdotted,mark=square*, mark options={solid, line width = 0.7pt},  mark size=1.5pt]
table[x index={0}, y index={4}] {Figures/QPSK_PER_vs_SNR_R05.txt};									
	

	\addplot [anthrazit,solid,mark=*, mark options={solid, line width = 0.7pt,fill=white},  mark size=1.5pt]
	table[x index={0}, y index={13}] {Figures/QPSK_PER_vs_SNR_R05.txt};

	
	\addplot [mittelblau,dashed,mark=*, mark options={solid, line width = 0.7pt},  mark size=1.5pt]
	table[x index={0}, y index={16}] {Figures/QPSK_PER_vs_SNR_R05.txt};


\end{semilogyaxis}
\end{tikzpicture}%
			\label{fig:PER_vs_rho_R05}
		}
		\\	
		\subfloat[BLER vs $M$ with RP, $i = 8$, and $\varrho = \sigma^2$ (SNR~${=0}$ dB).]{

\begin{tikzpicture}
	\begin{semilogyaxis}[
	width=0.475\textwidth,
	height=0.35\textwidth,
	y label style = {at={(-0.03,0.5)},anchor=north},
	xlabel={Number of BS antennas ($M$)},
	ylabel={BLER},
	extra x ticks={25},
	extra x tick labels={25},		
	xmin = 25,
	xmax = 400,
	ymin = 0.003,
	ymax = 1,
	line width = 1pt,
	grid=both,
	legend columns = 1,
	legend style={at={(axis cs: 400,1)},font=\footnotesize,line width=1pt,draw=black,mark size=0.1pt},
	]	
	
%
%
%
	
	\addlegendentry{\footnotesize \textsc{rp mr} pilot}	
	\addlegendimage{rot,densely dotted,mark=square*, mark options={solid, line width = 0.7pt,fill=white},  mark size=1.5pt}
	
	\addlegendentry{\footnotesize \textsc{rp s-mmse} pilot}	
	\addlegendimage{orange, loosely dotted, mark=star, mark options={solid, line width = 0.7pt},  mark size=1.8pt}
	
	\addlegendentry{\footnotesize \textsc{rp s-mmse miced}}	
	\addlegendimage{gelb,loosely dashdotted,mark=10-pointed star, mark options={solid, line width = 0.7pt},  mark size=1.8pt}
	
	\addlegendentry{\footnotesize \textsc{rp mr miced}}	
	\addlegendimage{apfelgruen,dashdotted,mark=square*, mark options={solid, line width = 0.7pt},  mark size=1.5pt}

	\draw (axis cs:150,0.007) node[draw=none,fill =white,align = center] {\footnotesize{ \textsc{rp s-mmse} pilot reuse-3  }};
	\draw[-,color = gray] (axis cs:150,0.011) -- (axis cs:124,0.038);	
	\draw[-,color = gray] plot [smooth , tension=4 ] coordinates{ (axis cs: 127,0.038) (axis cs: 118,0.042) (axis cs: 127,0.045)   };
	
	
\addplot [pink,  solid, mark=none] 					   
table[x index={0}, y index={8}] {Figures/QPSK_PER_vs_M_R05.txt};

\addplot [rot,densely dotted,mark=square*, mark options={solid, line width = 0.7pt,fill=white},  mark size=1.5pt]
table[x index={0}, y index={1}] {Figures/QPSK_PER_vs_M_R05.txt};

\addplot [orange, loosely dotted, mark=star, mark options={solid, line width = 0.7pt},  mark size=1.8pt]
table[x index={0}, y index={2}] {Figures/QPSK_PER_vs_M_R05.txt};				
	
\addplot [apfelgruen,dashdotted,mark=square*, mark options={solid, line width = 0.7pt},  mark size=1.5pt]
table[x index={0}, y index={4}] {Figures/QPSK_PER_vs_M_R05.txt};									
	
\addplot [gelb,loosely dashdotted,mark=10-pointed star, mark options={solid, line width = 0.7pt},  mark size=1.8pt]
table[x index={0}, y index={5}] {Figures/QPSK_PER_vs_M_R05.txt};

	\end{semilogyaxis}
	\end{tikzpicture}%
			\label{fig:PER_vs_M_R05_RP}
		}
		\subfloat[BLER vs $M$ with SP, $i = 8$, and $\varrho = \sigma^2$ (SNR~${=0}$ dB).]{

\begin{tikzpicture}
	\begin{semilogyaxis}[
	width=0.475\textwidth,
	height=0.35\textwidth,
	y label style = {at={(-0.03,0.5)},anchor=north},
	xlabel={Number of BS antennas ($M$)},
	ylabel={BLER},
	extra x ticks={25},
	extra x tick labels={25},		
	xmin = 25,
	xmax = 400,
	ymin = 0.003,
	ymax = 1,
	line width = 1pt,
	grid=both,
	legend columns = 1,
	legend style={at={(axis cs: 400,1)},font=\footnotesize,line width=1pt,draw=black,mark size=0.1pt},
	]	
	
%
%
%
%
\addlegendentry{\footnotesize \textsc{sp mr} pilot}	
\addlegendimage{anthrazit,solid,mark=*, mark options={solid, line width = 0.7pt,fill=white},  mark size=1.5pt}

\addlegendentry{\footnotesize \textsc{sp s-mmse} pilot}	
\addlegendimage{mittelgrau,solid,mark=otimes*, mark options={solid, line width = 0.7pt,fill=white},  mark size=1.8pt}

\addlegendentry{\footnotesize \textsc{sp s-mmse miced}}	
\addlegendimage{hellblau,loosely dashed,mark=oplus*, mark options={solid, line width = 0.7pt,fill=hellgrau},  mark size=1.8pt}

\addlegendentry{\footnotesize \textsc{sp mr miced}}	
\addlegendimage{mittelblau,dashed,mark=*, mark options={solid, line width = 0.7pt},  mark size=1.5pt}

	\draw (axis cs:150,0.007) node[draw=none,fill =white,align = center] {\footnotesize{ \textsc{rp s-mmse} pilot reuse-3  }};
	\draw[-,color = gray] (axis cs:150,0.011) -- (axis cs:124,0.038);	
	\draw[-,color = gray] plot [smooth , tension=4 ] coordinates{ (axis cs: 127,0.038) (axis cs: 118,0.042) (axis cs: 127,0.045)   };
	

\addplot [pink,  solid, mark=none] 					   
table[x index={0}, y index={8}] {Figures/QPSK_PER_vs_M_R05.txt};

\addplot [anthrazit,solid,mark=*, mark options={solid, line width = 0.7pt,fill=white},  mark size=1.5pt]
table[x index={0}, y index={13}] {Figures/QPSK_PER_vs_M_R05.txt};

\addplot [mittelgrau,solid,mark=otimes*, mark options={solid, line width = 0.7pt,fill=white},  mark size=1.8pt]
table[x index={0}, y index={14}] {Figures/QPSK_PER_vs_M_R05.txt};					
									
\addplot [mittelblau,dashed,mark=*, mark options={solid, line width = 0.7pt},  mark size=1.5pt]
table[x index={0}, y index={16}] {Figures/QPSK_PER_vs_M_R05.txt};

\addplot [hellblau,loosely dashed,mark=oplus*, mark options={solid, line width = 0.7pt,fill=hellgrau},  mark size=1.8pt]
table[x index={0}, y index={17}] {Figures/QPSK_PER_vs_M_R05.txt};

	\end{semilogyaxis}
	\end{tikzpicture}%
			\label{fig:PER_vs_M_R05_SP}
		}
	}\caption{BLER versus number of iterations, SNR, and number of BS antennas with ${\Tul = 200}$, $K = 10$, and $R_{cd}=1/2$. The BLER with RP, pilot-based channel estimation, S-MMSE combining, and pilot reuse 3 is included as a benchmark.  }
	\label{fig:PER_QPSK}
	\hrulefill
\end{figure*}
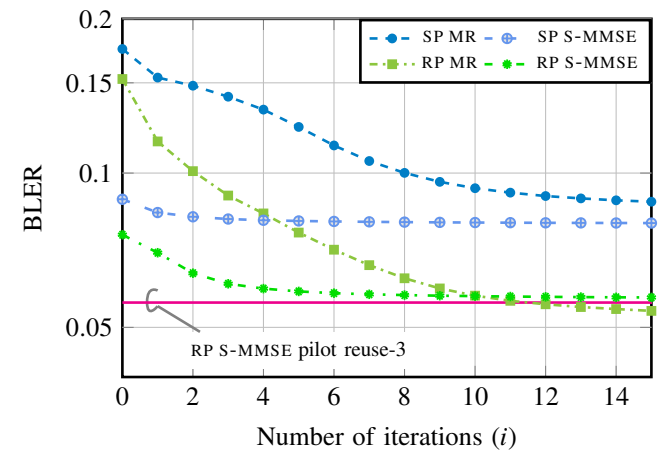
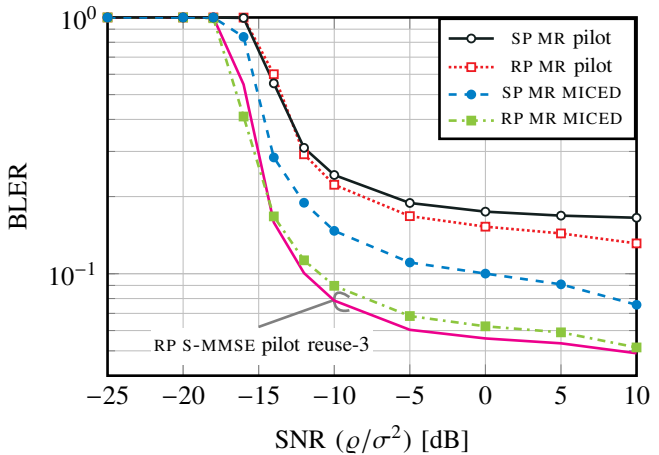
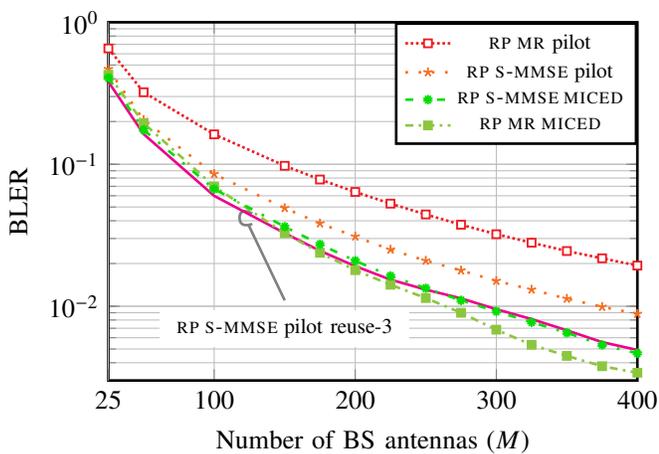
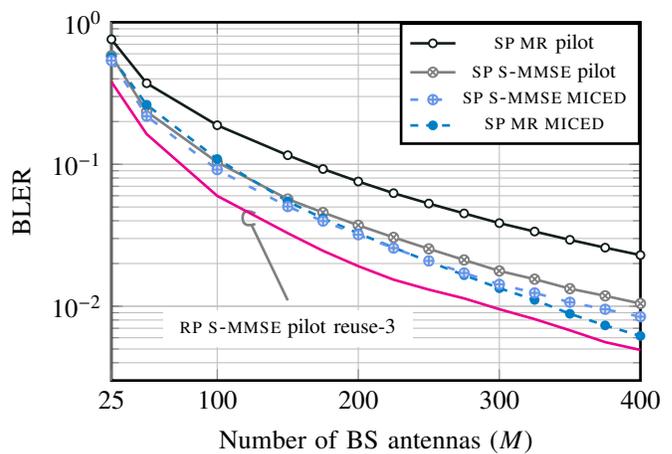

\begin{figure*}[!t]	
	\centering
	{\captionsetup{width=0.4\textwidth}	
		\subfloat[Achievable SE vs $\Tul$ with RP and $M = 100$.]{

\begin{tikzpicture}
	\begin{axis}[
	width=0.475\textwidth,
	height=0.35\textwidth,
	y label style = {at={(-0.03,0.5)},anchor=north},
	xlabel={Size of coherence block ($\Tul$)},
	ylabel={Achievable SE [bit/s/Hz]},
	xmin = 25,
	xmax = 300,
	ymin = 0.8,
	ymax = 1.47,
	line width = 1pt,
	grid=both,
	legend columns = 1,
	legend style={at={(axis cs: 300,1.09)},font=\footnotesize,line width=1pt,draw=black,mark size=0.1pt},
	]



\addlegendentry{\footnotesize \textsc{rp s-mmse miced}}	
\addlegendimage{gelb,loosely dashdotted,mark=10-pointed star, mark options={solid, line width = 0.7pt},  mark size=1.8pt}
\addlegendentry{\footnotesize \textsc{rp mr miced}}	
\addlegendimage{apfelgruen,dashdotted,mark=square*, mark options={solid, line width = 0.7pt},  mark size=1.5pt}
\addlegendentry{\footnotesize \textsc{rp s-mmse} pilot}	
\addlegendimage{orange, loosely dotted, mark=star, mark options={solid, line width = 0.7pt},  mark size=1.8pt}
\addlegendentry{\footnotesize \textsc{rp mr} pilot}	
\addlegendimage{rot,densely dotted,mark=square*, mark options={solid, line width = 0.7pt,fill=white},  mark size=1.5pt}

	\draw (axis cs:226,1.12) node[draw=none,fill =white,align = center] {\footnotesize{\textsc{s-mmse} pilot reuse-3}};
	\draw[-,color = gray] (axis cs:155,1.12) -- (axis cs:125,1.04);	
	\draw[-,color = gray] plot [smooth , tension=5 ] coordinates{ (axis cs: 126,1.04) (axis cs: 119,1.06) (axis cs: 126,1.07)   };

	\addplot [pink,  solid, mark=none] 					   
	table[x index={0}, y index={8}] {Figures/QPSK_MutInf_vs_tau_c_R075.txt};

	\addplot [rot,densely dotted,mark=square*, mark options={solid, line width = 0.7pt,fill=white},  mark size=1.5pt]
	table[x index={0}, y index={1}] {Figures/QPSK_MutInf_vs_tau_c_R075.txt};

	\addplot [orange, loosely dotted, mark=star, mark options={solid, line width = 0.7pt},  mark size=1.8pt]
	table[x index={0}, y index={2}] {Figures/QPSK_MutInf_vs_tau_c_R075.txt};				
	
	\addplot [apfelgruen,dashdotted,mark=square*, mark options={solid, line width = 0.7pt},  mark size=1.5pt]
	table[x index={0}, y index={4}] {Figures/QPSK_MutInf_vs_tau_c_R075.txt};

	\addplot [gelb,loosely dashdotted,mark=10-pointed star, mark options={solid, line width = 0.7pt},  mark size=1.8pt]
	table[x index={0}, y index={5}] {Figures/QPSK_MutInf_vs_tau_c_R075.txt};

	\end{axis}
	\end{tikzpicture}%
			\label{fig:SE_vs_tau_c_RP}
		}
		\subfloat[Achievable SE vs $\Tul$ with SP and $M = 100$.]{

\begin{tikzpicture}
	\begin{axis}[
	width=0.475\textwidth,
	height=0.35\textwidth,
	y label style = {at={(-0.03,0.5)},anchor=north},
	xlabel={Size of coherence block ($\Tul$)},
	ylabel={Achievable SE [bit/s/Hz]},
	xmin = 25,
	xmax = 300,
	ymin = 0.8,
	ymax = 1.47	,
	line width = 1pt,
	grid=both,
	legend columns = 1,
	legend style={at={(axis cs: 300,1.09)},font=\footnotesize,line width=1pt,draw=black,mark size=0.1pt},
	]

%
\addlegendentry{\footnotesize \textsc{sp s-mmse miced}}	
\addlegendimage{hellblau,loosely dashed,mark=oplus*, mark options={solid, line width = 0.7pt,fill=hellgrau},  mark size=1.8pt}%
\addlegendentry{\footnotesize \textsc{sp mr miced}}	
\addlegendimage{mittelblau,dashed,mark=*, mark options={solid, line width = 0.7pt},  mark size=1.5pt}
\addlegendentry{\footnotesize \textsc{sp s-mmse} pilot}	
\addlegendimage{mittelgrau,solid,mark=otimes*, mark options={solid, line width = 0.7pt,fill=white},  mark size=1.8pt}
\addlegendentry{\footnotesize \textsc{sp mr} pilot}	
\addlegendimage{anthrazit,solid,mark=*, mark options={solid, line width = 0.7pt,fill=white},  mark size=1.5pt}

			\draw (axis cs:226,1.12) node[draw=none,fill =white,align = center] {\footnotesize{\textsc{s-mmse} pilot reuse-3}};
			\draw[-,color = gray] (axis cs:155,1.12) -- (axis cs:125,1.04);	
			\draw[-,color = gray] plot [smooth , tension=5 ] coordinates{ (axis cs: 126,1.04) (axis cs: 119,1.06) (axis cs: 126,1.07)   };

%
%
%
%
%
%
%

	\addplot [pink,  solid, mark=none] 					   
	table[x index={0}, y index={8}] {Figures/QPSK_MutInf_vs_tau_c_R075.txt};

	\addplot [anthrazit,solid,mark=*, mark options={solid, line width = 0.7pt,fill=white},  mark size=1.5pt]
	table[x index={0}, y index={13}] {Figures/QPSK_MutInf_vs_tau_c_R075.txt};

	\addplot [mittelgrau,solid,mark=otimes*, mark options={solid, line width = 0.7pt,fill=white},  mark size=1.8pt]
	table[x index={0}, y index={14}] {Figures/QPSK_MutInf_vs_tau_c_R075.txt};

	\addplot [mittelblau,dashed,mark=*, mark options={solid, line width = 0.7pt},  mark size=1.5pt]
	table[x index={0}, y index={16}] {Figures/QPSK_MutInf_vs_tau_c_R075.txt};

	\addplot [hellblau,loosely dashed,mark=oplus*, mark options={solid, line width = 0.7pt,fill=hellgrau},  mark size=1.8pt]
	table[x index={0}, y index={17}] {Figures/QPSK_MutInf_vs_tau_c_R075.txt};

	\end{axis}
	\end{tikzpicture}%
			\label{fig:SE_vs_tau_c_SP}
		}
		\\
		\subfloat[Achievable rate vs $M$ with RP and $\Tul =200$.]{

\begin{tikzpicture}
	\begin{axis}[
	width=0.475\textwidth,
	height=0.35\textwidth,
	y label style = {at={(-0.03,0.5)},anchor=north},
	xlabel={Number of BS antennas ($M$)},
	ylabel={Achievable SE [bit/s/Hz]},
	xmin = 15,
	xmax = 200,
	ymin = 0.7,
	ymax = 1.5,
	line width = 1pt,
	grid=both,
	legend columns = 1,
	legend style={at={(axis cs: 200,1.05)},font=\footnotesize,line width=1pt,draw=black,mark size=0.1pt},
	]	

%
%
%
		\addlegendentry{\footnotesize \textsc{rp s-mmse miced}}	
		\addlegendimage{gelb,loosely dashdotted,mark=10-pointed star, mark options={solid, line width = 0.7pt},  mark size=1.8pt}
	\addlegendentry{\footnotesize \textsc{rp mr miced}}	
	\addlegendimage{apfelgruen,dashdotted,mark=square*, mark options={solid, line width = 0.7pt},  mark size=1.5pt}
	\addlegendentry{\footnotesize \textsc{rp s-mmse} pilot}	
	\addlegendimage{orange, loosely dotted, mark=star, mark options={solid, line width = 0.7pt},  mark size=1.8pt}
	\addlegendentry{\footnotesize \textsc{rp mr} pilot}	
	\addlegendimage{rot,densely dotted,mark=square*, mark options={solid, line width = 0.7pt,fill=white},  mark size=1.5pt}

	\draw (axis cs:148,1.1) node[draw=none,fill =white,align = center] {\footnotesize{\textsc{s-mmse} pilot reuse-3}};
	\draw[-,color = gray] (axis cs:145,1.13) -- (axis cs:150,1.21	);	
	\draw[-,color = gray] plot [smooth , tension=5 ] coordinates{ (axis cs: 152,1.22) (axis cs: 148,1.23) (axis cs: 152,1.25)   };

	\addplot [pink,  solid, mark=none] 					   
	table[x index={0}, y index={8}] {Figures/QPSK_MutInf_vs_M_R075.txt};

	\addplot [rot,densely dotted,mark=square*, mark options={solid, line width = 0.7pt,fill=white},  mark size=1.5pt]
	table[x index={0}, y index={1}] {Figures/QPSK_MutInf_vs_M_R075.txt};

	\addplot [orange, loosely dotted, mark=star, mark options={solid, line width = 0.7pt},  mark size=1.8pt]
	table[x index={0}, y index={2}] {Figures/QPSK_MutInf_vs_M_R075.txt};

	\addplot [gelb,loosely dashdotted,mark=10-pointed star, mark options={solid, line width = 0.7pt},  mark size=1.8pt]
	table[x index={0}, y index={5}] {Figures/QPSK_MutInf_vs_M_R075.txt};

		\addplot [apfelgruen,dashdotted,mark=square*, mark options={solid, line width = 0.7pt},  mark size=1.5pt]
		table[x index={0}, y index={4}] {Figures/QPSK_MutInf_vs_M_R075.txt};	
	
	\end{axis}
	\end{tikzpicture}%
			\label{fig:SE_vs_M_RP}
		}
		\subfloat[Achievable rate vs $M$ with SP and $\Tul =200$.]{

\begin{tikzpicture}
	\begin{axis}[
	width=0.475\textwidth,
	height=0.35\textwidth,
	y label style = {at={(-0.03,0.5)},anchor=north},
	xlabel={Number of BS antennas ($M$)},
	ylabel={Achievable SE [bit/s/Hz]},
	xmin = 15,
	xmax = 200,
	ymin = 0.7,
	ymax = 1.5,
	line width = 1pt,
	grid=both,
	legend columns = 1,
	legend style={at={(axis cs: 200,1.05)},font=\footnotesize,line width=1pt,draw=black,mark size=0.1pt},
	]	
%
%
%
%
\addlegendentry{\footnotesize \textsc{sp s-mmse miced}}	
\addlegendimage{hellblau,loosely dashed,mark=oplus*, mark options={solid, line width = 0.7pt,fill=hellgrau},  mark size=1.8pt}
\addlegendentry{\footnotesize \textsc{sp mr miced}}	
\addlegendimage{mittelblau,dashed,mark=*, mark options={solid, line width = 0.7pt},  mark size=1.5pt}
\addlegendentry{\footnotesize \textsc{sp s-mmse} pilot}	
\addlegendimage{mittelgrau,solid,mark=otimes*, mark options={solid, line width = 0.7pt,fill=white},  mark size=1.8pt}
\addlegendentry{\footnotesize \textsc{sp mr} pilot}	
\addlegendimage{anthrazit,solid,mark=*, mark options={solid, line width = 0.7pt,fill=white},  mark size=1.5pt}

	\draw (axis cs:148,1.1) node[draw=none,fill =white,align = center] {\footnotesize{\textsc{s-mmse} pilot reuse-3}};
	\draw[-,color = gray] (axis cs:145,1.13) -- (axis cs:150,1.21	);	
	\draw[-,color = gray] plot [smooth , tension=5 ] coordinates{ (axis cs: 152,1.22) (axis cs: 148,1.23) (axis cs: 152,1.25)   };

\addplot [pink,  solid, mark=none] 					   
table[x index={0}, y index={8}] {Figures/QPSK_MutInf_vs_M_R075.txt};

\addplot [anthrazit,solid,mark=*, mark options={solid, line width = 0.7pt,fill=white},  mark size=1.5pt]
table[x index={0}, y index={13}] {Figures/QPSK_MutInf_vs_M_R075.txt};

\addplot [mittelgrau,solid,mark=otimes*, mark options={solid, line width = 0.7pt,fill=white},  mark size=1.8pt]
table[x index={0}, y index={14}] {Figures/QPSK_MutInf_vs_M_R075.txt};

\addplot [hellblau,loosely dashed,mark=oplus*, mark options={solid, line width = 0.7pt,fill=hellgrau},  mark size=1.8pt]
table[x index={0}, y index={17}] {Figures/QPSK_MutInf_vs_M_R075.txt};	

\addplot [mittelblau,dashed,mark=*, mark options={solid, line width = 0.7pt},  mark size=1.5pt]
table[x index={0}, y index={16}] {Figures/QPSK_MutInf_vs_M_R075.txt};

	\end{axis}
	\end{tikzpicture}%
			\label{fig:SE_vs_M_SP}
		}
	}\caption{Achievable SE versus coherence block size and number of BS antennas with $i = 8$, $K = 10$, $\varrho = \sigma^2$ (SNR~${=0}$ dB), and $R_{cd}=3/4$. The SE with RP, pilot-based channel estimation, S-MMSE combining, and pilot reuse 3 is included as a benchmark.}
	\label{fig:SE_QPSK_tau_c_M}
	\hrulefill
\end{figure*}
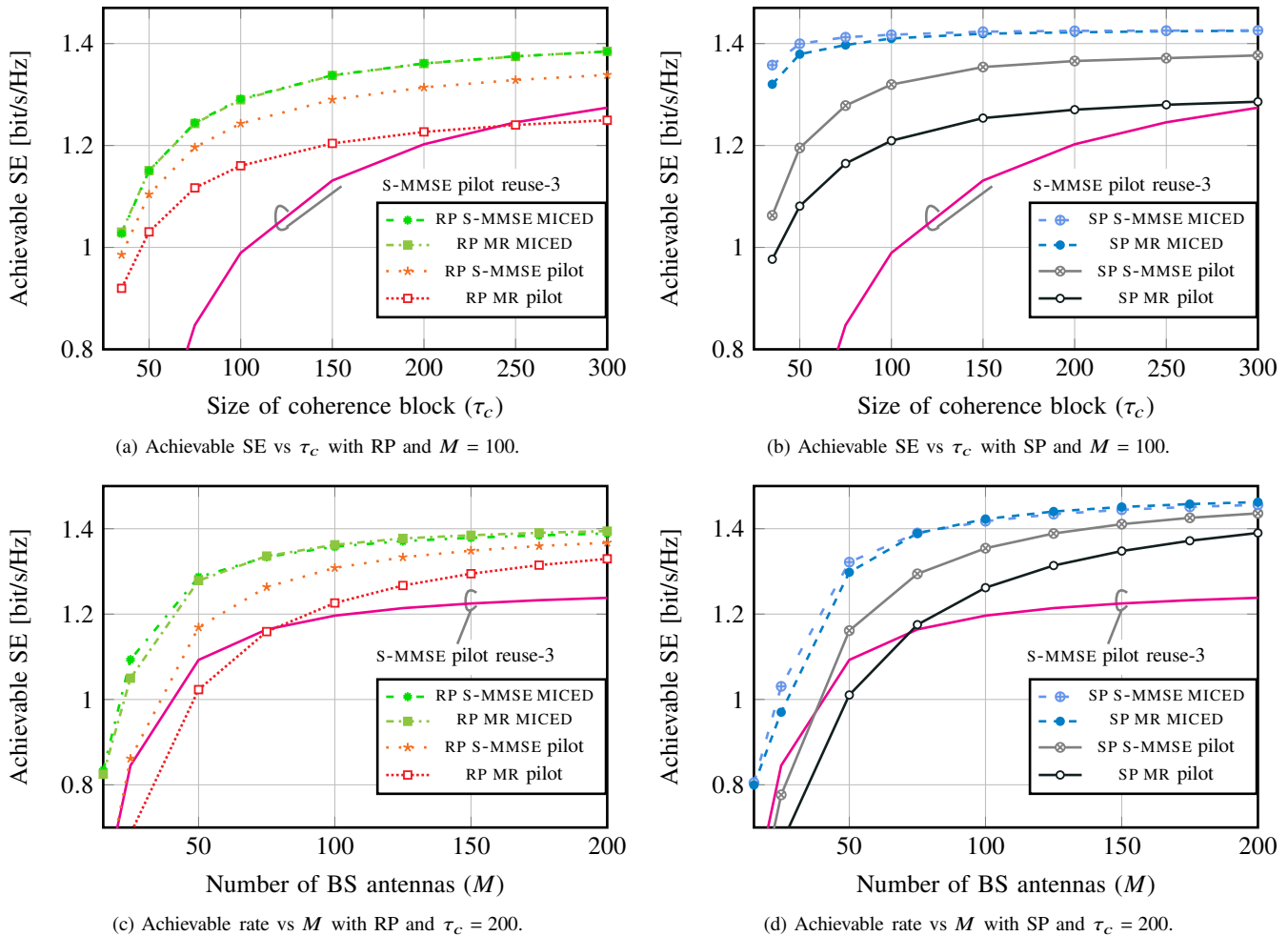
\subsection{Numerical examples with finite-alphabet symbols}
\label{subsec:Num_res_finite}
To illustrate the gains of the MICED algorithm in a practical system, the same simulation setup as in Section~\ref{sec:num_ex} is used but now considering LDPC codes and QPSK modulation. The choice of parity check matrix is done following the new radio (NR) 3GPP specifications \cite{LDPC_NR_3gpp}. Two code rates are evaluated: $R_{cd} \in \{1/2,\,3/4\}$ with a codeword length of $3840$ and $3888$ bits respectively. Note that for these simulations, the channel error correlation matrices in \eqref{eq:coCchObsDataAid} are computed numerically. This is done to obtain more accurate results compared to using the closed-form expressions Theorem~\ref{th:LB_covMatDataAid}, given that the data symbol distribution is no longer Gaussian. To evaluate the reliability of the transmitted data, the BLER is calculated assuming that each codeword corresponds to one block. The results of implementing the MICED algorithm with RP assume pilot reuse one ($\Tp = K$). In the case of SP, the pilot reuse is the closest to $\Tul/K$  that is allowed in an hexagonal grid. To establish a benchmark, the performance of standard pilot-based channel estimation with RP is included with pilot reuse 1 and 3. 

Figure~\ref{fig:PER_QPSK} depicts the BLER versus the number of iterations, SNR, and the number of BS antennas with a code rate $R_{cd} = 1/2$. Figure~\ref{fig:SE_QPSK_tau_c_M} shows the achievable SE versus the size of the coherence block and the number of BS antennas with a code rate of $R_{cd} = 3/4$. Note that when the BLER is low (e.g. $10^{-2}$) the resulting changes in the achievable SE are very small. Thus, different code rates for BLER and achievable SE curves are chosen to observe the potential difference between the evaluated methods within a few hundred BS antennas.  The achievable SE at the $i^{th}$ iteration is obtained from the mutual information between the input bits, and the soft symbol estimates at the output of the decoder. The number of encoded bits in a codeword is denoted as $N_\textsc{enc}$, and these encoded bits are staked into the vector $\bf{b}_{lk}$ such that $[\bf{b}_{lk}]_m\in \{0,\,1\}$ for $m \in \{1,\ldots,N_\textsc{enc}\}$. Assuming that the bits at the output of the decoder are independent, then the achievable SE is
\begin{align*}
&\rm{R}_{lk} 
=\frac{\Tuld N_b R_{cd}}{\Tul N_\textsc{enc}} \sum_{m=1}^{N_\textsc{enc}} \left( 1 \vphantom{\sum\limits_{d=0}^{1}}\right.
\\&\left.
 + \sum_{d=0}^{1} \rm{Pr}\left([\bf{b}_{lk}]_m = d | \bar{\bf{y}}_{lk}^\iter{}{i}\right)\log_2\left(\rm{Pr}\left([\bf{b}_{lk}]_m = d | \bar{\bf{y}}_{lk}^\iter{}{i}\right)\right) \right)
\stepcounter{equation} \tag{\theequation} \label{eq:ach_rate}
\end{align*}
where the probabilities in \eqref{eq:ach_rate} are obtained as in \eqref{eq:Prob_LLR}. Note that the achievable SE in \eqref{eq:ach_rate} accounts the overhead from coding and using dedicated pilot symbols in the case of RP.

Figure~\ref{fig:PER_vs_iter} depicts the BLER versus the number of iterations. It can be seen  after 8 iterations the results stabilize and therefore that is the number of iterations selected for the rest of the figures. In addition, as the iterations progress the MICED algorithm decreases the BLER with MR combining further than with S-MMSE when compared to its initial state at $i = 0$. Figure~\ref{fig:PER_vs_rho_R05} shows the BLER versus the average SNR per symbol indicating that the benefits of the MICED algorithm are achieve for both low and high SNR regimes. In Figures~\ref{fig:PER_vs_M_R05_RP}~and~~\ref{fig:PER_vs_M_R05_SP} the BLER is shown as a function of the number of BS antennas. It can also be seen that compared to pilot-based channel estimation, the use of the MICED algorithm is more beneficial for MR combining and when the number of antennas grows it outperforms S-MMSE. This effect is due to the presence of poor quality data estimates in the data-aided channel estimation, which in turn, make the interference suppression capabilities of S-MMSE less effective when compared to subtracting the estimated intracell interference (see third term of the effective noise in  \eqref{eq:data_est_RPfull} and \eqref{eq:data_est_SPfull}). Moreover, when using the MICED algorithm with RP, the BLER is lower compared to S-MMSE combining with pilot reuse 3 which means that greater reliability can be achieved despite the 3 times lower pilot overhead. By comparing the BLER with RP and SP, it can be seen that RP provides lower BLER than SP, however for practical number of BS antennas (e.g. $M = 100$) the difference between RP and SP is rather small.

Figure~\ref{fig:SE_QPSK_tau_c_M} shows the achievable SE as a function of the coherence block size and number of BS antennas. In this case, the use of the MICED algorithm provides the greatest gains for MR combining which is in line with the results in Section~\ref{sec:num_ex}. Furthermore, the SE with the MICED algorithm for MR and S-MMSE is very close which means that the benefit of the interference suppression with S-MMSE is comparable to removing the intracell interference  with MR (see third term of the effective noise in  \eqref{eq:data_est_RPfull} and \eqref{eq:data_est_SPfull}). The reason for this behavior, is the presence of inaccurate data estimates in the data-aided channel estimation process. Therefore, the performance of the MICED algorithm with S-MMSE combining can be further improved by controlling the use of data estimates based on their accuracy.

 In addition, Figure~\ref{fig:SE_QPSK_tau_c_M} shows that the MICED algorithm with SP achieves higher SE than RP since it does not have a pilot overhead (i.e., all symbols in the coherence block are used for data). The most benefit of the MICED algorithm is obtained for small coherence block size which corresponds to high mobility scenarios or higher carrier frequencies.

\begin{figure*}
	\begin{align*}
	\sum_{k'=1}^{K}\bf{R}_{llk'}\bb{E}\left\{\left|(\bf{u}_{lk}^{\iter{}{i}})^H\tilde{\bf{x}}_{lk'}^\iter{}{i}  \right|^2 \right\} 
	&\stackrel{\rm{(a_1)}}{=}
	\sum_{k'=1}^{K}\bf{R}_{llk'}\pulrp_{lk'}(1 - \sigma_{lk'}^{\iter{}{i}^2})\bb{E}\left\{\left\|\left[ \hat{\bf{S}}_{l}^\iter{}{i} \bf{D}_\rho^{\frac{1}{2}} \left(\hat{\bf{X}}_{l}^{\iter{}{i}^H}\hat{\bf{X}}_{l}^\iter{}{i}\right)^{-1}\right]_k\right\|^2 \right\}
	\\
	&
	\stackrel{\rm{(a_2)}}{=} 
	\sum_{k'=1}^{K}\bf{R}_{llk'}\frac{\pulrp_{lk'}(1 - \sigma_{lk'}^{\iter{}{i}^2})}{2\qulrp_{lk} \Tp}\bb{E}\left\{\left[\left(\vphantom{\frac{K^K}{K}}\right.\right. \right.
	\underbrace{ \frac{\Tp}{2}\bf{D}_{q\rho}  \left(\hat{\bf{S}}_{l}^{\iter{}{i}^H}\hat{\bf{S}}_{l}^\iter{}{i}\right)^{-1}
		+ \bf{I}_K +   \frac{1}{2\Tp}\bf{D}_{q\rho}^{-1} \hat{\bf{S}}_{l}^{\iter{}{i}^H}\hat{\bf{S}}_{l}^\iter{}{i} 
	}_{=\bf{T}_l^\iter{}{i}}\left.\left.\left.\vphantom{\frac{K^K}{K}}\right)^{-1} \right]_{kk} \right\}
	\\
	&\
	\stackrel{\rm{(a_3)}}{\succeq}
	\sum_{k'=1}^{K}\bf{R}_{llk'}\pulrp_{lk'}(1 - \sigma_{lk'}^{\iter{}{i}^2})\underbrace{\frac{1}{2\qulrp_{lk} \Tp}\bb{E}\left\{\left(\frac{\qulrp_{lk} \Tp}{2\pulrp_{lk}\sigma_{lk}^{\iter{}{i}^2} \chi_a }   + 1 + \frac{\pulrp_{lk} \sigma_{lk}^{\iter{}{i}^2}}{2 \qulrp_{lk} \Tp} \chi_b\right)^{-1}\right\}}_{=\sf{a}_{lk}^\iter{}{i}}
	\tag{34}\label{eq:uHxtilde}
	\end{align*}
	\begin{align*}
	\sum_{\ell\in \Phi\backslash l}\sum_{k'=1}^{K}\bf{R}_{l\ell k'}\bb{E}\left\{\left|\bf{u}_{lk}^{\iter{}{i}^H}\bf{x}_{\ell k'} \right|^2 \right\} 
	&=
	\sum_{\ell\in \Phi\backslash l}\sum_{k'=1}^{K}\bf{R}_{l\ell k'}\bb{E}\left\{\left|\left(\sqrt{\qulrp_{\ell k'}}\begin{bmatrix}
	\bfs{\phi}_{\ell k'} \\\bf{0}
	\end{bmatrix} + \sqrt{\pulrp_{\ell k'}}\begin{bmatrix}
	\bf{0} \\\bf{s}_{\ell k'}
	\end{bmatrix}\right)^H\!\!\! \bf{u}_{lk}^\iter{}{i} \right|^2\right\}
	\\
	\tag{35}\label{eq:succeq_intercellint}
	&=
	\sum_{\ell\in \Phi\backslash l}\sum_{k'=1}^{K}\bf{R}_{l\ell k'} \bb{E}\left\{\qulrp_{\ell k'}\left|\left[ \bfs{\phi}_{\ell k'}^H\bf{P}_{l}^\textsc{rp} \bf{D}_q^{\frac{1}{2}} \left(\hat{\bf{X}}_{l}^{\iter{}{i}^H}\hat{\bf{X}}_{l}^\iter{}{i}\right)^{-1}\right]_k\right|^2 + \pulrp_{\ell k'}\left\|\left[ \hat{\bf{S}}_{l}^\iter{}{i} \bf{D}_\rho^{\frac{1}{2}} \left(\hat{\bf{X}}_{l}^{\iter{}{i}^H}\hat{\bf{X}}_{l}^\iter{}{i}\right)^{-1}\right]_k\right\|^2 \right\} 
	\\
	&
	\tag{36}\label{eq:succ_b_1}
	\stackrel{\rm{(b_1)}}{\succeq}
	\sum_{(\ell,k')\in \cal{P}_{lk}^\textsc{rp}} \bf{R}_{l\ell k'}\qulrp_{\ell k'}\qulrp_{lk}\Tp^2\bb{E}\left\{\left|\left[ \left(\hat{\bf{X}}_{l}^{\iter{}{i}^H}\hat{\bf{X}}_{l}^\iter{}{i}\right)^{-1}\right]_{kk}\right|^2 \right\} + \sum_{\ell\in \Phi\backslash l}\sum_{k'=1}^{K}\bf{R}_{l\ell k'}\frac{\pulrp_{\ell k'} \bb{E}\left\{ \left[\left(\bf{T}_l^\iter{}{i}\right)^{-1}\right]_{kk} \right\}}{2\qulrp_{lk} \Tp}
	\\
	&
	\stackrel{\rm{(b_2)}}{\succeq}
	\sum_{(\ell,k')\in \cal{P}_{lk}^\textsc{rp}}\bf{R}_{l\ell k'}\qulrp_{\ell k'}\qulrp_{lk}\Tp^2\bb{E}\left\{\left(\qulrp_{lk}\Tp + \pulrp_{lk}\sigma_{lk}^{\iter{}{i}^2} \chi_b \right)^{-2}\right\} +  \sum_{\ell\in \Phi\backslash l}\sum_{k'=1}^{K}\bf{R}_{l\ell k'}\pulrp_{\ell k'}\sf{a}_{lk}^\iter{}{i}.
	\end{align*}
	\hrulefill
\end{figure*}

\section{Conclusion}
\label{sec:conc}
This article evaluates the use of iterative data-aided channel estimation in multicell Massive MIMO systems, where the partially decoded bits are used to improve the channel estimates and reduce the decoding errors at the receiver. The MICED algorithm is proposed and analyzed with RP and SP transmission methods along with MR and S-MMSE processing assuming spatially correlated channels. The results show that the MICED algorithm increases the SE and reduces the BLER compared to pilot-based channel estimation with both RP and SP. The highest SE is found when implementing the MICED algorithm with SP since the cost of the pilot overhead is removed and the data interference is mitigated by the data-aided channel estimation process. The MICED algorithm with SP is most beneficial in high mobility or high carrier frequencies scenarios with small coherence block size, outperforming RP in terms of SE. In addition, the MICED algorithm with SP allows for aggressive spatial multiplexing, increasing SE and facilitating implementation of other technologies like machine type communication.

The quality of data estimates plays a key roles when using linear combining that has interference suppression like S-MMSE. Thus, further improvements of the MICED algorithm can be attained when adding control mechanisms for the use of data estimates based on their quality.

\appendices

\begin{figure*}[!t]
\begin{align*}
\sum_{k'=1}^{K}\bf{R}_{llk'}\bb{E}\left\{\left|(\bf{u}_{lk}^{\iter{}{i}})^H\tilde{\bf{x}}_{lk'}^\iter{}{i}  \right|^2 \right\} 
&=
\sum_{k'=1}^{K}\bf{R}_{llk'} \pulsp_{lk'} \left(1 - \sigma_{lk'}^{\iter{}{i}^2}\right)\bb{E}\left\{\left[ \left(\hat{\bf{X}}_{l}^{\iter{}{i}^H}\hat{\bf{X}}_{l}^\iter{}{i}\right)^{-1}\right]_{kk} \right\}
\\
&\stackrel{\rm{(c_1)}}{\succeq} 
\sum_{k'=1}^{K}\bf{R}_{llk'}\pulsp_{lk'} \left(1 - \sigma_{lk'}^{\iter{}{i}^2}\right) 
\underbrace{\bb{E}\left\{\left(\qulsp_{lk} \Tul +\pulsp_{lk}\left\|\hat{\bf{s}}_{lk}^\iter{}{i}\right\|^2  + 
2\sqrt{\qulsp_{lk}\pulsp_{lk}}\Re\left\{  \bfs{\varphi}_{lk}^H\hat{\bf{s}}_{lk}^\iter{}{i}\right\} \right)^{-1} \right\}}_{=\sf{b}_{lk}^\iter{}{i}}.
\tag{37} \label{eq:CovSPChEstErr}
\end{align*}
\begin{align*}
\sum_{\ell\in \Phi\backslash l}\sum_{k'=1}^{K}\bf{R}_{l\ell k'}\bb{E}\left\{\left|\bf{u}_{lk}^{\iter{}{i}^H}\bf{x}_{\ell k'} \right|^2 \right\} 
&=\sum_{\ell\in \Phi\backslash l}\sum_{k'=1}^{K}\bf{R}_{l\ell k'}\bb{E}\left\{\left|\left(\sqrt{\qulsp_{\ell k'}}\bfs{\varphi}_{\ell k'} + \sqrt{\pulsp_{\ell k'}}\bf{s}_{\ell k'}\right)^H\bf{u}_{lk}^\iter{}{i} \right|^2\right\}
\\
\tag{38}\label{eq:Cov_intcell_1}
&\stackrel{\rm{(d_1)}}{\succeq}
\sum_{\ell\in \Phi\backslash l}\sum_{k'=1}^{K}\bf{R}_{l\ell k'}
\bb{E}\left\{ \qulsp_{\ell k'} \left|\left[  \bfs{\varphi}_{\ell k'}^H\bf{P}_l^\textsc{sp}\bf{D}_{q}^{\frac{1}{2}}\left(\hat{\bf{X}}_l^{\iter{}{i}^H}  \hat{\bf{X}}_l^\iter{}{i}  \right)^{-1}\right]_{k} \right|^2 
+
\pulsp_{\ell k'}\left[ \left(\hat{\bf{X}}_{l}^{\iter{}{i}^H}\hat{\bf{X}}_{l}^\iter{}{i}\right)^{-1}\right]_{kk}
\right\}
\\
&\stackrel{\rm{(d_2)}}{\succeq}
\sum_{(\ell, k')\in \cal{P}_{lk}^\textsc{sp}} \bf{R}_{l\ell k'}\qulsp_{\ell k'}\qulsp_{lk}\Tul^2 \bb{E}\left\{\left|\left[ \left(\hat{\bf{X}}_{l}^{\iter{}{i}^H}\hat{\bf{X}}_{l}^\iter{}{i}\right)^{-1}\right]_{kk}\right|^2
\right\}  + \sum_{\ell\in \Phi\backslash l}\sum_{k'=1}^{K}\bf{R}_{l\ell k'}\pulsp_{\ell k'}\sf{b}_{lk}^\iter{}{i}
\\
&\stackrel{\rm{(d_3)}}{\succeq}\!\!
\sum_{(\ell, k')\in \cal{P}_{lk}^\textsc{sp}}\! \!\!\!\bf{R}_{l\ell k'}\qulsp_{\ell k'}\qulsp_{lk}\Tul^2 \bb{E}\!\left\{\!\!\left(\qulsp_{lk} \Tul +\pulsp_{lk}\left\|\hat{\bf{s}}_{lk}^\iter{}{i}\right\|^2 \! + 
2\sqrt{\qulsp_{lk}\pulsp_{lk}}\Re\!\left\{  \bfs{\varphi}_{lk}^H\hat{\bf{s}}_{lk}^\iter{}{i}\right\} \right)^{\!-2}
\!\right\}\!  +\! \sum_{\ell\in \Phi\backslash l}\sum_{k'=1}^{K}\bf{R}_{l\ell k'}\pulsp_{\ell k'}\sf{b}_{lk}^\iter{}{i}
\end{align*}
\hrulefill
\end{figure*}

\section{Proof of data-aided correlation matrices}
\label{app:proofCovMat}
Consider a square matrix $\bf{A}$ that is positive semi-definite and constants $a,\,b \in \bb{R}$ such that $a \geq b\geq 0$, then it follows that $\bf{A}a \succeq \bf{A}b$. Note that all correlation matrices are positive semi-definite by definition, and since the expectations in \eqref{eq:coCchObsDataAid} are scalar quantities, taking lower bounds on the expectations would result in correlation matrices that fulfill Theorem~\ref{th:LB_covMatDataAid}. It is also worth mentioning that by assuming circularly symmetric complex Gaussian symbols (i.e., $\bf{s}_{lk}\sim\cal{CN}(\bf{0},\bf{I}_{\Tuld})$) the resulting MMSE data estimate and its error are statistically independent \cite{16_Marzetta_MAMIMO_book,17_Bjornson_MAMIMO_book}. To obtain the closed-form expression in  \eqref{eq:cov_chObs_dataRP} and \eqref{eq:CorMatInt_dataRP} with RP, the terms in \eqref{eq:coCchObsDataAid} are analyzed separately, let $\bf{D}_\rho = \rm{diag}(\pulrp_{l1},\ldots,\pulrp_{lK})$ and $\bf{D}_{q\rho} = \rm{diag}( \qulrp_{l1}/\pulrp_{l1} ,\ldots, \qulrp_{lK}/\pulrp_{lK})$. Then, the calculations in  \eqref{eq:uHxtilde} at the top of the page hold,
where $\rm{(a_1)}$ follows from the independence between data estimates and errors (i.e., $\bf{u}_{lk}^{\iter{}{i}}$ and $\tilde{\bf{x}}_{lk'}^\iter{}{i}$ are independent). The second equality $\rm{(a_2)}$ is obtained by expanding the Gramian of $\hat{\bf{X}}_{l}^\iter{}{i}$ and applying known properties of the matrix inverse operator. Note that $\bf{T}_l^\iter{}{i}$ is Hermitian and positive semi-definite, thus from the result in Lemma~\ref{lem:diaginv} of Appendix~\ref{app:matTh}, it holds that $[(\bf{T}_l^\iter{}{i})^{-1}]_{kk}\geq 1/[\bf{T}_l^\iter{}{i}]_{kk}$. In addition, the diagonal elements of $\hat{\bf{S}}_{l}^{\iter{}{i}^H}\hat{\bf{S}}_{l}^\iter{}{i}$ are independent and have a chi-squared distribution. Thus,  $\rm{(a_3)}$ holds for $ \chi_a\sim 0.5 \chi_{2(\Tuld+ 1-K)}^2$ and $ \chi_b\sim 0.5 \chi_{2\Tuld}^2$ \cite[App.~B]{16_Marzetta_MAMIMO_book}. Then, a final lower bound on $\sf{a}_{lk}^\iter{}{i}$ is obtained based on Jensen's inequality since $\bb{E}\{1/x\}\geq 1/\bb{E}\{x\}$, which yields the expression of the second term in \eqref{eq:cov_chObs_dataRP} and first term in \eqref{eq:CorMatInt_dataRP}.

For the second term in \eqref{eq:coCchObsDataAid} (intercell interference), note that the data symbols between UEs are independent, thus, $\bf{s}_{\ell k'}$ and $\bf{u}_{lk}^{\iter{}{i}}$ are independent for $\ell \neq l$. Let $\bf{P}_l^\textsc{rp} = [\bfs{\phi}_{l1},\ldots, \bfs{\phi}_{lK}]$, then the calculations in \eqref{eq:succeq_intercellint} and \eqref{eq:succ_b_1} at the top of the page hold.

Notice that interference from pilot symbols (see the first term in \eqref{eq:succeq_intercellint}) is only non-zero if $\rm{UE}_{\ell k'}$ shares a pilot with a UE in cell $l$, and in particular, the largest interference will come from UEs sharing the same pilots as $\rm{UE}_{lk}$. Thus, $\rm{(b_1)}$ holds by discarding the pilot interference that does not come from UEs sharing the same pilots as $\rm{UE}_{lk}$ (see the first term of \eqref{eq:succ_b_1}) and by applying the same method as in \eqref{eq:uHxtilde} for the data interference (see the second term of \eqref{eq:succ_b_1}). Then, $\rm{(b_2)}$ holds by applying Lemma~\ref{lem:diaginv} in Appendix~\ref{app:matTh} and taking the expectation over the inverse of the diagonal elements. Finally, by applying Jensen's inequality the expression for the pilot intercell interference (see second term in \eqref{eq:CorMatInt_dataRP}) is found. For the third term in \eqref{eq:coCchObsDataAid} (noise), the following result holds
	\begin{align*}
	\bb{E}\left\{\left\|\bf{u}_{lk}^\iter{}{i} \right\|^2 \right\} 
	&=\bb{E}\left\{\left[ \left(\hat{\bf{X}}_{l}^{\iter{}{i}^H}\hat{\bf{X}}_{l}^\iter{}{i}\right)^{-1}\right]_{kk} \right\}
	\\
	&
	\geq
	\bb{E}\left\{\left(\qulrp_{lk}\Tp + \pulrp_{lk}\sigma_{lk}^{\iter{}{i}^2} \chi_b \right)^{-1}\right\}, 
	\end{align*}
then, by means of the Jensen's inequality the third term in \eqref{eq:CorMatInt_dataRP} is obtained.

In the case of the expression in \eqref{eq:cov_chObs_dataSP} and \eqref{eq:CorMatInt_dataSP} with SP, notice that $\tilde{\bf{x}}_{lk'}^\iter{}{i} = \sqrt{\pulsp_{lk'}}\tilde{\bf{s}}_{lk'}^\iter{}{i}$, then, because of independence between data estimates and error, the result in \eqref{eq:CovSPChEstErr} holds.
Here, $\rm{(c_1)}$ holds by taking the inverse of diagonal elements based on Lemma~\ref{lem:diaginv} in Appendix~\ref{app:matTh}, and the expressions in the second term of \eqref{eq:cov_chObs_dataSP} and first term in \eqref{eq:CorMatInt_dataSP} follow from Jensen's inequality. Notice that the term $\sf{b}_{lk}^\iter{}{i}$ in \eqref{eq:CovSPChEstErr} is also used to calculate the noise influence (i.e., the last term in  \eqref{eq:coCchObsDataAid}). For the intercell interference, let $\bf{D}_q = \rm{diag}(\qulsp_{l1},\ldots,\qulsp_{lK})$, then the result in \eqref{eq:Cov_intcell_1} holds
where $\rm{(d_1)}$ follows from discarding the cross products between pilot and data symbols (see the first term in \eqref{eq:Cov_intcell_1}), and $\rm{(d2)}$ holds by discarding the pilot interference that is caused by UEs that do not share the same pilot as $\rm{UE}_{lk}$. Then similarly to the result in \eqref{eq:CovSPChEstErr}, by applying Jensen's inequality to the expectation of the inverse diagonal elements of the Gramian of $\hat{\bf{X}}_l^\iter{}{i}$, the expression in the second term of \eqref{eq:CorMatInt_dataSP} is found.

\section{}
\label{app:matTh}
\begin{lemma}
\label{lem:diaginv}
Let $\bf{A}\! \in\! \bb{C}^{N \times N}$  be Hermitian (i.e., $\bf{A} = \bf{A}^H$) and positive semi-definite, then it holds that 
\addtocounter{equation}{5}
\begin{equation}
\left[\left(\bf{A}\right)^{-1}\right]_{kk} \geq  \frac{1}{\left[\bf{A}\right]_{kk}}.
\end{equation}
\end{lemma}
\begin{IEEEproof}
The matrix $\bf{A}$ can be expressed as 
$\bf{A} = 
\begin{bsmallmatrix}
a & \bf{b}^H\\
\bf{b} & \bf{C}
\end{bsmallmatrix}
$
where $a \in \bb{R}$,   $\bf{b} \in \bb{C}^{(N-1) \times 1}$, and $\bf{C} \in \bb{C}^{(N-1) \times (N-1)}$. By applying results from the inverse of a partitioned matrix and the Sherman–Morrison–Woodbury formula \cite[Ch.~0]{Horn2012Matrixbook} it follows that
\begin{align*}
\left[\left(\bf{A}\right)^{-1}\right]_{11} 
&= \left(a - \bf{b}^H\bf{C}^{-1}\bf{b}\right)^{-1} 
\\
&
 = \frac{1}{a} +\frac{1}{a^2} \bf{b}^H\left(\bf{C} - \frac{\bf{b}\bf{b}^H}{a}\right)^{-1}\bf{b}   \stackrel{\rm{(e_1)}}{\geq} \frac{1}{a}
\end{align*}
where $\rm{(e_1)}$ holds by discarding the second positive term since $\bf{C} - \frac{\bf{b}\bf{b}^H}{a} \succ 0$ given that $\bf{A}$ is Hermitian and positive semi-definite \cite[Ch.~7]{Horn2012Matrixbook}. Let $\bfs{\Pi} \in \bb{R}^{N\times N}$ be a permutation matrix that moves the $k^{th}$ row of a matrix towards the first position such that $\left[ \bf{A} \right]_{kk} = \left[\bfs{\Pi} \bf{A}\bfs{\Pi}^H\right]_{11}$ and $\bfs{\Pi}^{-1} = \bfs{\Pi}^H$, it follows that
\begin{align*}
\left[ \bf{A}^{-1} \right]_{kk} &= \left[\bfs{\Pi} \bf{A}^{-1}\bfs{\Pi}^H\right]_{11} = \left[\left(\bfs{\Pi} \bf{A}\bfs{\Pi}^H\right)^{-1}\right]_{11}
\\
&
\geq \frac{1}{\left[\bfs{\Pi} \bf{A}\bfs{\Pi}^H\right]_{11}} = \frac{1}{\left[ \bf{A} \right]_{kk}}.
\end{align*}
\end{IEEEproof}

\bibliographystyle{IEEEtran}

\begin{thebibliography}{10}
\providecommand{\url}[1]{#1}
\csname url@samestyle\endcsname
\providecommand{\newblock}{\relax}
\providecommand{\bibinfo}[2]{#2}
\providecommand{\BIBentrySTDinterwordspacing}{\spaceskip=0pt\relax}
\providecommand{\BIBentryALTinterwordstretchfactor}{4}
\providecommand{\BIBentryALTinterwordspacing}{\spaceskip=\fontdimen2\font plus
	\BIBentryALTinterwordstretchfactor\fontdimen3\font minus
	\fontdimen4\font\relax}
\providecommand{\BIBforeignlanguage}[2]{{%
		\expandafter\ifx\csname l@#1\endcsname\relax
		\typeout{** WARNING: IEEEtran.bst: No hyphenation pattern has been}%
		\typeout{** loaded for the language `#1'. Using the pattern for}%
		\typeout{** the default language instead.}%
		\else
		\language=\csname l@#1\endcsname
		\fi
		#2}}
\providecommand{\BIBdecl}{\relax}
\BIBdecl

\bibitem{Hien_bounds}
H.~Q. Ngo, E.~G. Larsson, and T.~L. Marzetta, ``Energy and spectral efficiency
of very large multiuser {MIMO} systems,'' \emph{IEEE Trans. Commun.},
vol.~61, no.~4, pp. 1436--1449, Apr. 2013.

\bibitem{16_Marzetta_MAMIMO_book}
T.~L. Marzetta, E.~G. Larsson, H.~Yang, and H.~Q. Ngo, \emph{Fundamentals of
	{M}assive {MIMO}}.\hskip 1em plus 0.5em minus 0.4em\relax Cambridge Press,
2016.

\bibitem{17_Bjornson_MAMIMO_book}
E.~Bj\"ornson, J.~Hoydis, and L.~Sanguinetti, ``Massive {MIMO} networks:
Spectral, energy, and hardware efficiency,'' \emph{Foundations and Trends®
	in Signal Processing}, vol.~11, no. 3-4, pp. 154--655, Nov. 2017.

\bibitem{Emil_pilot_SE}
E.~Bj\"{o}rnson, E.~Larsson, and M.~Debbah, ``Massive {MIMO} for maximal
spectral efficiency: How many users and pilots should be allocated?''
\emph{IEEE Trans. Wireless Commun.}, vol.~15, no.~2, pp. 1293--1308, Feb.
2016.

\bibitem{Mueller2014b}
R.~R. M\"{u}ller, L.~Cottatellucci, and M.~Vehkaper\"{a}, ``Blind pilot
decontamination,'' \emph{IEEE J. Sel. Topics Signal Process.}, vol.~8, no.~5,
pp. 773--786, Oct 2014.

\bibitem{Yin2016a}
H.~Yin, L.~Cottatellucci, D.~Gesbert, R.~R. M\"{u}ller, and G.~He, ``Robust
pilot decontamination based on joint angle and power domain discrimination,''
\emph{IEEE Trans. Signal Process.}, vol.~64, no.~11, pp. 2990--3003, 2016.

\bibitem{Julia_pilot_decont}
J.~Vinogradova, E.~Bj\"{o}rnson, and E.~G. Larsson, ``On the separability of
signal and interference-plus-noise subspaces in blind pilot
decontamination,'' in \emph{Proc. IEEE ICASSP}, Mar. 2016, pp. 3421--3425.

\bibitem{Hien_EVD_pilot}
H.~Q. Ngo and E.~G. Larsson, ``{EVD}-based channel estimation in multicell
multiuser {MIMO} systems with very large antenna arrays,'' in \emph{Proc.
	IEEE ICASSP}, Mar. 2012, pp. 3249--3252.

\bibitem{MAMI_low_nr_ant}
H.~Huh, G.~Caire, H.~C. Papadopoulos, and S.~A. Ramprashad, ``Achieving
"massive {MIMO}" spectral efficiency with a not-so-large number of
antennas,'' \emph{IEEE Trans. Wireless Commun.}, vol.~11, no.~9, pp.
3226--3239, Sep. 2012.

\bibitem{coord_ch_est}
H.~Yin, D.~Gesbert, M.~Filippou, and Y.~Liu, ``A coordinated approach to
channel estimation in large-scale multiple-antenna systems,'' \emph{IEEE J.
	Sel. Areas Commun.}, vol.~31, no.~2, pp. 264--273, Feb. 2013.

\bibitem{18_Bjornson_unlimited_cap}
E.~Bj{\"{o}}rnson, J.~Hoydis, and L.~Sanguinetti, ``Massive {MIMO} has
unlimited capacity,'' \emph{IEEE Trans. Wireless Commun.}, vol.~17, no.~1,
pp. 574--590, Jan. 2018.

\bibitem{T_marzetta_total_EE}
H.~Yang and T.~L. Marzetta, ``Total energy efficiency of cellular large scale
antenna system multiple access mobile networks,'' in \emph{Proc. IEEE
	OnlineGreenComm}, Oct. 2013, pp. 27--32.

\bibitem{T_marzetta_non_asym}
Y.~Li, Y.-H. Nam, B.~L. Ng, and J.~Zhang, ``A non-asymptotic throughput for
massive {MIMO} cellular uplink with pilot reuse,'' in \emph{Proc. IEEE
	GLOBECOM}, Dec. 2012, pp. 4500--4504.

\bibitem{Emil_pilot_cluster}
R.~Mochaourab, E.~Bj{\"{o}}rnson, and M.~Bengtsson, ``Adaptive pilot clustering
in heterogeneous massive {MIMO} networks,'' \emph{IEEE Trans. Wireless
	Commun.}, vol.~15, no.~8, pp. 5555--5568, Aug. 2016.

\bibitem{Hoeher99_ch_est_SP}
P.~Hoeher and F.~Tufvesson, ``Channel estimation with superimposed pilot
sequence,'' in \emph{Proc. IEEE GLOBECOM}, Dec. 1999, pp. 2162--2166.

\bibitem{SP_stat_fading_MIMO_2017}
A.~T. Asyhari and S.~ten Brink, ``Orthogonal or superimposed pilots? a
rate-efficient channel estimation strategy for stationary {MIMO} fading
channels,'' \emph{IEEE Trans. Wireless Commun.}, vol.~16, no.~5, pp.
2776--2789, May 2017.

\bibitem{SIP_part1_KU_SA}
K.~Upadhya, S.~A. Vorobyov, and M.~Vehkaper\"a, ``Superimposed pilots are
superior for mitigating pilot contamination in massive {MIMO},'' \emph{IEEE
	Trans. Signal Process.}, vol.~65, no.~11, pp. 2917--2932, Jun. 2017.

\bibitem{VT_SP_approx_2016}
H.~Zhang, S.~Gao, D.~Li, H.~Chen, and L.~Yang, ``On superimposed pilot for
channel estimation in multicell multiuser {MIMO} uplink: Large system
analysis,'' \emph{IEEE Trans. Veh. Technol.}, vol.~65, no.~3, pp. 1492--1505,
Mar. 2016.

\bibitem{Verenzuela2018a}
D.~Verenzuela, E.~Bj\"{o}rnson, and L.~Sanguinetti, ``Spectral and energy
efficiency of superimposed pilots in uplink massive {MIMO},'' \emph{IEEE
	Trans. Wireless Commun.}, vol.~17, no.~11, pp. 7099--7115, Nov 2018.

\bibitem{Zhu2003a}
H.~Zhu, B.~Farhang-Boroujeny, and C.~Schlegel, ``Pilot embedding for joint
channel estimation and data detection in {MIMO} communication systems,''
\emph{IEEE Commun. Letters}, vol.~7, no.~1, pp. 30--32, Jan 2003.

\bibitem{Khalighi2008a}
M.~A. Khalighi and S.~Bourennane, ``Semiblind single-carrier {MIMO} channel
estimation using overlay pilots,'' \emph{IEEE Trans. Veh. Technol.}, vol.~57,
no.~3, pp. 1951--1956, May 2008.

\bibitem{Ma2014a}
J.~Ma and L.~Ping, ``Data-aided channel estimation in large antenna systems,''
\emph{IEEE Trans. Signal Process.}, vol.~62, no.~12, pp. 3111--3124, Jun
2014.

\bibitem{D_Tse_wireless_book}
D.~Tse and P.~Viswanath, \emph{Fundamentals of wireless Communication}.\hskip
1em plus 0.5em minus 0.4em\relax Cambridge Press, 2005.

\bibitem{Ma2017a_MultiUE}
J.~{Ma}, C.~{Liang}, C.~{Xu}, and L.~{Ping}, ``On orthogonal and superimposed
pilot schemes in massive {MIMO} {NOMA} systems,'' \emph{IEEE J. Sel Areas
	Commun.}, vol.~35, no.~12, pp. 2696--2707, Dec 2017.

\bibitem{S_Jacobsson_Durisi_ADCs_UL}
S.~Jacobsson, G.~Durisi, M.~Coldrey, U.~Gustavsson, and C.~Studer, ``Throughput
analysis of massive {MIMO} uplink with low-resolution {ADC}s,'' \emph{IEEE
	Trans. Wireless Commun.}, vol.~16, no.~6, pp. 4038--4051, Jun. 2017.

\bibitem{Bjorson2019_MaMIMO_2}
\BIBentryALTinterwordspacing
L.~{Sanguinetti}, E.~{Bj{\"o}rnson}, and J.~{Hoydis}, ``Towards massive {MIMO}
2.0: Understanding spatial correlation, interference suppression, and pilot
contamination,'' \emph{arXiv e-prints}, Apr. 2019. [Online]. Available:
\url{https://arxiv.org/abs/1904.03406}
\BIBentrySTDinterwordspacing

\bibitem{steve_M_kay}
S.~M. Kay, \emph{Fundamentals of Statistical Signal Processing: Estimation
	Theory}.\hskip 1em plus 0.5em minus 0.4em\relax Prentice Hall, 1993.

\bibitem{Mollen2016a}
C.~Moll\'{e}n, E.~G. Larsson, and T.~Eriksson, ``Waveforms for the massive
{MIMO} downlink: Amplifier efficiency, distortion, and performance,''
\emph{IEEE Trans. Commun.}, vol.~64, no.~12, pp. 5050--5063, Dec 2016.

\bibitem{Morelli2001a}
M.~{Morelli} and U.~{Mengali}, ``A comparison of pilot-aided channel estimation
methods for {OFDM} systems,'' \emph{IEEE Trans. Signal Process.}, vol.~49,
no.~12, pp. 3065--3073, Dec. 2001.

\bibitem{VanderPerre2018_DSPMaMIMO}
L.~{Van der Perre}, L.~{Liu}, and E.~G. {Larsson}, ``Efficient {DSP} and
circuit architectures for massive {MIMO}: State of the art and future
directions,'' \emph{IEEE Trans. Signal Process.}, vol.~66, no.~18, pp.
4717--4736, Sep. 2018.

\bibitem{Malkowsky2017_LuMaMIMO}
S.~{Malkowsky}, J.~{Vieira}, L.~{Liu}, P.~{Harris}, K.~{Nieman}, N.~{Kundargi},
I.~C. {Wong}, F.~{Tufvesson}, V.~{Öwall}, and O.~{Edfors}, ``The world’s
first real-time testbed for massive {MIMO}: Design, implementation, and
validation,'' \emph{IEEE Access}, vol.~5, pp. 9073--9088, 2017.

\bibitem{Malkowsky2016_Lat_LuMaMi}
S.~{Malkowsky}, J.~{Vieira}, K.~{Nieman}, N.~{Kundargi}, I.~{Wong},
V.~{Öwall}, O.~{Edfors}, F.~{Tufvesson}, and L.~{Liu}, ``Implementation of
low-latency signal processing and data shuffling for {TDD} massive {MIMO}
systems,'' in \emph{IEEE Proc. SiPS}, Oct 2016, pp. 260--265.

\bibitem{SPPwctrl_Verenzuela2018a}
D.~{Verenzuela}, A.~{Bergström}, and E.~{Björnson}, ``Optimal power control
for superimposed pilots in uplink massive {MIMO} systems,'' in \emph{Proc.
	52nd Asilomar}, Oct. 2018, pp. 499--503.

\bibitem{LDPC_NR_3gpp}
``{NR} multiplexing and channel coding (release 15),'' Tech. Rep., 2018, 3GPP
TS 38.212.

\bibitem{Horn2012Matrixbook}
R.~A. Horn and C.~R. Johnson, \emph{Matrix Analysis}.\hskip 1em plus 0.5em
minus 0.4em\relax Cambridge Press, 2012.

\end{thebibliography}

\end{document}